\documentclass[aps,prb,10pt,twocolumn,superscriptaddress]{revtex4-1}

\usepackage{amsmath,amssymb,times,bbm,xcolor,mathtools,subfigure,graphicx,hyperref}
\graphicspath{{./Figures/}}
\hypersetup{breaklinks, colorlinks=false,
	pdftitle={Experimental Quantum Compressed Sensing for a 7-Qubit System},
}

\definecolor{jens}{rgb}{.2,0.7,.9}

\begin{document}

\title{Experimental quantum compressed sensing for a seven-qubit system}

\author{C.\ A.\ Riofr\'io}
\affiliation{Dahlem Center for Complex Quantum Systems, Freie Universit{\"a}t Berlin, D-14195 Berlin, Germany}

\author{D.\ Gross}
\affiliation{Institute for Theoretical Physics, University of Cologne, D-50937 Cologne, Germany}

\author{S.\ T.\ Flammia}
\affiliation{Centre for Engineered Quantum Systems, School of Physics,
The University of Sydney, Sydney, NSW, Australia}

\author{T.\ Monz}
\author{D.\ Nigg}
\author{R.\ Blatt}
\affiliation{Institut f{\"u}r Experimentalphysik, Universit{\"a}t Innsbruck, Technikerstrasse 25, A-6020 Innsbruck, Austria}

\author{J.\ Eisert}
\affiliation{Dahlem Center for Complex Quantum Systems, Freie Universit{\"a}t Berlin, D-14195 Berlin, Germany}

\date{\today}

\begin{abstract}
	Well-controlled quantum devices with their increasing system size 
	face a new roadblock hindering further development of quantum technologies: 
	The effort of quantum tomography---the characterization of
	processes and states within a quantum device---scales
	unfavorably to the point that state-of-the-art systems
	can no longer be treated. 
	Quantum compressed sensing mitigates this problem by reconstructing
	the state from an incomplete set of observables.
	In this work, we present an experimental implementation of
	compressed tomography of a seven qubit system---the largest-scale
	realization to date---and we introduce new 
	numerical methods in order to scale the reconstruction
	to this dimension. 
	Originally, compressed sensing has been advocated
	for density matrices with few non-zero eigenvalues. 
	Here, we argue that the low-rank estimates provided by compressed
	sensing can be appropriate even in the general case. The reason is that
	statistical noise often allows only for the leading eigenvectors to
	be reliably reconstructed: We find that the remaining eigenvectors
	behave in a way consistent with a random matrix model that carries
	no information about the true state. 
	We report a reconstruction of
	quantum states from a topological color code of seven qubits,
	prepared in a trapped ion architecture, based on tomographically
	incomplete data involving $127$ Pauli basis measurement settings
	only, repeated $100$ times each. 
\end{abstract}
\maketitle


Recent years have seen rapid progress in the development of quantum
technologies, with precisely controlled quantum systems reaching ever
larger system sizes.  Specifically, for systems of trapped ions,
precisely controlled arrays of tens and more individual ions have been
engineered and manipulated in their quantum
state,\cite{BlattColorCode, Lanyon_etal11, IonCrystal, BlattSimulator}
while architectures such as superconducting qubits
\cite{Superconducting, Ofek2016} and neutral
atoms,\cite{Saffman2015,Browaeys} among many others, are also
developing rapidly.  These technological and scientific developments
have enabled 
implementations of small-scale quantum
simulators,\cite{Lanyon_etal11, IonCrystal, BlattSimulator} small
measurement-based quantum computations,\cite{MBQCIon}, 
proof-of-principle gate-based quantum computations,\cite{Lieven,
Andreas, BlattColorCode, DeutschJozsa}
and
quantum error correction, e.g.\ based on topological color
codes.\cite{BlattColorCode} 

As a result of this fast development, a new roadblock is an increasing 
concern: The fact that the Hilbert space dimension scales
exponentially 
means that traditional methods for the experimental characterization
of the processes and states that have been implemented becomes
infeasible even for intermediate system sizes. This is problematic,
since such systems are relevant as building blocks for emerging quantum
technologies. 
To mitigate this problem, it has been suggested to use various
structural properties of natural quantum systems---e.g.\ high purity,
symmetries, sparsity in a known basis, or entanglement area laws---in
order to reduce the effort of characterization.\cite{Compressed,
MPSTomo, Compressed2, FidelityEstimation, WhiteCompressedSensing,
Wick, QuantumFieldTomography}

In this work, we demonstrate that this approach is reaching maturity
by implementing an experimental reconstruction of the state of a
7-qubit system from an informationally incomplete set of measurements.
To achieve this, we are relying on the technique of \emph{compressed
sensing}. This theory has emerged over the past decade in the field of
classical data analysis.\cite{CompressedSensingIntroCandes, CompressedSensingIntroRauhut} It is now routinely used to estimate vectors or matrices from
incomplete information, with manifold applications in such diverse
fields as image processing, seismology, wireless communication, and
many more.\cite{CompressedSensingIntroRauhut, Eldar2012} Compressed
sensing for low-rank matrices has been adapted as a tool for quantum
system characterization (also referred to as \emph{quantum
tomography}) in a series of works.\cite{Compressed, Compressed2,
KuengRauhut}  A particularly appealing feature of \emph{compressed quantum
tomography} (the combination of compressed sensing and quantum tomography)
is the fact that there is no need to make any \emph{a
priori} assumptions about the true quantum state.\cite{Carpentier2015,
Compressed2} 

Quantum compressed sensing is most effective on 
density matrices with quickly decaying eigenvalues. Such a matrix can
be well-approximated by one having a rank $r$ that is much smaller than the
dimension $d$ of the Hilbert space. A rank-$r$ matrix depends on 
only $O(rd)$ parameters, significantly fewer than the $d^2$
parameters required in general. In quantum information experiments,
the goal is often to prepare a pure state, described by a rank-1
density operator. Noise effects will typically require one to include
more than just one eigenvalue to obtain a good approximation of the
true density matrix. However, in highly controlled experiments, the number of 
additional eigenvalues required to obtain an accurate state estimate 
is expected to be small. In this context, the theory of compressed sensing
showed for the first time that the reduced number of parameters is
reflected in a reduced effort in both measurements and computation
required for tomographic reconstruction. Indeed, it has been
rigorously shown that an (approximate) rank-$r$ density matrix can be
recovered from $O(r d \log^2d)$ experimentally measured parameters.\cite{Compressed} 
This performance -- close to the absolute lower bound of $O(rd)$ 
--
can even be achieved when the eigenbasis is completely unknown.\cite{Compressed}
A variety of computationally efficient estimators
have been proposed to achieve recovery in practice, and we will
revisit this topic in more detail below when we describe the
numerical implementation we have used for this experiment.

While important steps towards quantum compressed sensing protocols have been implemented
before,\cite{CompressedMartinis, CompressedWeinfurter,
WhiteCompressedSensing} we report here the first implementation of
compressed state reconstruction in an intermediate-sized quantum
system. For the purposes of this work, we refer to a quantum system as
being intermediate-sized if it has $5$--$10$ physical qubits. 
This is the range where quantum error correction of one- and two-qubit logical 
gates becomes possible and full state reconstruction methods are most useful.
We do so on the basis of a platform of seven trapped ions, which are
prepared in a state of a topological color code.\cite{Bombin2006}

A secondary objective of this work is to argue that compressed
tomography, while originally developed for density matrices with a
small number of dominating eigenvalues,
can also be appropriate in situations where 
the unknown true density matrix is not, in fact, of low rank. This
counterintuitive conclusion follows from the finding that in
realistic regimes, the statistical signal-to-noise ratio is such
that only the leading eigenvectors of the density matrix can be
reliably reconstructed. Indeed, we find that the tail of least-significant eigenvectors behaves in
ways consistent with a random matrix model, which means that
reporting more than the first few eigenvectors 
reveals no information about the true state and thus
amounts to overfitting. 
To make this insight more concrete, we formulate a task 
that is very much reminiscent of 
\emph{support identification} in compressed sensing, referring to 
the problem of deciding which estimated eigenspaces should be included in an
estimate (see, e.g., ref.\ \cite{SI}), which we may call \emph{quantum support identification}. 
We give heuristics for identifying the relevant support,
based on comparing the behavior of the estimate with a random matrix
model. Our findings here are consistent with a recent approach that
recommends \emph{spectral thresholding} for statistical 
reasons;\cite{guta} and another work that shows that statistical noise in state reconstruction
protocols can manifest itself by giving rise to random-matrix like
behavior.\cite{Knips2015} 

We also observe that the estimators introduced in the context of
compressed sensing reconstruct the leading eigenvectors more
faithfully than more traditional approaches, at the price of being
less faithful on the spectral tail. 
This suggests that one should employ the former if one is more
interested in learning \emph{coherent errors} (i.e.\ the way in which
the first eigenvector deviates from its target), while the latter are
better-suited to analyze \emph{incoherent noise processes} that drive
up the rank. It should be noted that these observations are
qualitative rather than mathematically precise at this point.

The rest of this work is organized as follows: We first discuss the physical system at hand and 
state how the raw data is obtained. We then introduce the estimators used to recover the density
operator. Subsequently, we turn to the discussion of quantum identification and to what is called
model selection in the literature. We here also discuss the main results of this work and elaborate on the
outcomes of the actual reconstruction from experimental data. We conclude by presenting 
further perspective arising from our approach.

\subsection*{Physical system and raw data}

We begin by explaining the physical architecture of trapped ions that
serves as the platform for this endeavor. In the considered ion-trap
quantum computer, $^{40}$Ca$^+$ ions are stored in a linear Paul trap.
Each physical qubit is encoded in $S_{1/2}(m=-1/2)=\lvert 1\rangle$ and the metastable,
excited state corresponding to $D_{5/2}(m=-1/2)=\lvert 0 \rangle$. Manipulation of the qubit is
performed by laser pulses resonant (or close to resonant) to the atomic
transitions of $^{40}$Ca$^+$. The universal set of quantum
gates is implemented using three types of operations: 
collective operations of the form 
$\exp\bigl(-i ({\theta}/{2}) S_\phi\bigr)$ with 
\begin{equation}
	S_\phi = \sum_{l=1}^L  \bigl(\cos(\phi)X_l +
	\sin(\phi)Y_l \bigr)\,,
\end{equation}
and entangling operations of the form $\exp\bigl(-i ({\theta}/{4}) S_\phi^2\bigr)$,
reflecting the entangling M{\o}lmer-S{\o}renson interaction.\cite{ms_gate} 
Here $X_l, Y_l, Z_l$ are the Pauli
operators of qubit $l$, $\theta=\Omega t$ is determined by the Rabi
frequency $\Omega$ and laser pulse duration $t>0$, and $\phi$ is determined by
the relative phase between qubit and laser. The third type of operations are generated by single qubit phase
rotations induced by localized AC-Stark shifts. More details of this experimental setup are
covered in ref.~\citenum{order_finding_phips}.

Within this experimental setting involving $L=7$ qubits, 
quantum states have been prepared to the best of the experimental
knowledge which, however, is limited by statistical noise and systematic
errors.  The quantum states are described mathematically by 
density operators $\rho\in \mathbbm{H}_d(\mathbbm{C})$ 
(Hermitian $d\times d$ matrices) for $d=2^L$
that satisfy $\mathrm{tr}(\rho)=1$ and $\rho\geq 0$. In all of the experiments,
the aim was to prepare a pure state vector which is contained in the code 
space, which is a two-dimensional subspace of the Hilbert space of seven qubits spanned by 
$|\bar{0}\rangle$ and $|\bar{1}\rangle$. Here, the state vectors $|\bar{0}\rangle$ and 
$|\bar{1}\rangle$ span the code space and are joint $+1$ eigenstates of 
the set of stabilizer operators that define the code. 
The stabilizer operators are given explicitly in ref.~\citenum{BlattColorCode}. 
The particular basis for the code space is chosen by 
picking $|\bar{0}\rangle$ and $|\bar{1}\rangle$ to be the eigenvectors of 
$Z_1\otimes \dots \otimes Z_L$ with eigenvalues $+1$ and $-1$ respectively. 
The states that the ideal experiment would prepare will be referred to as
\emph{anticipated states} in what follows.
Both $|\bar{0}\rangle$ and $|\bar{1}\rangle $ are code words of a Calderbank-Shor-Steane code~\cite{CalderbankShor,SteaneCode}
originating from the theory of quantum error correction designed to protect fragile quantum information against unwanted 
local noise. At the same time, they can be seen as the smallest fully
functional instances of a \emph{topological color
code},\cite{Bombin2006} which are topological quantum error-correcting
codes defined on physical systems supported on two-dimensional
lattices. 

For each state, a set of $n=127$ Pauli basis \emph{measurement
settings} is chosen. (An informationally complete set would contain
$3^L=2187$ settings).
Each measurement setting $j$ is characterized by a choice of a local
Pauli matrix
\begin{equation} 
	W_l^{(j)}\in \{X_l,Y_l,Z_l\}, \qquad l=1,\dots, L. 
\end{equation}
for each of the $L$ qubits. The $l$th qubit is measured in the
eigenbasis of $W_l^{(j)}$. There are two possible outcomes for each
qubit, and therefore a total of $2^L$ possible outcomes per
experiment.
Each specific outcome $k$ is associated with a projection operator
\begin{equation} 
	P^{(j)}_k = |v_k^{(j)}\rangle \langle v_k^{(j)}|, \qquad k=1,
	\dots 2^L,
\end{equation} 
where $|v_k^{(j)}\rangle$ is a tensor product of eigenvectors of the
$W^{(j)}_l$.

For each measurement setting, the
measurement is repeated $m=100$ times and the statistics of measurement
outcomes is recorded. From the relative frequencies of outcomes $k$, 
the probability
$\mathrm{tr}(\rho P^{(j)}_k )$ is estimated.
Because of the relatively small number of repetitions of the measurements per setting, given $2^L$ potential outcomes,
many of the possible outcomes will not appear even once. 

Let us denote the
measurement settings that have been chosen as
$V\subset W$, where $W$ is the set of all possible measurement settings. 
We define the \emph{sampling operator}
$\mathcal{A}:\mathbbm{H}_d(\mathbbm{C})\rightarrow \mathbbm{R}^{n d}$ as
\begin{equation}\label{eq:PauliBasisMeasurementMap}
	\mathcal{A}(\rho) = \bigl(\mathrm{tr}(\rho P^{(1)}_1), \mathrm{tr}(\rho P^{(1)}_2), \dots, \mathrm{tr}(\rho P^{(n)}_d )\bigr),
\end{equation}
with $n= |V| $ the number of chosen settings.
That is, the sampling operator is the linear map that simply returns
the list of expectation values of the observables
$P^{(j)}_k$ measured in the state $\rho$.
The data taken are of the type
\begin{equation}
	\mathbf{y}= \mathcal{A}(\rho)+\mathbf{z}(\rho),
\end{equation}
where the zero-mean random vector $\mathbf{z}(\rho)$ captures the statistical noise.
The outcomes for any given basis follow a multinomial distribution,
from which one obtains the expression
\begin{equation}
	\frac{1}{m} \mathrm{tr}(\rho P^{(j)}_k) \bigl(1- \mathrm{tr}(\rho P^{(j)}_k) \bigr) ,
\end{equation}
for the second moment of each given component of $\mathbf{y}$.

For completeness, we note that the \emph{Pauli basis measurements}
considered here differ from the \emph{Pauli correlation measuemrents}
that were the basis of some previous works on compressed
sensing.\cite{Compressed} Pauli correlation measurements are of the
form 
$\mathrm{tr}\bigl(\rho (W_{1}^{(j)}\otimes \dots \otimes
W_L^{(j)})\bigr)$, where again the $W_l^{(j)}$ are Pauli matrices
acting on the $l$th qubit. These correlators associate \emph{one}
expectation value with each choice of local Pauli matrices and appear
e.g.\ as syndrome measurements in quantum error correction. As
detailed above, the \emph{basis measurements} yield $2^L$ parameters
per choice of local Pauli matrices. This is the number of ways of
picking one of the two eigenvectors of each Pauli matrix. Basis
measurements, which thus give much more detailed information per
setting, appear naturally in the ion trap architecture used for this
work.
One can recover Pauli correlations from basis measurements via the
relationship
\begin{equation}
	\mathrm{tr}\bigl(\rho (W_{1}^{(j)}\otimes \dots \otimes W_L^{(j)})\bigr) = \sum_{k=1}^{d}
	(-1)^{\chi(k)}
	\mathrm{tr}(\rho P_k^{(j)}),
\end{equation}
where $\chi(k)$ denotes the parity of the binary 
representation of the integer $k$.

\subsection*{Estimators and state reconstruction}

In statistics, an \emph{estimator} is a rule for 
mapping observed data 
(here, outcomes ${\mathbf y}$) to an estimate for an unknown
quantity (here, a density matrix $\rho$).
At the heart of the discussion here is an
estimator that is particularly common in the compressed sensing
literature. This is the so-called
\emph{matrix Lasso}~\cite{Compressed2} defined as
 \begin{equation}\label{eq:lasso}
 	\min_{X}\,  \|\mathbf{y}-\mathcal{A}(X)\|_2^2 + \mu \|X\|_*,
\end{equation}
where $\mu \geq 0$ is a regularization parameter and $\|.\|_*$ is
the nuclear or matrix trace norm. 
(The trace norm can be shown to be the tightest convex relaxation of
``rank''. Thus, the regularization term both encourages low-rank
solutions and, due to convexity, can be minimized efficiently 
\cite{CompressedSensingIntroRauhut}).
The estimator does make sense for $\mu=0$ if one adds the additional
constraint that the result be positive
semi-definite.\cite{KuengRauhut}
This \emph{positivity-constrained least squares
estimator} will be at the focus of attention in our approach. From its implementation,
we expect 
this estimator
to be specifically suited to the regime of intermediate and large
quantum systems and comparably little data in which we are interested.

While the estimator from eq.\ (\ref{eq:lasso}) is formally efficient in the
sense that it can be solved in polynomial time in the size of the input,
additional theoretical efforts are required to arrive at an
implementation that performs well in practice in the regime of
intermediate and large quantum systems.
To achieve this, we here introduce a tailor-made implementation for
the $\mu=0$ case.
In this approach, we parametrize the quantum state $\rho$ as
\begin{equation}
	\rho=	Q^\dagger Q\,,
\end{equation}
for some $r\times d$ complex matrix $Q$, where $r$ controls the rank of $\rho$. 
We then consider
\begin{equation}\label{eq:optimization}
\min_{Q}\,  g(Q)=\min_{Q}\,  \|\mathbf{y}-\mathcal{A}(Q^\dagger Q)\|^2_2,
\end{equation}
with $\|.\|_2$ is the vector $2$-norm.

Using this parameterization of $\rho$ ensures that it is positive
semidefinite by construction. 
The optimization problem itself is then solved using a gradient method. 
A gradient flow on the basis of $\rho$ directly would introduce
negative eigenvalues in every step. In contrast, we 
optimize over $Q$,
using the fact that we can analytically compute the gradient 
\begin{equation}
	\nabla_{Q}g(Q) = 4Q\mathcal{A}^\dagger\bigl(\mathcal{A}(Q^\dagger Q)-\mathbf{y}\bigr),
\end{equation}
of the objective function in eq.~(\ref{eq:optimization}). This way, we dispense with the unnecessary
and computationally expensive projection step that 
would otherwise be needed to enforce positivity.
 
This simplification significantly improves the computational effort 
as compared to earlier estimators that made use of an iterative 
gradient method based on $\rho$. We refer to this 
\emph{gradient method based on a manifestly positive parametrization of states}
as \texttt{GRAD}. We present details of this algorithm in the appendix.

\begin{figure*}[t]
\subfigure[\, Trace norm minimization estimate ($F = 0.98$),
corresponding to the
result of the matrix Lasso in eq.~\eqref{eq:lasso} in the limit $\mu\rightarrow \infty$.]{%
  \includegraphics[width=.65\columnwidth]{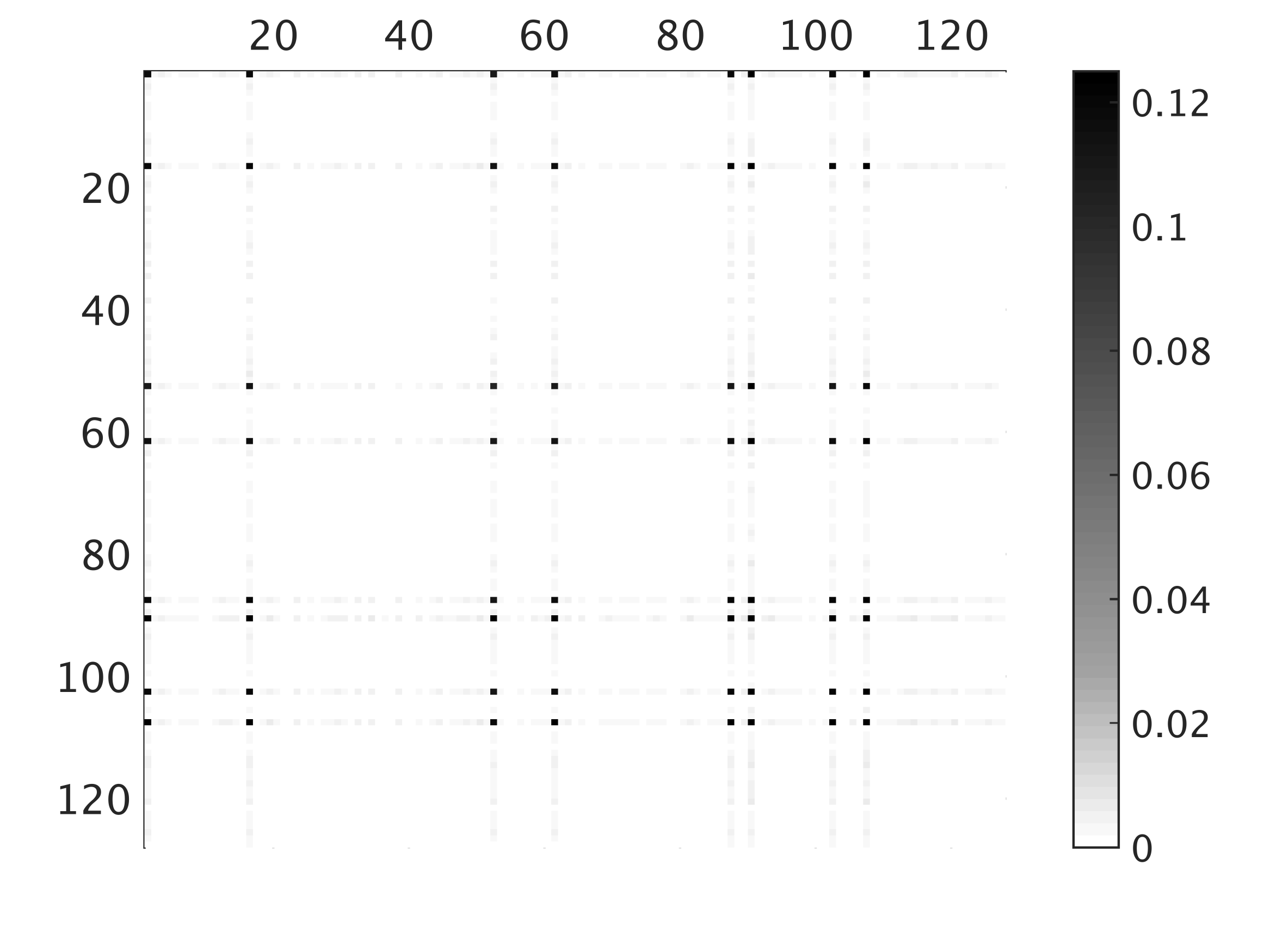}}
\subfigure[ \, Least squares estimate ($F = 0.30$), corresponding to eq.~\eqref{eq:lasso} with $\mu=0$.]{%
  \includegraphics[width=.65\columnwidth]{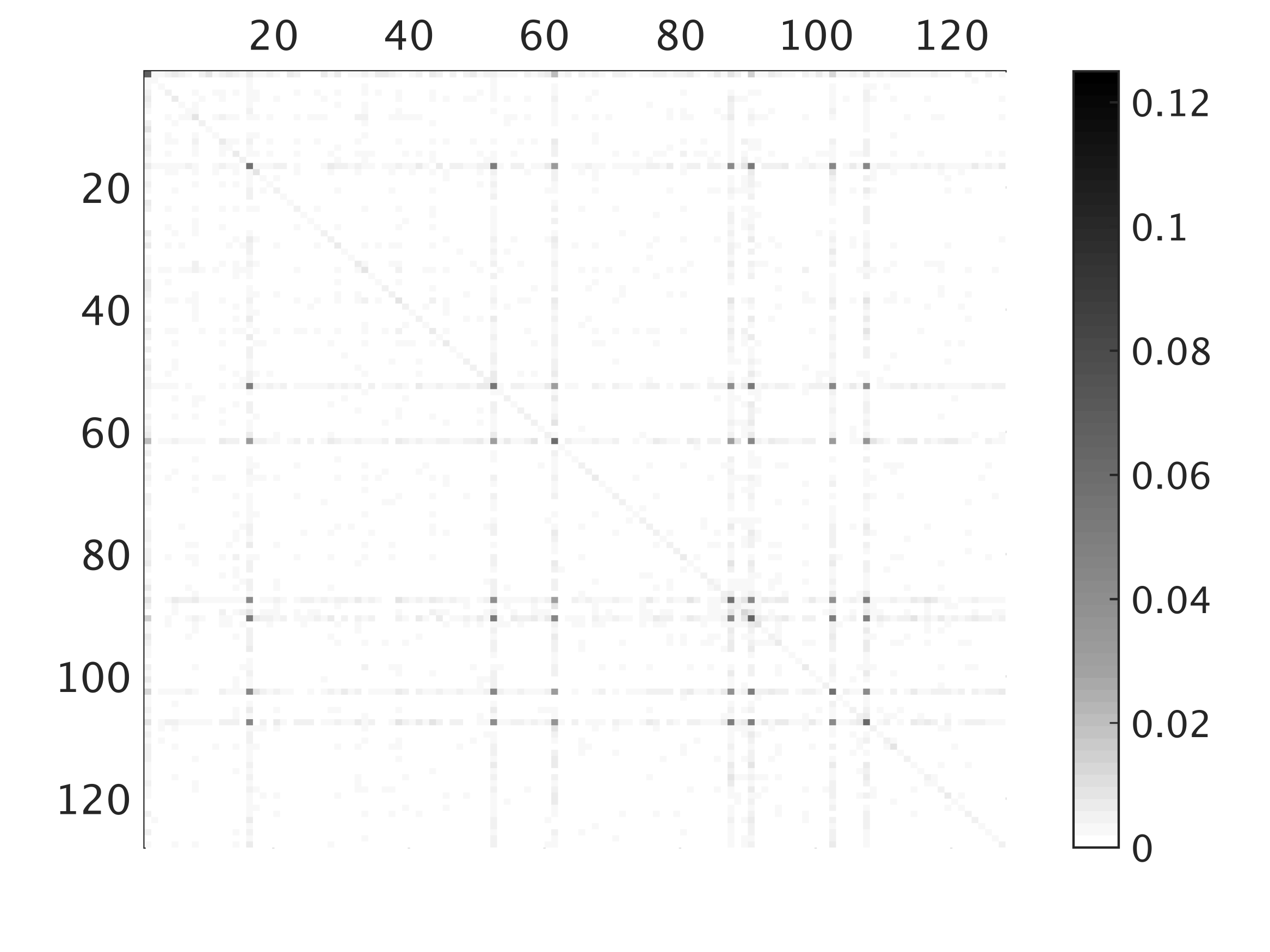}}
\quad
\subfigure[\, Rank $21$ leading subspace projection of the least squares estimate ($F = 0.32$) obtained by our 
spectral thresholding method.]{%
  \includegraphics[width=.65\columnwidth]{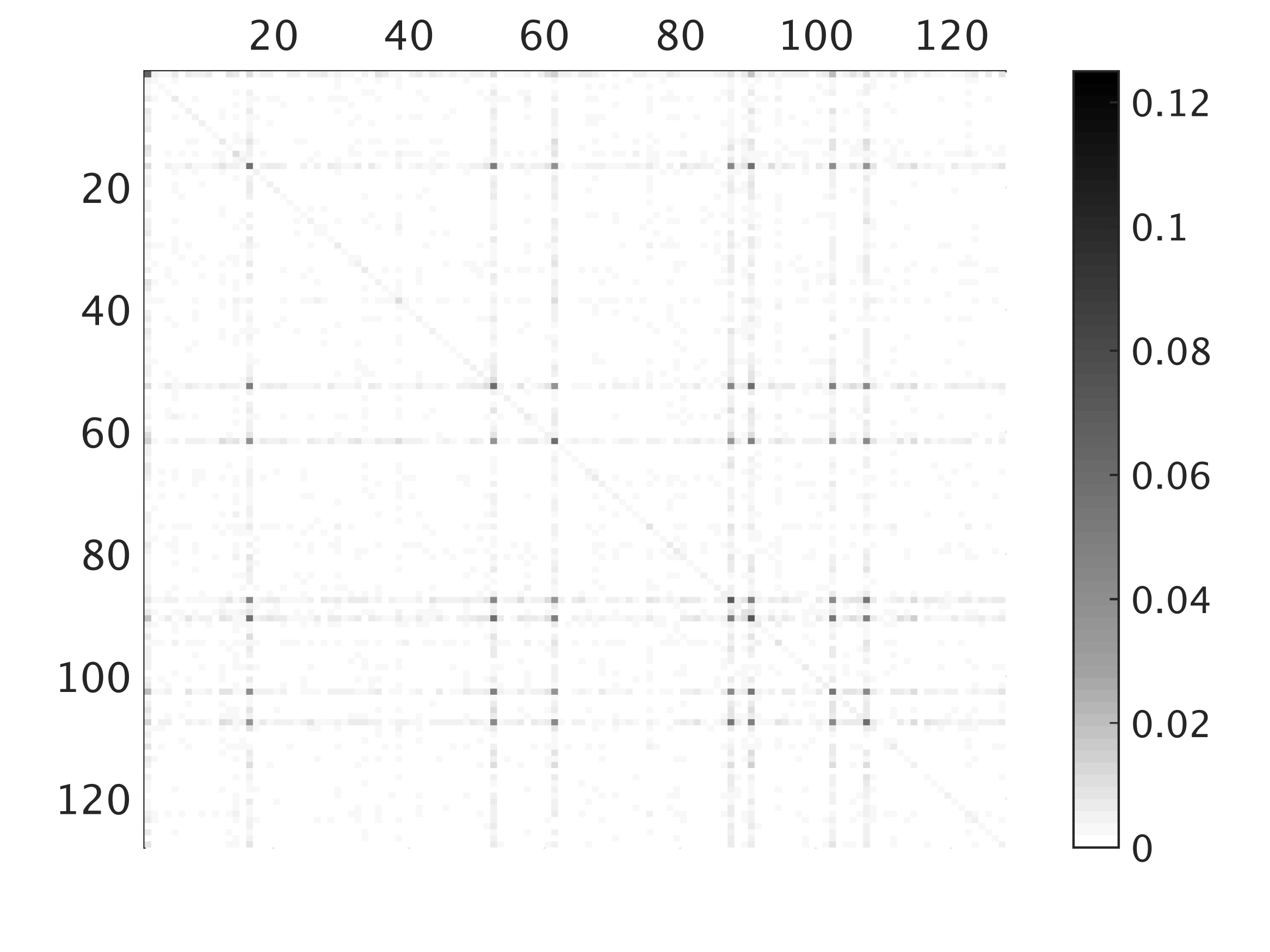}}
\caption{Example of quantum state reconstruction for the logical $|\bar{0}\rangle$
state vector. The plots are 2-D plots of the absolute values of the entries of the density matrix 
in the standard basis with magnitude represented by the grey scale. The axes are labeled by 
the computational basis vectors. 
For reasons of clarity, the basis vectors are numbered as 
$x\in\{1,2,\ldots,d\}$, where $|\chi(x-1)\rangle$ is the state vector in the standard 
computational basis, $\chi(x-1)$ being the binary representation of 
$x-1$. So $x=1$ corresponds to $|0,\dots, 0,0\rangle$, $x=2$ to  $|0,\dots, 0,1\rangle$ 
and so on. 
The performance of the reconstruction is measured by the fidelity 
$F = \langle\bar{0}|\hat{\rho}|\bar{0}\rangle$, where $\hat{\rho}$ is the estimated state. 
While all three estimators produce roughly similar looking estimates, they differ greatly in the fidelity with the anticipated state.}
\label{Data1}
\end{figure*}

The \texttt{GRAD} method allowed us to analyze data from informationally
incomplete measurements on a $7$-qubit trapped ion experiment.
In each experiment we have performed, the \emph{anticipated states}
were taken from the code space
$\mathrm{span}(|\bar{0}\rangle,|\bar{1}\rangle)$ of the topological
color code. In fig.\ \ref{Data1},
we present a graphical representation of an instance of such a
reconstruction. It depicts the anticipated state, the
reconstructed one based on the matrix Lasso from eq.\
\eqref{eq:lasso} with $\mu\to \infty$, 
the one obtained by positivity-constrained least squares
(using $\mu=0$), and an estimator that we call \emph{spectral thresholding} estimator, 
to be discussed in more detail below and for reasons that will become clear then, 
where only the highest eigenvalues have been kept.

From Fig.\ \ref{Data1} we see that the estimators give
rise to valid and faithful reconstructions of the anticipated state, in that the
reconstructed states are close in fidelity to the anticipated states,
though some estimators report higher fidelities than others. 
The computational runtime for the estimators is moreover quite modest, or the order of a few hours,
and of a few minutes for the \texttt{GRAD} estimator. This analysis can be seen as a first experimental implementation
of quantum state tomography based on a compressed sensing methodology
for high-dimensional quantum systems. 

If we compare a typical figure of merit for the quality of a state reconstruction---the fidelity to 
the anticipated state---then we notice the glaring feature that the three reported 
reconstruction fidelities differ from $F=0.98$ (trace norm minimization estimate) down to $F=0.30$ (least squares estimate) 
on the same data. We hypothesize that this difference is due to a combination of limited data (applicable to all estimators) 
and the fact that the Lasso estimator with $\mu\to \infty$ (the trace norm minimization estimate) gives a much higher penalty to mixed states than the other two 
estimators. Because this trace norm minimization estimator will favor reconstructing nearly 
pure states, it will not faithfully estimate the tail of the spectrum when data is scarce, but it 
is expected to be better at diagnosing \emph{coherent} errors, as it will return the dominant pure state. 
By contrast, the other estimators sacrifice purity to better match the spectrum, which causes them to have a poor fidelity with 
the anticipated state. Thus, these estimators might be better for diagnosing 
\emph{incoherent} noise, as these methods retain visibility for excessive noise in the system.

\begin{figure*}[t]
\subfigure[\,Trace norm estimate.]{%
  \includegraphics[width=.65\columnwidth]{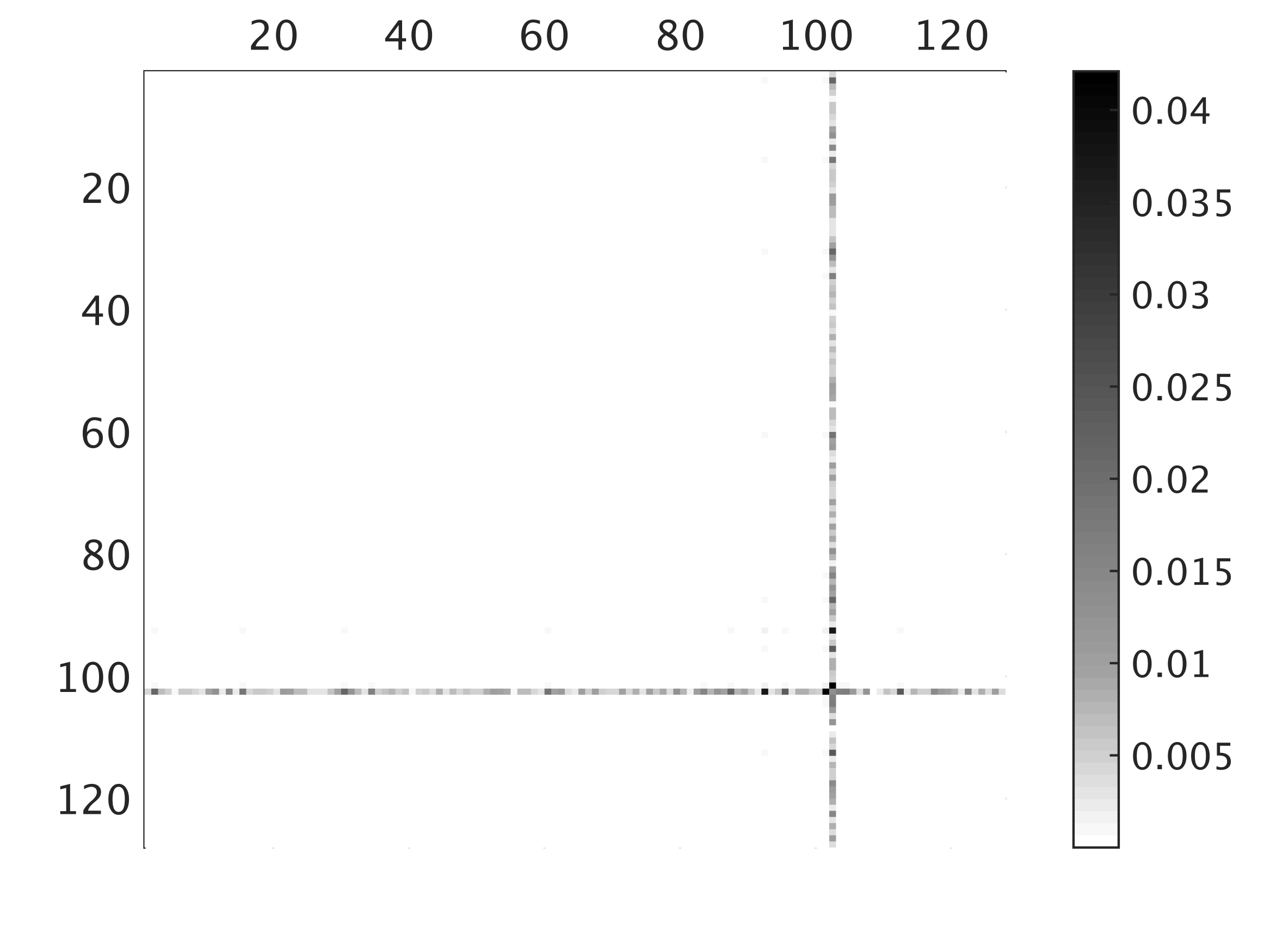}}
\quad
\subfigure[\, Least squares estimate.]{%
  \includegraphics[width=.65\columnwidth]{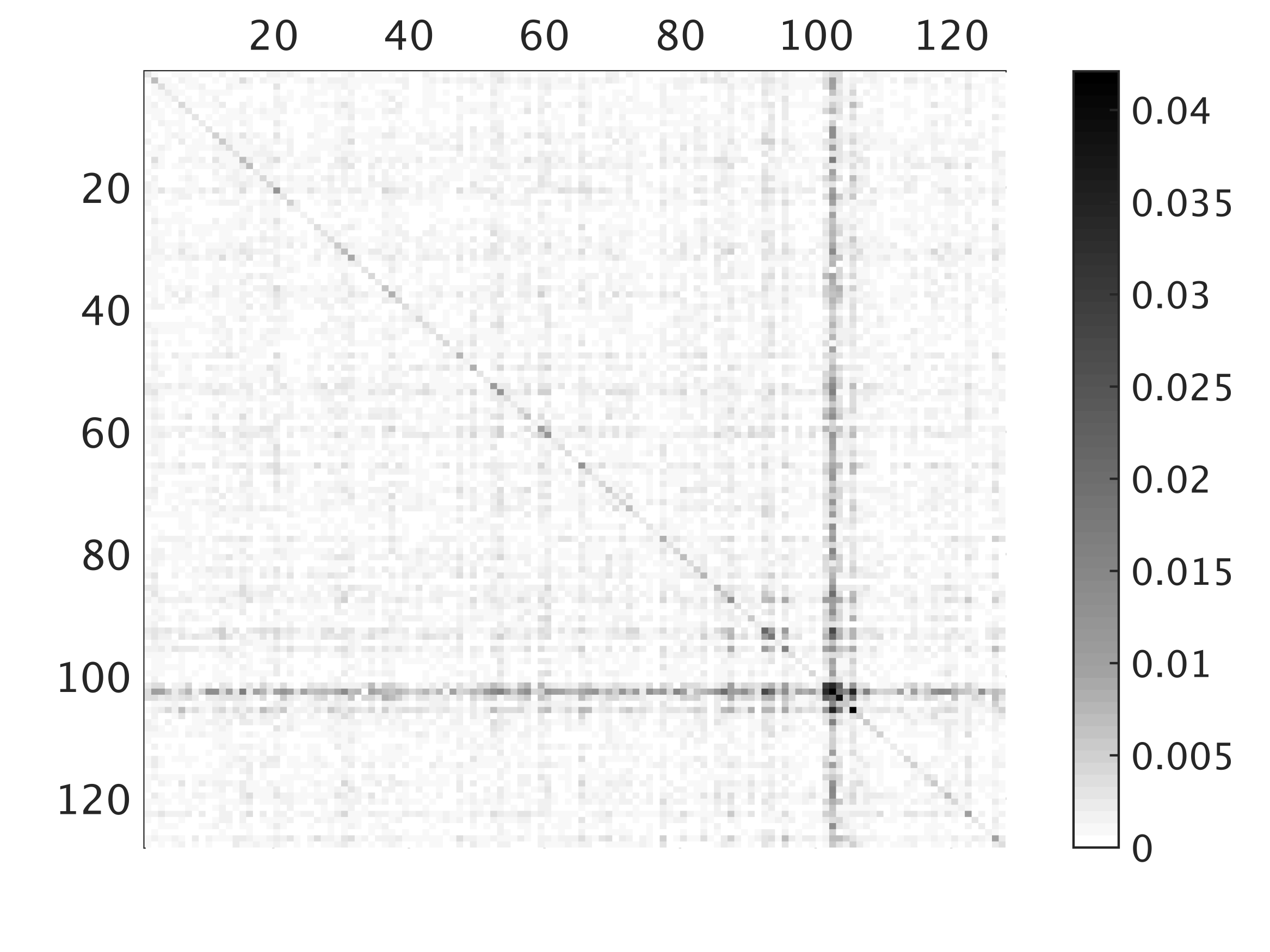}}
\quad
\subfigure[\, Diagonal element comparison.]{%
  \includegraphics[width=.65\columnwidth]{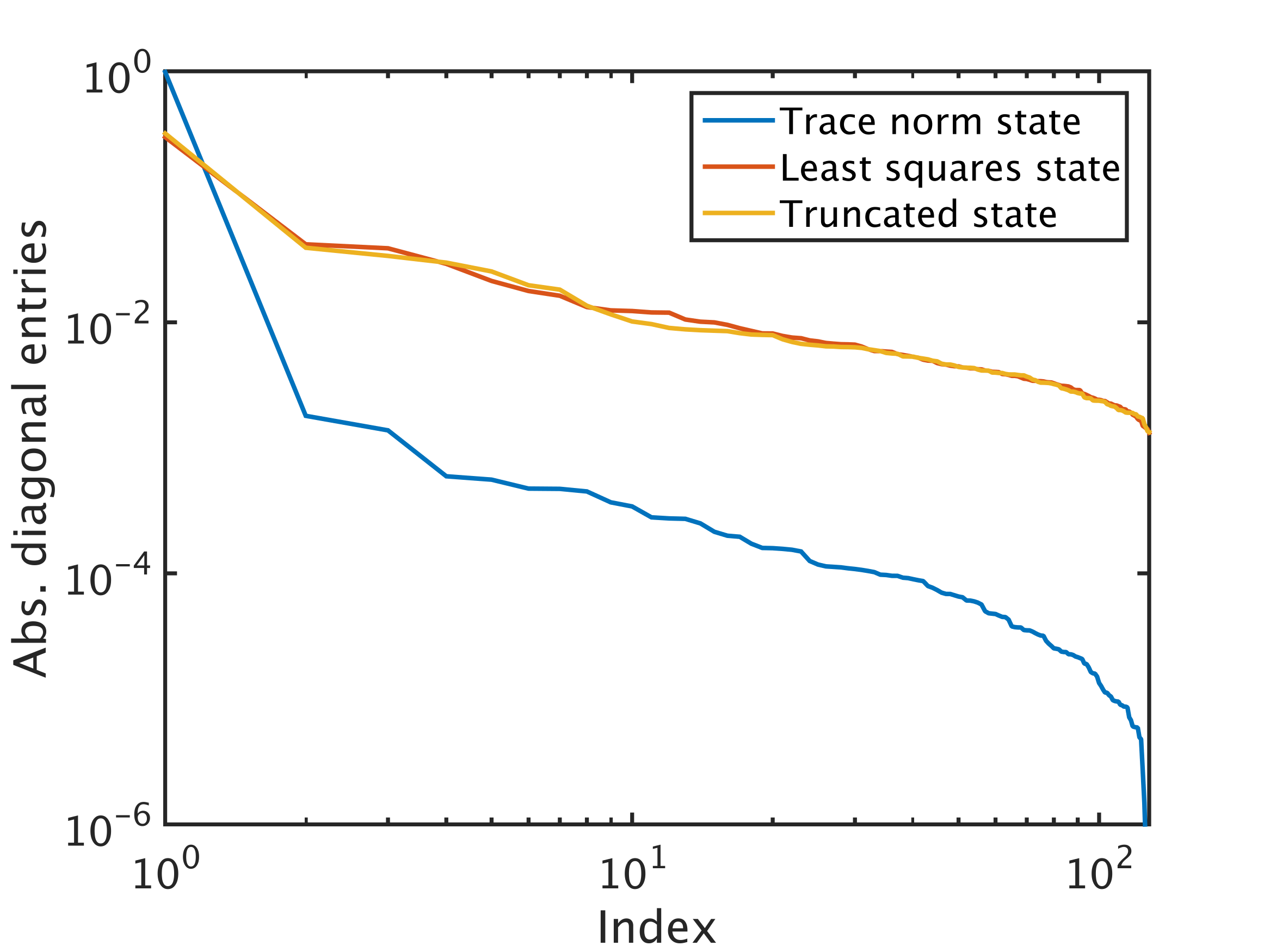}}
\caption{(a,b) 2-D plots of the absolute values of the entries of the difference between the 
anticipated state and the reconstructed state density matrices in the stabilizer basis of the 
anticipated state for the logical $|\bar{0}\rangle$ state vector. In this basis, the anticipated 
state is exactly diagonal with only one nonzero entry in the diagonal. While only the trace norm 
and least squares estimates are shown, the spectral thresholding estimate is very similar 
to (b) and is omitted.
(c) In the same basis, we plot the diagonal elements of the reconstructed density matrices in order of decreasing magnitude. The log-log plot shows that after a rapid initial decay, most of the diagonal elements follow an exponential decay curve. The trace norm estimator has almost all its support in few diagonal elements and thus is biased heavily towards pure states, while the least squares and spectral thresholding estimators have much heavier tails, despite still exhibiting exponential decay.}
\label{Diff2}
\end{figure*}

We can gather evidence for our hypothesis by looking at the diagonal matrix elements of the 
reconstructed states in the anticipated basis, meaning the stabilizer basis that includes the 
anticipated state. In fig.~\ref{Diff2}(a-b) we see the absolute values of the matrix elements 
of the reconstructed states using this basis. Here it is much clearer that the trace norm 
estimate is detecting coherent noise, while the least squares estimate (and the spectral 
thresholding estimate, not shown) achieve a more mixed reconstruction. In fact, 
fig.~\ref{Diff2}(c) shows that in every case the majority of the diagonal elements are decaying 
exponentially when ordered in decreasing magnitude, but with a much more rapid initial decay 
for the trace norm estimator. Although this constitutes evidence for our hypothesis, much more 
work should be done to determine if there is any advantage to using different estimators to 
highlight different features of the noise.

\subsection*{Quantum support identification}

The traditional goal of quantum state tomography is to estimate the
\emph{true density matrix} of the system---i.e.\ the one that would
result in the limit of infinitely many measurements, when all
statistical uncertainties have vanished (assuming no drift or other systematic errors). 
We will now argue that in a high-dimensional setting, with limited data, it
may be neither possible nor desirable to obtain a complete estimate of
the true state. 

It is not necessarily desirable, because it is unclear that a
high-dimensional matrix would provide either interpretable or
actionable information. Consider a typical use case for tomography, where 
the difference between the anticipated
state and the leading eigenvectors encodes useful information about the
dominating error sources. The eigenvectors associated with the first
few eigenvalues contain the most useful information about noise effects, 
and based in these inputs an experimentalist can adjust the apparatus to
achieve a higher fidelity in future runs. However, it is unclear which
action would possibly follow from knowing, say, the exact form of the
100th eigenvector.

At the same time, the data obtained may also not be sufficient to
estimate all the parameters of the full density matrix to a sensible
accuracy. Indeed, trying to fit too many degrees of freedom to noisy
data results in \emph{overfitting}, where the estimate depends
strongly on statistical fluctuations and only to a small degree on the
true state. To combat this, \emph{model
selection} methods give rules for selecting a
lower-dimensional model if the amount and variability of the data do
not allow for a reconstruction of the full set of unknown parameters.\cite{AIC}

In the context of quantum state estimation, \emph{spectral
thresholding} has been proposed as a model selection method and
theoretically analyzed in the regime of informationally complete
measurements.\cite{guta} Spectral thresholding here means that a
lower-dimensional model is selected by setting all eigenvalues of the
estimate to zero that are below a threshold value that depends on the
dimension of the Hilbert space and the variance of the individual
measurements.\cite{guta} 

Here we propose a new heuristic for selecting which eigenvalues of an
estimate to keep and which to discard as not meaningful. While it
lacks the rigorous guarantees of ref.~\citenum{guta}, it is applicable in more
general situations.  It is based on a transparent criterion:
Parameters of an estimated density matrix should not be reported if
they behave in ways consistent with a random matrix model---i.e.\ if
they can be explained as resulting from a purely random noise without
any signal. This approach is consistent with recent unrelated findings
that the spectrum of highly noisy quantum tomographic estimates
resembles the spectrum of random matrix models.\cite{Knips2015}

Technically, for a given data set ${\mathbf y}^{(1)}$, the spectral
decomposition of the positive semi-definite estimate ${\mathbf
y}^{(1)}\mapsto  f({\mathbf y}^{(1)})$ can be written as
\begin{equation}
	\hat \rho^{(1)} := f({\mathbf y}^{(1)}) =  \sum_{j=1}^{d} \lambda^{(1)}_j E^{(1)}_j,
\end{equation}
with decreasingly ordered eigenvalues $\{\lambda_j^{(1)}\}$ and corresponding eigenprojections $\{E^{(1)}_j\}$.
When insufficient data are taken in an experiment, not all eigenprojections can be characterized equally
well. Only for some eigenprojections will one have provided sufficient data. They
concomitantly 
will have low uncertainties and thus will be common to different
estimates of the same state based on different realizations of the experiment,
while the other directions will fluctuate wildly based on the particular data obtained.
Generating a different data set ${\mathbf y}^{(2)}$ using the bootstrapping techniques
detailed below, we arrive at the estimate $f({\mathbf y}^{(2)})$ with decomposition
\begin{equation}
	\hat \rho^{(2)} := f({\mathbf y}^{(2)}) =  \sum_{j=1}^{d} \lambda^{(2)}_j E^{(2)}_j.
\end{equation}
Our figure of merit is based on the Hilbert-Schmidt scalar product of the
eigenprojections
\begin{equation}\label{Eq:Overlap}
	M_j({\mathbf y}^{(1)}, {\mathbf y}^{(2)}) =  \text{tr}( E^{(1)}_jE^{(2)}_j) ,
\end{equation}
where $j=1,\ldots,d$. In the informationally incomplete regime we are in,
this quantity will show a strong overlap only between the dominant eigenvectors.
For the eigenvectors of the complement, the
overlaps resemble the overlap of state vectors chosen randomly from the unitarily invariant Haar measure.
In the light of this, the spectral thresholding
parameter $k_\mathrm{max}$ is taken to be
\begin{equation}\label{Eq:SpectralThresholding}
	k := \max \left\{j:
	\mathbbm{E} \left(
	M_j({\mathbf y}^{(1)}, {\mathbf y}^{(2)}) \right)> e_d
	\right\},
\end{equation}
in expectation over pairs  $({\mathbf y}^{(1)}, {\mathbf y}^{(2)}) $, where
the threshold $e_d$ is chosen as
$e_d = \mathbbm{E}(x) + \text{var}(x)^{1/2}$
for the random variable defined as $x(U) = |\langle \psi| U|\psi\rangle|^2 $
as overlaps between Haar random state vectors from $\mathbbm{C}^d$.
Specifically, a random matrix theory computation (see supplementary material) gives,
\begin{equation}\label{Eq:threshold}
	e_d = \frac{1}{d}+ \left(
	\frac{2}{d(d+1)}- \frac{1}{d^2}
	\right)^{1/2}.
\end{equation}
Based on such a significance threshold,  for the estimate based on the data,
we return the spectrally thresholded state $\rho_k$ with a normalization $c>0$, where
\begin{equation}
	\rho_k =  c \sum_{j=1}^{k} \lambda^{(1)}_j E^{(1)}_j \,.
\end{equation}

Let us now define the protocol we follow to provide an estimate that has low enough rank to be compatible with few data and yet avoid overfitting. For this, we briefly review the concept of bootstrapping. We consider two types of bootstrapping: parametric and non-parametric bootstrapping. In parametric bootstrapping, from the reconstructed density matrix, one simulates the experimental measurements (sampled according to the appropriate noise statistics) and for each sample data realization one computes a new estimated density matrix.  In non-parametric bootstrapping, however, the measured frequencies are assumed as the true probabilities, which in turn are used to simulate (sample) new data sets which are used, as before, to compute an ensemble of estimated density matrices. In both cases, one uses the ensemble of recovered density matrices to gain confidence on the reconstructed state.

The way we proceed is the following: From the experimentally measured frequencies we do either parametric or non-parametric bootstrapping to generate an ensemble of estimated density matrices. We find their spectral decomposition and order their eigenvalues and eigenvectors in descending order as explained above. Then, for all possible pairs of estimated density matrices, we compute the mean of Eq.~\eqref{Eq:Overlap} for all $j$. Finally, we report as the rank of the reconstructed state the largest $j$ for which the quantity $M_j$ has an overlap grater than the threshold computed in Eq.~\eqref{Eq:threshold}. The results are shown in part (d) of Fig.~\ref{Data1} and in the supplementary material.

\subsection*{Conclusion and perspectives}

Quantum tomography---the task of reconstructing unknown states from data---is a key primitive
in quantum technologies. At its heart, it aims at 
providing actionable advice upon which the experimenter can make the appropriate modifications to an experimental setup. 
It goes beyond mere certification of the correctness of an anticipated preparation of a quantum 
state~\cite{FidelityEstimation, Compressed2, Leandro}: By learning in what way the actually prepared state 
deviates from the anticipated one, one can modify the apparatus appropriately to improve performance in future runs.

The purpose of the present work is two-fold. On the one hand, it presents a successful first 
compressed sensing tomography implementation on a moderately sized quantum experiment, using estimators and reconstruction 
techniques that are efficient in the Hilbert space dimension. 
In this way, it demonstrates the potential of using the 
machinery of the ``big data'' paradigm to assess quantum systems close to the limit of what is 
experimentally feasible. More conceptually, on the other hand, 
we discuss ideas of \emph{quantum support identification}, 
related to the question of what quantum 
state tomography can actually mean in the regime of informationally incomplete data for 
intermediately sized quantum systems. We advocate a 
paradigm that only those low-rank states should be 
reported that have a statistical basis. It is the hope that in both ways we can inspire 
further work on the certification and reconstruction of quantum states and processes for 
increasingly large quantum systems, 
overcoming the roadblock against further development in quantum technologies.\\

\subsection*{Acknowledgements}
We would like to thank A.\ Steffens, M.\ Kliesch, and C.\ Ferrie for discussions and comments on the manuscript. 
This work has been supported by the Templeton Foundation, 
the EU (RAQUEL, AQuS), the ERC (TAQ), 
the Freie Universit\"{a}t Berlin within the Excellence Initiative of the
German Research Foundation, the DFG (SPP 1798 CoSIP, 
EI 519/7-1, EI 519/9-1 and
GRO 4334/2-1), by the Austrian
Science Fund (FWF), through the SFB FoQus (FWF
Project No.\ F4002-N16), the Institut f\"{u}r Quanteninformation
GmbH, and the BMBF (Q.com). This work was also supported by
the Australian Research Council via EQuS project number CE11001013, 
and by the US Army Research
Office grant numbers W911NF-14-1-0098 and W911NF-14-1-0103 within the
QCVV program, and by the Australia-Germany Joint Research Co-operation Scheme. 
Furthermore, this research was funded by the Office of
the Director of National Intelligence (ODNI), Intelligence Advanced
Research Projects Activity (IARPA), through the Army Research Office
grant W911NF-10-1-0284. All statements of fact, opinion or conclusions
contained herein are those of the authors and should not be construed
as representing the official views or policies of IARPA, the ODNI, or
the US Government. STF also acknowledges support from an 
Australian Research Council Future Fellowship FT130101744.

\bibliography{TomographyBib}

\begin{thebibliography}{43}%
\makeatletter
\providecommand \@ifxundefined [1]{%
 \@ifx{#1\undefined}
}%
\providecommand \@ifnum [1]{%
 \ifnum #1\expandafter \@firstoftwo
 \else \expandafter \@secondoftwo
 \fi
}%
\providecommand \@ifx [1]{%
 \ifx #1\expandafter \@firstoftwo
 \else \expandafter \@secondoftwo
 \fi
}%
\providecommand \natexlab [1]{#1}%
\providecommand \enquote  [1]{``#1''}%
\providecommand \bibnamefont  [1]{#1}%
\providecommand \bibfnamefont [1]{#1}%
\providecommand \citenamefont [1]{#1}%
\providecommand \href@noop [0]{\@secondoftwo}%
\providecommand \href [0]{\begingroup \@sanitize@url \@href}%
\providecommand \@href[1]{\@@startlink{#1}\@@href}%
\providecommand \@@href[1]{\endgroup#1\@@endlink}%
\providecommand \@sanitize@url [0]{\catcode `\\12\catcode `\$12\catcode
  `\&12\catcode `\#12\catcode `\^12\catcode `\_12\catcode `\%12\relax}%
\providecommand \@@startlink[1]{}%
\providecommand \@@endlink[0]{}%
\providecommand \url  [0]{\begingroup\@sanitize@url \@url }%
\providecommand \@url [1]{\endgroup\@href {#1}{\urlprefix }}%
\providecommand \urlprefix  [0]{URL }%
\providecommand \Eprint [0]{\href }%
\providecommand \doibase [0]{http://dx.doi.org/}%
\providecommand \selectlanguage [0]{\@gobble}%
\providecommand \bibinfo  [0]{\@secondoftwo}%
\providecommand \bibfield  [0]{\@secondoftwo}%
\providecommand \translation [1]{[#1]}%
\providecommand \BibitemOpen [0]{}%
\providecommand \bibitemStop [0]{}%
\providecommand \bibitemNoStop [0]{.\EOS\space}%
\providecommand \EOS [0]{\spacefactor3000\relax}%
\providecommand \BibitemShut  [1]{\csname bibitem#1\endcsname}%
\let\auto@bib@innerbib\@empty
\bibitem [{\citenamefont {Nigg}\ \emph {et~al.}(2014)\citenamefont {Nigg},
  \citenamefont {Mueller}, \citenamefont {Martinez}, \citenamefont {Schindler},
  \citenamefont {Hennrich}, \citenamefont {Monz}, \citenamefont
  {Martin-Delgado},\ and\ \citenamefont {Blatt}}]{BlattColorCode}%
  \BibitemOpen
  \bibfield  {author} {\bibinfo {author} {\bibfnamefont {D.}~\bibnamefont
  {Nigg}}, \bibinfo {author} {\bibfnamefont {M.}~\bibnamefont {Mueller}},
  \bibinfo {author} {\bibfnamefont {E.~A.}\ \bibnamefont {Martinez}}, \bibinfo
  {author} {\bibfnamefont {P.}~\bibnamefont {Schindler}}, \bibinfo {author}
  {\bibfnamefont {M.}~\bibnamefont {Hennrich}}, \bibinfo {author}
  {\bibfnamefont {T.}~\bibnamefont {Monz}}, \bibinfo {author} {\bibfnamefont
  {M.~A.}\ \bibnamefont {Martin-Delgado}}, \ and\ \bibinfo {author}
  {\bibfnamefont {R.}~\bibnamefont {Blatt}},\ }\href@noop {} {\bibfield
  {journal} {\bibinfo  {journal} {Science}\ }\textbf {\bibinfo {volume}
  {345}},\ \bibinfo {pages} {302} (\bibinfo {year} {2014})}\BibitemShut
  {NoStop}%
\bibitem [{\citenamefont {Lanyon}\ \emph {et~al.}(2011)\citenamefont {Lanyon},
  \citenamefont {Hempel}, \citenamefont {Nigg}, \citenamefont {M\"uller},
  \citenamefont {Gerritsma}, \citenamefont {Z\"ahringer}, \citenamefont
  {Schindler}, \citenamefont {Barreiro}, \citenamefont {Rambach}, \citenamefont
  {Kirchmair}, \citenamefont {Hennrich}, \citenamefont {Zoller}, \citenamefont
  {Blatt},\ and\ \citenamefont {Roos}}]{Lanyon_etal11}%
  \BibitemOpen
  \bibfield  {author} {\bibinfo {author} {\bibfnamefont {B.~P.}\ \bibnamefont
  {Lanyon}}, \bibinfo {author} {\bibfnamefont {C.}~\bibnamefont {Hempel}},
  \bibinfo {author} {\bibfnamefont {D.}~\bibnamefont {Nigg}}, \bibinfo {author}
  {\bibfnamefont {M.}~\bibnamefont {M\"uller}}, \bibinfo {author}
  {\bibfnamefont {R.}~\bibnamefont {Gerritsma}}, \bibinfo {author}
  {\bibfnamefont {F.}~\bibnamefont {Z\"ahringer}}, \bibinfo {author}
  {\bibfnamefont {P.}~\bibnamefont {Schindler}}, \bibinfo {author}
  {\bibfnamefont {J.~T.}\ \bibnamefont {Barreiro}}, \bibinfo {author}
  {\bibfnamefont {M.}~\bibnamefont {Rambach}}, \bibinfo {author} {\bibfnamefont
  {G.}~\bibnamefont {Kirchmair}}, \bibinfo {author} {\bibfnamefont
  {M.}~\bibnamefont {Hennrich}}, \bibinfo {author} {\bibfnamefont
  {P.}~\bibnamefont {Zoller}}, \bibinfo {author} {\bibfnamefont
  {R.}~\bibnamefont {Blatt}}, \ and\ \bibinfo {author} {\bibfnamefont {C.~F.}\
  \bibnamefont {Roos}},\ }\href@noop {} {\bibfield  {journal} {\bibinfo
  {journal} {Science}\ }\textbf {\bibinfo {volume} {334}},\ \bibinfo {pages}
  {57} (\bibinfo {year} {2011})}\BibitemShut {NoStop}%
\bibitem [{\citenamefont {Kaufmann}\ \emph {et~al.}(2012)\citenamefont
  {Kaufmann}, \citenamefont {Ulm}, \citenamefont {Jacob}, \citenamefont
  {Poschinger}, \citenamefont {Landa}, \citenamefont {Retzker}, \citenamefont
  {Plenio},\ and\ \citenamefont {Schmidt-Kaler}}]{IonCrystal}%
  \BibitemOpen
  \bibfield  {author} {\bibinfo {author} {\bibfnamefont {H.}~\bibnamefont
  {Kaufmann}}, \bibinfo {author} {\bibfnamefont {S.}~\bibnamefont {Ulm}},
  \bibinfo {author} {\bibfnamefont {G.}~\bibnamefont {Jacob}}, \bibinfo
  {author} {\bibfnamefont {U.~G.}\ \bibnamefont {Poschinger}}, \bibinfo
  {author} {\bibfnamefont {H.}~\bibnamefont {Landa}}, \bibinfo {author}
  {\bibfnamefont {A.}~\bibnamefont {Retzker}}, \bibinfo {author} {\bibfnamefont
  {M.~B.}\ \bibnamefont {Plenio}}, \ and\ \bibinfo {author} {\bibfnamefont
  {F.}~\bibnamefont {Schmidt-Kaler}},\ }\href@noop {} {\bibfield  {journal}
  {\bibinfo  {journal} {Phys. Rev. Lett.}\ }\textbf {\bibinfo {volume} {109}},\
  \bibinfo {pages} {263003} (\bibinfo {year} {2012})}\BibitemShut {NoStop}%
\bibitem [{\citenamefont {Blatt}\ and\ \citenamefont
  {Roos}(2012)}]{BlattSimulator}%
  \BibitemOpen
  \bibfield  {author} {\bibinfo {author} {\bibfnamefont {R.}~\bibnamefont
  {Blatt}}\ and\ \bibinfo {author} {\bibfnamefont {C.~F.}\ \bibnamefont
  {Roos}},\ }\href@noop {} {\bibfield  {journal} {\bibinfo  {journal} {Nature
  Phys.}\ }\textbf {\bibinfo {volume} {8}},\ \bibinfo {pages} {277} (\bibinfo
  {year} {2012})}\BibitemShut {NoStop}%
\bibitem [{\citenamefont {Barends}\ \emph {et~al.}(2014)\citenamefont
  {Barends}, \citenamefont {Kelly}, \citenamefont {Megrant}, \citenamefont
  {Veitia}, \citenamefont {Sank}, \citenamefont {Jeffrey}, \citenamefont
  {White}, \citenamefont {Mutus}, \citenamefont {Fowler}, \citenamefont
  {Campbell}, \citenamefont {Chen}, \citenamefont {Chen}, \citenamefont
  {Chiaro}, \citenamefont {Dunsworth}, \citenamefont {Neill}, \citenamefont
  {O'Malley}, \citenamefont {Roushan}, \citenamefont {Vainsencher},
  \citenamefont {Wenner}, \citenamefont {Korotkov}, \citenamefont {Cleland},\
  and\ \citenamefont {Martinis}}]{Superconducting}%
  \BibitemOpen
  \bibfield  {author} {\bibinfo {author} {\bibfnamefont {R.}~\bibnamefont
  {Barends}}, \bibinfo {author} {\bibfnamefont {J.}~\bibnamefont {Kelly}},
  \bibinfo {author} {\bibfnamefont {A.}~\bibnamefont {Megrant}}, \bibinfo
  {author} {\bibfnamefont {A.}~\bibnamefont {Veitia}}, \bibinfo {author}
  {\bibfnamefont {D.}~\bibnamefont {Sank}}, \bibinfo {author} {\bibfnamefont
  {E.}~\bibnamefont {Jeffrey}}, \bibinfo {author} {\bibfnamefont {T.~C.}\
  \bibnamefont {White}}, \bibinfo {author} {\bibfnamefont {J.}~\bibnamefont
  {Mutus}}, \bibinfo {author} {\bibfnamefont {A.~G.}\ \bibnamefont {Fowler}},
  \bibinfo {author} {\bibfnamefont {B.}~\bibnamefont {Campbell}}, \bibinfo
  {author} {\bibfnamefont {Y.}~\bibnamefont {Chen}}, \bibinfo {author}
  {\bibfnamefont {Z.}~\bibnamefont {Chen}}, \bibinfo {author} {\bibfnamefont
  {B.}~\bibnamefont {Chiaro}}, \bibinfo {author} {\bibfnamefont
  {A.}~\bibnamefont {Dunsworth}}, \bibinfo {author} {\bibfnamefont
  {C.}~\bibnamefont {Neill}}, \bibinfo {author} {\bibfnamefont
  {P.}~\bibnamefont {O'Malley}}, \bibinfo {author} {\bibfnamefont
  {P.}~\bibnamefont {Roushan}}, \bibinfo {author} {\bibfnamefont
  {A.}~\bibnamefont {Vainsencher}}, \bibinfo {author} {\bibfnamefont
  {J.}~\bibnamefont {Wenner}}, \bibinfo {author} {\bibfnamefont {A.~N.}\
  \bibnamefont {Korotkov}}, \bibinfo {author} {\bibfnamefont {A.~N.}\
  \bibnamefont {Cleland}}, \ and\ \bibinfo {author} {\bibfnamefont {J.~M.}\
  \bibnamefont {Martinis}},\ }\href@noop {} {\bibfield  {journal} {\bibinfo
  {journal} {Nature}\ }\textbf {\bibinfo {volume} {508}},\ \bibinfo {pages}
  {500} (\bibinfo {year} {2014})}\BibitemShut {NoStop}%
\bibitem [{\citenamefont {Ofek}\ \emph {et~al.}(2016)\citenamefont {Ofek},
  \citenamefont {Petrenko}, \citenamefont {Heeres}, \citenamefont {Reinhold},
  \citenamefont {Leghtas}, \citenamefont {Vlastakis}, \citenamefont {Liu},
  \citenamefont {Frunzio}, \citenamefont {Girvin}, \citenamefont {Jiang},
  \citenamefont {Mirrahimi}, \citenamefont {Devoret},\ and\ \citenamefont
  {Schoelkopf}}]{Ofek2016}%
  \BibitemOpen
  \bibfield  {author} {\bibinfo {author} {\bibfnamefont {N.}~\bibnamefont
  {Ofek}}, \bibinfo {author} {\bibfnamefont {A.}~\bibnamefont {Petrenko}},
  \bibinfo {author} {\bibfnamefont {R.}~\bibnamefont {Heeres}}, \bibinfo
  {author} {\bibfnamefont {P.}~\bibnamefont {Reinhold}}, \bibinfo {author}
  {\bibfnamefont {Z.}~\bibnamefont {Leghtas}}, \bibinfo {author} {\bibfnamefont
  {B.}~\bibnamefont {Vlastakis}}, \bibinfo {author} {\bibfnamefont
  {Y.}~\bibnamefont {Liu}}, \bibinfo {author} {\bibfnamefont {L.}~\bibnamefont
  {Frunzio}}, \bibinfo {author} {\bibfnamefont {S.~M.}\ \bibnamefont {Girvin}},
  \bibinfo {author} {\bibfnamefont {L.}~\bibnamefont {Jiang}}, \bibinfo
  {author} {\bibfnamefont {M.}~\bibnamefont {Mirrahimi}}, \bibinfo {author}
  {\bibfnamefont {M.~H.}\ \bibnamefont {Devoret}}, \ and\ \bibinfo {author}
  {\bibfnamefont {R.~J.}\ \bibnamefont {Schoelkopf}},\ }\href@noop {} {\
  (\bibinfo {year} {2016})},\ \Eprint {http://arxiv.org/abs/1602.04768}
  {arXiv:1602.04768} \BibitemShut {NoStop}%
\bibitem [{\citenamefont {Maller}\ \emph {et~al.}(2015)\citenamefont {Maller},
  \citenamefont {Lichtman}, \citenamefont {Xia}, \citenamefont {Sun},
  \citenamefont {Piotrowicz}, \citenamefont {Carr}, \citenamefont {Isenhower},\
  and\ \citenamefont {Saffman}}]{Saffman2015}%
  \BibitemOpen
  \bibfield  {author} {\bibinfo {author} {\bibfnamefont {K.~M.}\ \bibnamefont
  {Maller}}, \bibinfo {author} {\bibfnamefont {M.~T.}\ \bibnamefont
  {Lichtman}}, \bibinfo {author} {\bibfnamefont {T.}~\bibnamefont {Xia}},
  \bibinfo {author} {\bibfnamefont {Y.}~\bibnamefont {Sun}}, \bibinfo {author}
  {\bibfnamefont {M.~J.}\ \bibnamefont {Piotrowicz}}, \bibinfo {author}
  {\bibfnamefont {A.~W.}\ \bibnamefont {Carr}}, \bibinfo {author}
  {\bibfnamefont {L.}~\bibnamefont {Isenhower}}, \ and\ \bibinfo {author}
  {\bibfnamefont {M.}~\bibnamefont {Saffman}},\ }\href@noop {} {\bibfield
  {journal} {\bibinfo  {journal} {Phys. Rev. A}\ }\textbf {\bibinfo {volume}
  {92}},\ \bibinfo {pages} {022336} (\bibinfo {year} {2015})}\BibitemShut
  {NoStop}%
\bibitem [{\citenamefont {Nogrette}\ \emph {et~al.}(2014)\citenamefont
  {Nogrette}, \citenamefont {Labuhn}, \citenamefont {Ravets}, \citenamefont
  {Barredo}, \citenamefont {B\'eguin}, \citenamefont {Vernier}, \citenamefont
  {Lahaye},\ and\ \citenamefont {Browaeys}}]{Browaeys}%
  \BibitemOpen
  \bibfield  {author} {\bibinfo {author} {\bibfnamefont {F.}~\bibnamefont
  {Nogrette}}, \bibinfo {author} {\bibfnamefont {H.}~\bibnamefont {Labuhn}},
  \bibinfo {author} {\bibfnamefont {S.}~\bibnamefont {Ravets}}, \bibinfo
  {author} {\bibfnamefont {D.}~\bibnamefont {Barredo}}, \bibinfo {author}
  {\bibfnamefont {L.}~\bibnamefont {B\'eguin}}, \bibinfo {author}
  {\bibfnamefont {A.}~\bibnamefont {Vernier}}, \bibinfo {author} {\bibfnamefont
  {T.}~\bibnamefont {Lahaye}}, \ and\ \bibinfo {author} {\bibfnamefont
  {A.}~\bibnamefont {Browaeys}},\ }\href@noop {} {\bibfield  {journal}
  {\bibinfo  {journal} {Phys. Rev. X}\ }\textbf {\bibinfo {volume} {4}},\
  \bibinfo {pages} {021034} (\bibinfo {year} {2014})}\BibitemShut {NoStop}%
\bibitem [{\citenamefont {Lanyon}\ \emph {et~al.}(2013)\citenamefont {Lanyon},
  \citenamefont {Jurcevic}, \citenamefont {Zwerger}, \citenamefont {Hempel},
  \citenamefont {Martinez}, \citenamefont {D{\"u}r}, \citenamefont {Briegel},
  \citenamefont {Blatt},\ and\ \citenamefont {Roos}}]{MBQCIon}%
  \BibitemOpen
  \bibfield  {author} {\bibinfo {author} {\bibfnamefont {B.~P.}\ \bibnamefont
  {Lanyon}}, \bibinfo {author} {\bibfnamefont {P.}~\bibnamefont {Jurcevic}},
  \bibinfo {author} {\bibfnamefont {M.}~\bibnamefont {Zwerger}}, \bibinfo
  {author} {\bibfnamefont {C.}~\bibnamefont {Hempel}}, \bibinfo {author}
  {\bibfnamefont {E.~A.}\ \bibnamefont {Martinez}}, \bibinfo {author}
  {\bibfnamefont {W.}~\bibnamefont {D{\"u}r}}, \bibinfo {author} {\bibfnamefont
  {H.~J.}\ \bibnamefont {Briegel}}, \bibinfo {author} {\bibfnamefont
  {R.}~\bibnamefont {Blatt}}, \ and\ \bibinfo {author} {\bibfnamefont {C.~F.}\
  \bibnamefont {Roos}},\ }\href@noop {} {\bibfield  {journal} {\bibinfo
  {journal} {Phys. Rev. Lett.}\ }\textbf {\bibinfo {volume} {111}},\ \bibinfo
  {pages} {210501} (\bibinfo {year} {2013})}\BibitemShut {NoStop}%
\bibitem [{\citenamefont {Vandersypen}\ \emph {et~al.}(2001)\citenamefont
  {Vandersypen}, \citenamefont {Steffen}, \citenamefont {Breyta}, \citenamefont
  {Yannoni}, \citenamefont {Sherwood},\ and\ \citenamefont {Chuang}}]{Lieven}%
  \BibitemOpen
  \bibfield  {author} {\bibinfo {author} {\bibfnamefont {L.~M.~K.}\
  \bibnamefont {Vandersypen}}, \bibinfo {author} {\bibfnamefont
  {M.}~\bibnamefont {Steffen}}, \bibinfo {author} {\bibfnamefont
  {G.}~\bibnamefont {Breyta}}, \bibinfo {author} {\bibfnamefont {C.~S.}\
  \bibnamefont {Yannoni}}, \bibinfo {author} {\bibfnamefont {M.~H.}\
  \bibnamefont {Sherwood}}, \ and\ \bibinfo {author} {\bibfnamefont {I.~L.}\
  \bibnamefont {Chuang}},\ }\href@noop {} {\bibfield  {journal} {\bibinfo
  {journal} {Nature}\ }\textbf {\bibinfo {volume} {414}},\ \bibinfo {pages}
  {883} (\bibinfo {year} {2001})}\BibitemShut {NoStop}%
\bibitem [{\citenamefont {Fedorov}\ \emph {et~al.}(2012)\citenamefont
  {Fedorov}, \citenamefont {Steffen}, \citenamefont {Baur}, \citenamefont
  {da~Silva},\ and\ \citenamefont {Wallraff}}]{Andreas}%
  \BibitemOpen
  \bibfield  {author} {\bibinfo {author} {\bibfnamefont {A.}~\bibnamefont
  {Fedorov}}, \bibinfo {author} {\bibfnamefont {L.}~\bibnamefont {Steffen}},
  \bibinfo {author} {\bibfnamefont {M.}~\bibnamefont {Baur}}, \bibinfo {author}
  {\bibfnamefont {M.~P.}\ \bibnamefont {da~Silva}}, \ and\ \bibinfo {author}
  {\bibfnamefont {A.}~\bibnamefont {Wallraff}},\ }\href@noop {} {\bibfield
  {journal} {\bibinfo  {journal} {Nature}\ }\textbf {\bibinfo {volume} {481}}
  (\bibinfo {year} {2012})}\BibitemShut {NoStop}%
\bibitem [{\citenamefont {Gulde}\ \emph {et~al.}(2003)\citenamefont {Gulde},
  \citenamefont {Riebe}, \citenamefont {Lancaster}, \citenamefont {Becher},
  \citenamefont {Eschner}, \citenamefont {H{\"a}ffner}, \citenamefont
  {Schmidt-Kaler}, \citenamefont {Chuang},\ and\ \citenamefont
  {Blatt}}]{DeutschJozsa}%
  \BibitemOpen
  \bibfield  {author} {\bibinfo {author} {\bibfnamefont {S.}~\bibnamefont
  {Gulde}}, \bibinfo {author} {\bibfnamefont {M.}~\bibnamefont {Riebe}},
  \bibinfo {author} {\bibfnamefont {G.~P.~T.}\ \bibnamefont {Lancaster}},
  \bibinfo {author} {\bibfnamefont {C.}~\bibnamefont {Becher}}, \bibinfo
  {author} {\bibfnamefont {J.}~\bibnamefont {Eschner}}, \bibinfo {author}
  {\bibfnamefont {H.}~\bibnamefont {H{\"a}ffner}}, \bibinfo {author}
  {\bibfnamefont {F.}~\bibnamefont {Schmidt-Kaler}}, \bibinfo {author}
  {\bibfnamefont {I.~L.}\ \bibnamefont {Chuang}}, \ and\ \bibinfo {author}
  {\bibfnamefont {R.}~\bibnamefont {Blatt}},\ }\href@noop {} {\bibfield
  {journal} {\bibinfo  {journal} {Nature}\ }\textbf {\bibinfo {volume} {421}},\
  \bibinfo {pages} {48} (\bibinfo {year} {2003})}\BibitemShut {NoStop}%
\bibitem [{\citenamefont {Gross}\ \emph {et~al.}(2010)\citenamefont {Gross},
  \citenamefont {Liu}, \citenamefont {Flammia}, \citenamefont {Becker},\ and\
  \citenamefont {Eisert}}]{Compressed}%
  \BibitemOpen
  \bibfield  {author} {\bibinfo {author} {\bibfnamefont {D.}~\bibnamefont
  {Gross}}, \bibinfo {author} {\bibfnamefont {Y.-K.}\ \bibnamefont {Liu}},
  \bibinfo {author} {\bibfnamefont {S.~T.}\ \bibnamefont {Flammia}}, \bibinfo
  {author} {\bibfnamefont {S.}~\bibnamefont {Becker}}, \ and\ \bibinfo {author}
  {\bibfnamefont {J.}~\bibnamefont {Eisert}},\ }\href@noop {} {\bibfield
  {journal} {\bibinfo  {journal} {Phys. Rev. Lett.}\ }\textbf {\bibinfo
  {volume} {105}},\ \bibinfo {pages} {150401} (\bibinfo {year}
  {2010})}\BibitemShut {NoStop}%
\bibitem [{\citenamefont {Cramer}\ \emph {et~al.}(2010)\citenamefont {Cramer},
  \citenamefont {Plenio}, \citenamefont {Flammia}, \citenamefont {Somma},
  \citenamefont {Gross}, \citenamefont {Bartlett}, \citenamefont
  {Landon-Cardinal}, \citenamefont {Poulin},\ and\ \citenamefont
  {Liu}}]{MPSTomo}%
  \BibitemOpen
  \bibfield  {author} {\bibinfo {author} {\bibfnamefont {M.}~\bibnamefont
  {Cramer}}, \bibinfo {author} {\bibfnamefont {M.~B.}\ \bibnamefont {Plenio}},
  \bibinfo {author} {\bibfnamefont {S.~T.}\ \bibnamefont {Flammia}}, \bibinfo
  {author} {\bibfnamefont {R.}~\bibnamefont {Somma}}, \bibinfo {author}
  {\bibfnamefont {D.}~\bibnamefont {Gross}}, \bibinfo {author} {\bibfnamefont
  {S.~D.}\ \bibnamefont {Bartlett}}, \bibinfo {author} {\bibfnamefont
  {O.}~\bibnamefont {Landon-Cardinal}}, \bibinfo {author} {\bibfnamefont
  {D.}~\bibnamefont {Poulin}}, \ and\ \bibinfo {author} {\bibfnamefont {Y.-K.}\
  \bibnamefont {Liu}},\ }\href@noop {} {\bibfield  {journal} {\bibinfo
  {journal} {Nature Comm.}\ }\textbf {\bibinfo {volume} {1}},\ \bibinfo {pages}
  {149} (\bibinfo {year} {2010})}\BibitemShut {NoStop}%
\bibitem [{\citenamefont {Flammia}\ \emph {et~al.}(2012)\citenamefont
  {Flammia}, \citenamefont {Gross}, \citenamefont {Liu},\ and\ \citenamefont
  {Eisert}}]{Compressed2}%
  \BibitemOpen
  \bibfield  {author} {\bibinfo {author} {\bibfnamefont {S.~T.}\ \bibnamefont
  {Flammia}}, \bibinfo {author} {\bibfnamefont {D.}~\bibnamefont {Gross}},
  \bibinfo {author} {\bibfnamefont {Y.-K.}\ \bibnamefont {Liu}}, \ and\
  \bibinfo {author} {\bibfnamefont {J.}~\bibnamefont {Eisert}},\ }\href@noop {}
  {\bibfield  {journal} {\bibinfo  {journal} {New J. Phys.}\ }\textbf {\bibinfo
  {volume} {14}},\ \bibinfo {pages} {095022} (\bibinfo {year}
  {2012})}\BibitemShut {NoStop}%
\bibitem [{\citenamefont {Flammia}\ and\ \citenamefont
  {Liu}(2011)}]{FidelityEstimation}%
  \BibitemOpen
  \bibfield  {author} {\bibinfo {author} {\bibfnamefont {S.~T.}\ \bibnamefont
  {Flammia}}\ and\ \bibinfo {author} {\bibfnamefont {Y.-K.}\ \bibnamefont
  {Liu}},\ }\href@noop {} {\bibfield  {journal} {\bibinfo  {journal} {Phys.
  Rev. Lett.}\ }\textbf {\bibinfo {volume} {106}},\ \bibinfo {pages} {230501}
  (\bibinfo {year} {2011})}\BibitemShut {NoStop}%
\bibitem [{\citenamefont {Shabani}\ \emph {et~al.}(2011)\citenamefont
  {Shabani}, \citenamefont {Kosut}, \citenamefont {Mohseni}, \citenamefont
  {Rabitz}, \citenamefont {Broome}, \citenamefont {Almeida}, \citenamefont
  {Fedrizzi},\ and\ \citenamefont {White}}]{WhiteCompressedSensing}%
  \BibitemOpen
  \bibfield  {author} {\bibinfo {author} {\bibfnamefont {A.}~\bibnamefont
  {Shabani}}, \bibinfo {author} {\bibfnamefont {R.~L.}\ \bibnamefont {Kosut}},
  \bibinfo {author} {\bibfnamefont {M.}~\bibnamefont {Mohseni}}, \bibinfo
  {author} {\bibfnamefont {H.}~\bibnamefont {Rabitz}}, \bibinfo {author}
  {\bibfnamefont {M.~A.}\ \bibnamefont {Broome}}, \bibinfo {author}
  {\bibfnamefont {M.~P.}\ \bibnamefont {Almeida}}, \bibinfo {author}
  {\bibfnamefont {A.}~\bibnamefont {Fedrizzi}}, \ and\ \bibinfo {author}
  {\bibfnamefont {A.~G.}\ \bibnamefont {White}},\ }\href@noop {} {\bibfield
  {journal} {\bibinfo  {journal} {Phys. Rev. Lett.}\ }\textbf {\bibinfo
  {volume} {106}},\ \bibinfo {pages} {100401} (\bibinfo {year}
  {2011})}\BibitemShut {NoStop}%
\bibitem [{\citenamefont {H{\"u}bener}\ \emph {et~al.}(2013)\citenamefont
  {H{\"u}bener}, \citenamefont {Mari},\ and\ \citenamefont {Eisert}}]{Wick}%
  \BibitemOpen
  \bibfield  {author} {\bibinfo {author} {\bibfnamefont {R.}~\bibnamefont
  {H{\"u}bener}}, \bibinfo {author} {\bibfnamefont {A.}~\bibnamefont {Mari}}, \
  and\ \bibinfo {author} {\bibfnamefont {J.}~\bibnamefont {Eisert}},\
  }\href@noop {} {\bibfield  {journal} {\bibinfo  {journal} {Phys. Rev. Lett.}\
  }\textbf {\bibinfo {volume} {110}},\ \bibinfo {pages} {040401} (\bibinfo
  {year} {2013})}\BibitemShut {NoStop}%
\bibitem [{\citenamefont {Steffens}\ \emph {et~al.}(2015)\citenamefont
  {Steffens}, \citenamefont {Friesdorf}, \citenamefont {Langen}, \citenamefont
  {Rauer}, \citenamefont {Schweigler}, \citenamefont {H{\"u}bener},
  \citenamefont {Schmiedmayer}, \citenamefont {Riofrio},\ and\ \citenamefont
  {Eisert}}]{QuantumFieldTomography}%
  \BibitemOpen
  \bibfield  {author} {\bibinfo {author} {\bibfnamefont {A.}~\bibnamefont
  {Steffens}}, \bibinfo {author} {\bibfnamefont {M.}~\bibnamefont {Friesdorf}},
  \bibinfo {author} {\bibfnamefont {T.}~\bibnamefont {Langen}}, \bibinfo
  {author} {\bibfnamefont {B.}~\bibnamefont {Rauer}}, \bibinfo {author}
  {\bibfnamefont {T.}~\bibnamefont {Schweigler}}, \bibinfo {author}
  {\bibfnamefont {R.}~\bibnamefont {H{\"u}bener}}, \bibinfo {author}
  {\bibfnamefont {J.}~\bibnamefont {Schmiedmayer}}, \bibinfo {author}
  {\bibfnamefont {C.~A.}\ \bibnamefont {Riofrio}}, \ and\ \bibinfo {author}
  {\bibfnamefont {J.}~\bibnamefont {Eisert}},\ }\href@noop {} {\bibfield
  {journal} {\bibinfo  {journal} {Nature Comm.}\ }\textbf {\bibinfo {volume}
  {6}},\ \bibinfo {pages} {7663} (\bibinfo {year} {2015})}\BibitemShut
  {NoStop}%
\bibitem [{\citenamefont {Candes}\ and\ \citenamefont
  {Wakin}(2008)}]{CompressedSensingIntroCandes}%
  \BibitemOpen
  \bibfield  {author} {\bibinfo {author} {\bibfnamefont {E.}~\bibnamefont
  {Candes}}\ and\ \bibinfo {author} {\bibfnamefont {M.}~\bibnamefont {Wakin}},\
  }\href {\doibase 10.1109/msp.2007.914731} {\bibfield  {journal} {\bibinfo
  {journal} {{IEEE} Signal Process. Mag.}\ }\textbf {\bibinfo {volume} {25}},\
  \bibinfo {pages} {21} (\bibinfo {year} {2008})}\BibitemShut {NoStop}%
\bibitem [{\citenamefont {Foucart}\ and\ \citenamefont
  {Rauhut}(2013)}]{CompressedSensingIntroRauhut}%
  \BibitemOpen
  \bibfield  {author} {\bibinfo {author} {\bibfnamefont {S.}~\bibnamefont
  {Foucart}}\ and\ \bibinfo {author} {\bibfnamefont {H.}~\bibnamefont
  {Rauhut}},\ }\href@noop {} {\emph {\bibinfo {title} {A mathematical
  introduction to compressive sensing}}}\ (\bibinfo  {publisher} {Springer},\
  \bibinfo {address} {Heidelberg},\ \bibinfo {year} {2013})\BibitemShut
  {NoStop}%
\bibitem [{\citenamefont {Eldar}(2012)}]{Eldar2012}%
  \BibitemOpen
  \bibfield  {author} {\bibinfo {author} {\bibfnamefont {Y.}~\bibnamefont
  {Eldar}},\ }\href@noop {} {\emph {\bibinfo {title} {Compressed sensing:
  theory and applications}}}\ (\bibinfo  {publisher} {Cambridge University
  Press},\ \bibinfo {address} {Cambridge New York},\ \bibinfo {year}
  {2012})\BibitemShut {NoStop}%
\bibitem [{\citenamefont {Kueng}\ \emph {et~al.}(2015)\citenamefont {Kueng},
  \citenamefont {Rauhut},\ and\ \citenamefont {Terstiege}}]{KuengRauhut}%
  \BibitemOpen
  \bibfield  {author} {\bibinfo {author} {\bibfnamefont {R.}~\bibnamefont
  {Kueng}}, \bibinfo {author} {\bibfnamefont {H.}~\bibnamefont {Rauhut}}, \
  and\ \bibinfo {author} {\bibfnamefont {U.}~\bibnamefont {Terstiege}},\ }\href
  {\doibase 10.1016/j.acha.2015.07.007} {\bibfield  {journal} {\bibinfo
  {journal} {Applied and Computational Harmonic Analysis}\ } (\bibinfo {year}
  {2015}),\ 10.1016/j.acha.2015.07.007}\BibitemShut {NoStop}%
\bibitem [{\citenamefont {Carpentier}\ \emph {et~al.}(2015)\citenamefont
  {Carpentier}, \citenamefont {Eisert}, \citenamefont {Gross},\ and\
  \citenamefont {Nickl}}]{Carpentier2015}%
  \BibitemOpen
  \bibfield  {author} {\bibinfo {author} {\bibfnamefont {A.}~\bibnamefont
  {Carpentier}}, \bibinfo {author} {\bibfnamefont {J.}~\bibnamefont {Eisert}},
  \bibinfo {author} {\bibfnamefont {D.}~\bibnamefont {Gross}}, \ and\ \bibinfo
  {author} {\bibfnamefont {R.}~\bibnamefont {Nickl}},\ }\href@noop {}
  {\bibfield  {journal} {\bibinfo  {journal} {arXiv:1504.03234v2}\ } (\bibinfo
  {year} {2015})}\BibitemShut {NoStop}%
\bibitem [{\citenamefont {Rodionov}\ \emph {et~al.}(2014)\citenamefont
  {Rodionov}, \citenamefont {Veitia}, \citenamefont {Barends}, \citenamefont
  {Kelly}, \citenamefont {Sank}, \citenamefont {Wenner}, \citenamefont
  {Martinis}, \citenamefont {Kosut},\ and\ \citenamefont
  {Korotkov}}]{CompressedMartinis}%
  \BibitemOpen
  \bibfield  {author} {\bibinfo {author} {\bibfnamefont {A.~V.}\ \bibnamefont
  {Rodionov}}, \bibinfo {author} {\bibfnamefont {A.}~\bibnamefont {Veitia}},
  \bibinfo {author} {\bibfnamefont {R.}~\bibnamefont {Barends}}, \bibinfo
  {author} {\bibfnamefont {J.}~\bibnamefont {Kelly}}, \bibinfo {author}
  {\bibfnamefont {D.}~\bibnamefont {Sank}}, \bibinfo {author} {\bibfnamefont
  {J.}~\bibnamefont {Wenner}}, \bibinfo {author} {\bibfnamefont {J.~M.}\
  \bibnamefont {Martinis}}, \bibinfo {author} {\bibfnamefont {R.~L.}\
  \bibnamefont {Kosut}}, \ and\ \bibinfo {author} {\bibfnamefont {A.~N.}\
  \bibnamefont {Korotkov}},\ }\href@noop {} {\bibfield  {journal} {\bibinfo
  {journal} {Phys. Rev. B}\ }\textbf {\bibinfo {volume} {90}},\ \bibinfo
  {pages} {144504} (\bibinfo {year} {2014})}\BibitemShut {NoStop}%
\bibitem [{\citenamefont {Schwemmer}\ \emph {et~al.}(2014)\citenamefont
  {Schwemmer}, \citenamefont {Toth}, \citenamefont {Niggebaum}, \citenamefont
  {Moroder}, \citenamefont {Gross}, \citenamefont {G{\"u}hne},\ and\
  \citenamefont {Weinfurter}}]{CompressedWeinfurter}%
  \BibitemOpen
  \bibfield  {author} {\bibinfo {author} {\bibfnamefont {C.}~\bibnamefont
  {Schwemmer}}, \bibinfo {author} {\bibfnamefont {G.}~\bibnamefont {Toth}},
  \bibinfo {author} {\bibfnamefont {A.}~\bibnamefont {Niggebaum}}, \bibinfo
  {author} {\bibfnamefont {T.}~\bibnamefont {Moroder}}, \bibinfo {author}
  {\bibfnamefont {D.}~\bibnamefont {Gross}}, \bibinfo {author} {\bibfnamefont
  {O.}~\bibnamefont {G{\"u}hne}}, \ and\ \bibinfo {author} {\bibfnamefont
  {H.}~\bibnamefont {Weinfurter}},\ }\href@noop {} {\bibfield  {journal}
  {\bibinfo  {journal} {Phys. Rev. Lett.}\ }\textbf {\bibinfo {volume} {113}},\
  \bibinfo {pages} {040503} (\bibinfo {year} {2014})}\BibitemShut {NoStop}%
\bibitem [{\citenamefont {Bombin}\ and\ \citenamefont
  {Martin-Delgado}(2006)}]{Bombin2006}%
  \BibitemOpen
  \bibfield  {author} {\bibinfo {author} {\bibfnamefont {H.}~\bibnamefont
  {Bombin}}\ and\ \bibinfo {author} {\bibfnamefont {M.~A.}\ \bibnamefont
  {Martin-Delgado}},\ }\href {\doibase 10.1103/PhysRevLett.97.180501}
  {\bibfield  {journal} {\bibinfo  {journal} {Phys. Rev. Lett.}\ }\textbf
  {\bibinfo {volume} {97}},\ \bibinfo {pages} {180501} (\bibinfo {year}
  {2006})}\BibitemShut {NoStop}%
\bibitem [{\citenamefont {Waters}\ \emph {et~al.}(2011)\citenamefont {Waters},
  \citenamefont {Sankaranarayanan},\ and\ \citenamefont {Baraniuk}}]{SI}%
  \BibitemOpen
  \bibfield  {author} {\bibinfo {author} {\bibfnamefont {A.~E.}\ \bibnamefont
  {Waters}}, \bibinfo {author} {\bibfnamefont {A.~C.}\ \bibnamefont
  {Sankaranarayanan}}, \ and\ \bibinfo {author} {\bibfnamefont
  {R.}~\bibnamefont {Baraniuk}},\ }in\ \href@noop {} {\emph {\bibinfo
  {booktitle} {Advances in Neural Information Processing Systems 24}}},\
  \bibinfo {editor} {edited by\ \bibinfo {editor} {\bibfnamefont
  {J.}~\bibnamefont {Shawe-Taylor}}, \bibinfo {editor} {\bibfnamefont {R.~S.}\
  \bibnamefont {Zemel}}, \bibinfo {editor} {\bibfnamefont {P.~L.}\ \bibnamefont
  {Bartlett}}, \bibinfo {editor} {\bibfnamefont {F.}~\bibnamefont {Pereira}}, \
  and\ \bibinfo {editor} {\bibfnamefont {K.~Q.}\ \bibnamefont {Weinberger}}}\
  (\bibinfo  {publisher} {Curran Associates, Inc.},\ \bibinfo {year} {2011})\
  pp.\ \bibinfo {pages} {1089--1097}\BibitemShut {NoStop}%
\bibitem [{\citenamefont {Guta}\ \emph {et~al.}(2012)\citenamefont {Guta},
  \citenamefont {Kypraios},\ and\ \citenamefont {Dryden}}]{guta}%
  \BibitemOpen
  \bibfield  {author} {\bibinfo {author} {\bibfnamefont {M.}~\bibnamefont
  {Guta}}, \bibinfo {author} {\bibfnamefont {T.}~\bibnamefont {Kypraios}}, \
  and\ \bibinfo {author} {\bibfnamefont {I.}~\bibnamefont {Dryden}},\
  }\href@noop {} {\bibfield  {journal} {\bibinfo  {journal} {New J. Phys.}\
  }\textbf {\bibinfo {volume} {14}},\ \bibinfo {pages} {105002} (\bibinfo
  {year} {2012})}\BibitemShut {NoStop}%
\bibitem [{\citenamefont {Knips}\ \emph {et~al.}(2015)\citenamefont {Knips},
  \citenamefont {Schwemmer}, \citenamefont {Klein}, \citenamefont {Reuter},
  \citenamefont {T\'{o}th},\ and\ \citenamefont {Weinfurter}}]{Knips2015}%
  \BibitemOpen
  \bibfield  {author} {\bibinfo {author} {\bibfnamefont {L.}~\bibnamefont
  {Knips}}, \bibinfo {author} {\bibfnamefont {C.}~\bibnamefont {Schwemmer}},
  \bibinfo {author} {\bibfnamefont {N.}~\bibnamefont {Klein}}, \bibinfo
  {author} {\bibfnamefont {J.}~\bibnamefont {Reuter}}, \bibinfo {author}
  {\bibfnamefont {G.}~\bibnamefont {T\'{o}th}}, \ and\ \bibinfo {author}
  {\bibfnamefont {H.}~\bibnamefont {Weinfurter}},\ }\href@noop {} {\bibfield
  {journal} {\bibinfo  {journal} {arXiv:1512.06866v1}\ } (\bibinfo {year}
  {2015})}\BibitemShut {NoStop}%
\bibitem [{\citenamefont {S{\o}rensen}\ and\ \citenamefont
  {M{\o}lmer}(1999)}]{ms_gate}%
  \BibitemOpen
  \bibfield  {author} {\bibinfo {author} {\bibfnamefont {A.}~\bibnamefont
  {S{\o}rensen}}\ and\ \bibinfo {author} {\bibfnamefont {K.}~\bibnamefont
  {M{\o}lmer}},\ }\href {\doibase
  http://dx.doi.org/10.1103/PhysRevLett.82.1971} {\bibfield  {journal}
  {\bibinfo  {journal} {Phys. Rev. Lett.}\ }\textbf {\bibinfo {volume} {82}},\
  \bibinfo {pages} {1971} (\bibinfo {year} {1999})}\BibitemShut {NoStop}%
\bibitem [{\citenamefont {Schindler}\ \emph {et~al.}(2013)\citenamefont
  {Schindler}, \citenamefont {Nigg}, \citenamefont {Monz}, \citenamefont
  {Barreiro}, \citenamefont {Martinez}, \citenamefont {Wang}, \citenamefont
  {Quint}, \citenamefont {Brandl}, \citenamefont {Nebendahl}, \citenamefont
  {Roos}, \citenamefont {Chwalla}, \citenamefont {Hennrich},\ and\
  \citenamefont {Blatt}}]{order_finding_phips}%
  \BibitemOpen
  \bibfield  {author} {\bibinfo {author} {\bibfnamefont {P.}~\bibnamefont
  {Schindler}}, \bibinfo {author} {\bibfnamefont {D.}~\bibnamefont {Nigg}},
  \bibinfo {author} {\bibfnamefont {T.}~\bibnamefont {Monz}}, \bibinfo {author}
  {\bibfnamefont {J.~T.}\ \bibnamefont {Barreiro}}, \bibinfo {author}
  {\bibfnamefont {E.}~\bibnamefont {Martinez}}, \bibinfo {author}
  {\bibfnamefont {S.~X.}\ \bibnamefont {Wang}}, \bibinfo {author}
  {\bibfnamefont {S.}~\bibnamefont {Quint}}, \bibinfo {author} {\bibfnamefont
  {M.~F.}\ \bibnamefont {Brandl}}, \bibinfo {author} {\bibfnamefont
  {V.}~\bibnamefont {Nebendahl}}, \bibinfo {author} {\bibfnamefont {C.~F.}\
  \bibnamefont {Roos}}, \bibinfo {author} {\bibfnamefont {M.}~\bibnamefont
  {Chwalla}}, \bibinfo {author} {\bibfnamefont {M.}~\bibnamefont {Hennrich}}, \
  and\ \bibinfo {author} {\bibfnamefont {R.}~\bibnamefont {Blatt}},\ }\href
  {\doibase 10.1088/1367-2630/15/12/123012} {\bibfield  {journal} {\bibinfo
  {journal} {New J. Phys.}\ }\textbf {\bibinfo {volume} {15}},\ \bibinfo
  {pages} {123012} (\bibinfo {year} {2013})}\BibitemShut {NoStop}%
\bibitem [{\citenamefont {Calderbank}\ and\ \citenamefont
  {Shor}(1996)}]{CalderbankShor}%
  \BibitemOpen
  \bibfield  {author} {\bibinfo {author} {\bibfnamefont {A.~R.}\ \bibnamefont
  {Calderbank}}\ and\ \bibinfo {author} {\bibfnamefont {P.~W.}\ \bibnamefont
  {Shor}},\ }\href@noop {} {\bibfield  {journal} {\bibinfo  {journal} {Phys.
  Rev. A}\ }\textbf {\bibinfo {volume} {54}},\ \bibinfo {pages} {1098}
  (\bibinfo {year} {1996})}\BibitemShut {NoStop}%
\bibitem [{\citenamefont {Steane}(1996)}]{SteaneCode}%
  \BibitemOpen
  \bibfield  {author} {\bibinfo {author} {\bibfnamefont {A.~M.}\ \bibnamefont
  {Steane}},\ }\href@noop {} {\bibfield  {journal} {\bibinfo  {journal} {Phys.
  Rev. Lett.}\ }\textbf {\bibinfo {volume} {77}},\ \bibinfo {pages} {793}
  (\bibinfo {year} {1996})}\BibitemShut {NoStop}%
\bibitem [{\citenamefont {Akaike}(1974)}]{AIC}%
  \BibitemOpen
  \bibfield  {author} {\bibinfo {author} {\bibfnamefont {H.}~\bibnamefont
  {Akaike}},\ }\href@noop {} {\bibfield  {journal} {\bibinfo  {journal} {IEEE
  Trans. Aut. Contr.}\ }\textbf {\bibinfo {volume} {19}},\ \bibinfo {pages}
  {716} (\bibinfo {year} {1974})}\BibitemShut {NoStop}%
\bibitem [{\citenamefont {Aolita}\ \emph {et~al.}(2015)\citenamefont {Aolita},
  \citenamefont {Gogolin}, \citenamefont {Kliesch},\ and\ \citenamefont
  {Eisert}}]{Leandro}%
  \BibitemOpen
  \bibfield  {author} {\bibinfo {author} {\bibfnamefont {L.}~\bibnamefont
  {Aolita}}, \bibinfo {author} {\bibfnamefont {C.}~\bibnamefont {Gogolin}},
  \bibinfo {author} {\bibfnamefont {M.}~\bibnamefont {Kliesch}}, \ and\
  \bibinfo {author} {\bibfnamefont {J.}~\bibnamefont {Eisert}},\ }\href@noop {}
  {\bibfield  {journal} {\bibinfo  {journal} {Nature Comm.}\ }\textbf {\bibinfo
  {volume} {6}},\ \bibinfo {pages} {8498} (\bibinfo {year} {2015})}\BibitemShut
  {NoStop}%
\bibitem [{\citenamefont {Burer}\ and\ \citenamefont
  {Monteiro}(2003)}]{NonlinearProgramming}%
  \BibitemOpen
  \bibfield  {author} {\bibinfo {author} {\bibfnamefont {S.}~\bibnamefont
  {Burer}}\ and\ \bibinfo {author} {\bibfnamefont {R.~D.~C.}\ \bibnamefont
  {Monteiro}},\ }\href@noop {} {\bibfield  {journal} {\bibinfo  {journal}
  {Math. Program. B}\ }\textbf {\bibinfo {volume} {95}},\ \bibinfo {pages}
  {329} (\bibinfo {year} {2003})}\BibitemShut {NoStop}%
\bibitem [{\citenamefont {Kyrillidis}\ \emph {et~al.}(2012)\citenamefont
  {Kyrillidis}, \citenamefont {Becker}, \citenamefont {Cevher},\ and\
  \citenamefont {Koch}}]{Volkan}%
  \BibitemOpen
  \bibfield  {author} {\bibinfo {author} {\bibfnamefont {A.}~\bibnamefont
  {Kyrillidis}}, \bibinfo {author} {\bibfnamefont {S.}~\bibnamefont {Becker}},
  \bibinfo {author} {\bibfnamefont {V.}~\bibnamefont {Cevher}}, \ and\ \bibinfo
  {author} {\bibfnamefont {C.}~\bibnamefont {Koch}},\ }\href@noop {} {\bibfield
   {journal} {\bibinfo  {journal} {arXiv:1206.1529}\ } (\bibinfo {year}
  {2012})}\BibitemShut {NoStop}%
\bibitem [{\citenamefont {Becker}\ \emph {et~al.}()\citenamefont {Becker},
  \citenamefont {Cevher},\ and\ \citenamefont {Kyrillidis}}]{RandomizedSVP}%
  \BibitemOpen
  \bibfield  {author} {\bibinfo {author} {\bibfnamefont {S.}~\bibnamefont
  {Becker}}, \bibinfo {author} {\bibfnamefont {V.}~\bibnamefont {Cevher}}, \
  and\ \bibinfo {author} {\bibfnamefont {A.}~\bibnamefont {Kyrillidis}},\
  }\href@noop {} {\bibinfo  {journal} {arXiv:1303.0167}\ }\BibitemShut
  {NoStop}%
\bibitem [{\citenamefont {Weingarten}(1978)}]{Weingarten}%
  \BibitemOpen
\bibfield  {journal} {  }\bibfield  {author} {\bibinfo {author} {\bibfnamefont
  {D.}~\bibnamefont {Weingarten}},\ }\href@noop {} {\bibfield  {journal}
  {\bibinfo  {journal} {J. Math. Phys.}\ }\textbf {\bibinfo {volume} {19}},\
  \bibinfo {pages} {999} (\bibinfo {year} {1978})}\BibitemShut {NoStop}%
\bibitem [{\citenamefont {Suess}\ \emph {et~al.}(2016)\citenamefont {Suess},
  \citenamefont {Rudnicki},\ and\ \citenamefont {Gross}}]{Suess}%
  \BibitemOpen
  \bibfield  {author} {\bibinfo {author} {\bibfnamefont {D.}~\bibnamefont
  {Suess}}, \bibinfo {author} {\bibfnamefont {L.}~\bibnamefont {Rudnicki}}, \
  and\ \bibinfo {author} {\bibfnamefont {D.}~\bibnamefont {Gross}},\
  }\href@noop {} {} (\bibinfo {year} {2016}),\ \bibinfo {note}
  {arXiv:1608.00374}\BibitemShut {NoStop}%
\bibitem [{\citenamefont {Candes}(2008)}]{Can08}%
  \BibitemOpen
  \bibfield  {author} {\bibinfo {author} {\bibfnamefont {E.~J.}\ \bibnamefont
  {Candes}},\ }\href@noop {} {\bibfield  {journal} {\bibinfo  {journal} {Compte
  Rendus de l'Academie des Sciences}\ }\textbf {\bibinfo {volume} {346}},\
  \bibinfo {pages} {589} (\bibinfo {year} {2008})}\BibitemShut {NoStop}%
\bibitem [{\citenamefont {Butucea}\ \emph {et~al.}(2015)\citenamefont
  {Butucea}, \citenamefont {Guta},\ and\ \citenamefont {Kypraios}}]{Guta2}%
  \BibitemOpen
  \bibfield  {author} {\bibinfo {author} {\bibfnamefont {C.}~\bibnamefont
  {Butucea}}, \bibinfo {author} {\bibfnamefont {M.}~\bibnamefont {Guta}}, \
  and\ \bibinfo {author} {\bibfnamefont {T.}~\bibnamefont {Kypraios}},\
  }\href@noop {} {} (\bibinfo {year} {2015}),\ \bibinfo {note}
  {arXiv:1504.08295}\BibitemShut {NoStop}%
\end{thebibliography}%

\section*{Supplementary material}

\subsection*{Grad estimator}
In this supplementary material, we present details of the reconstruction sketched in the main text, beginning with the estimators we use.
In our approach, we parametrize the density matrix as
\begin{equation}
	\rho=	Q^\dagger Q,
\end{equation}
which makes it manifestly positive semidefinite. We then solve
\begin{equation}
	\min_{Q}\,  g(Q)=\min_{Q}\,  \|\mathbf{y}-\mathcal{A}(Q^\dagger Q)\|^2_2,
\end{equation}
with $\|.\|_2$ being the vector $2$-norm.
We do this using a gradient search algorithm, but notably for $Q$ and not for $\rho$ itself.
This is the key feature of this approach. This estimator derives from the idea presented in ref.~\citenum{NonlinearProgramming}.
The basic iteration step in a sequence of $\{Q_i\}$ is
\begin{equation}
Q_{i+1} = Q_{i}-\alpha_i\nabla_{Q}g(Q_i).
\end{equation}
For the moment $\alpha_i>0$ is chosen to be a sufficiently small step size, but this can surely be refined to a conjugate gradient
method if absolutely necessary, and can hence be
 refined to increase convergence speed.
In our case, the actual gradient can be computed. Note that $g(Q)$ can be written as
\begin{equation}
g(Q)  = \sum_k(y_k-{\rm Tr}(P_kQ^\dagger Q))^2,
\end{equation}
 and its gradient
\begin{eqnarray}
	\nabla_{Q}g(Q) &= &-2\sum_k(y_k-{\rm Tr}(P_kQ^\dagger Q))\nabla_{Q}{\rm Tr}(QP_kQ^\dagger)\nonumber\\
	&=&4Q\mathcal{A}^\dagger(\mathcal{A}(Q^\dagger Q)-\mathbf{y}),
\end{eqnarray}
as stated in the main text. Here, we have used the standard matrix identity $\nabla_{X}{\rm Tr}(XBX^\dagger) = XB^\dagger+XB$ for the particular case in which $B$ is Hermitian. 
We then iterate the previous equation until reaching convergence. The state is renormalized
at the end, as the trace is not constrained in this way.
This is an extremely fast and elegant way to incorporate positivity of $Q^\dagger Q$.

It is worth mentioning that this approach significantly improves earlier ideas deriving from Refs.~\citenum{Volkan,RandomizedSVP},
in which a gradient method for the state $\rho$ was combined with a suitable projection. Specifically,
\begin{equation}
	\min_{X}\,  f(X)=\min_{X}\,  \|\mathbf{y}-\mathcal{A}(X)\|^2_2,\quad\text{s.t.}\quad X\ge 0
\end{equation}
was solved by
moving away from the semidefinite program and solving the optimization problem using a gradient search algorithm. The basic iteration is
\begin{equation}
	X_{i+1} = \mathcal{P}(X_{i}-\alpha_i\nabla_{X}f(X_i)),
\end{equation}
where here $\nabla_{X}$ is the gradient operator with respect to matrix $X$, and $\mathcal{P}(X)$ is a projector that makes the estimated state positive
semidefinite.
The gradient is explicitly
\begin{equation}
\nabla_{X}f(X) = 2\mathcal{A}^\dagger(\mathcal{A}(X)-\mathbf{y}).
\end{equation}
While this approach also works, the projection significantly slows down the algorithm, hence the need for our method that directly incorporates positivity.

\subsection*{Random matrices}
In this supplementary material, we present results from random matrix
theory on expected overlaps of random vectors. Specifically,
for an arbitrary vector $|\psi\rangle\in \mathbbm{C}^d$, we consider
the random variable defined by
\begin{equation}
	x(U) = |\langle\psi |U| \psi\rangle|^2
\end{equation}
and moments thereof with respect to the Haar measure. This quantity
is easily identified as the overlap of two random vectors from $\mathbbm{C}^d$.
We compute first and second moments thereof. They can
be computed making use of the powerful Weingarten function
formalism \cite{Weingarten}. We find in terms of a Weingarten function $Wg$,
\begin{eqnarray}
	\int_{U(d)} dU  |\langle\psi |U| \psi\rangle|^2 = Wg((1),d) = \frac{1}{d}.
\end{eqnarray}
The second moments can be expressed as
\begin{eqnarray}
	\int_{U(d)} dU  |\langle\psi |U| \psi\rangle|^4 &=& 2 Wg((1,2),d) + 2 Wg((2,1),d)
	\nonumber\\
	&=& \frac{2}{d^2-1} - \frac{2}{d(d^2-1)} \nonumber\\
	&=&
	\frac{2}{d(d+1)}
\end{eqnarray}
using suitable Weingarten functions.
The sum over all permutations on two symbols in the relationship between
Haar averages and Weingarten functions,
$(1,2)\mapsto (1,2)$ and $(1,2)\mapsto (2,1)$,
then simply gives rise to the above two terms.
These result in the expression for the variance
\begin{equation}
	\mathrm{var}(x) = \frac{2}{d(d+1)} - \frac{1}{d^2}.
\end{equation}
In the main text, the quantity
\begin{eqnarray}
	e_d &=& \mathbbm{E}(x) + \mathrm{var(x)}^{1/2}\nonumber\\
	&=& \frac{1}{d}+ \left(
	\frac{2}{d(d+1)}- \frac{1}{d^2}
	\right)^{1/2}
\end{eqnarray}
has been derived from this.

\subsection*{Further estimators}

In this section, we review some of the estimators used in this work.
The first estimator, referred to as \texttt{LS-SDP}, is a \emph{least squares estimator with positivity constraint as a semi-definite program} (SDP). It solves
\begin{equation}
\min_{X}\,  \|\mathbf{y}-\mathcal{A}(X)\|^2_2,\quad\text{s.t.}\quad X\ge 0.
\end{equation}
As with other SDP approaches, $\mathcal{A}(X)$ has to be computed in matrix form, which produces an unfavorable scaling of effort and memory resources in the system size.
As before, we can use this for the $L=7$ qubit problem if the number of measurement settings is not too large.
That is to say, for $n=127$ it is still usable.
The positivity constraint on $X$ helps the estimation process, based on the intuition that the set of feasible density operators lies at the intersection of those operators compatible with the data and the positive cone.
In practice, one can perform very good estimation with this algorithm, even with informationally incomplete measurements, if the actual 
state is not too mixed and hence close to the boundary of state space. There is empirical evidence for this observation, which 
can also made precise \cite{Suess}. 

The second estimator is the \emph{trace norm minimizer} referred to as 
\texttt{TNM}: This is an estimator based on a \emph{trace minimization with a positivity constraint} as an SDP.
Here, we solve the following problem
\begin{equation}
\min_{X}\,  \|X\|_* = \mathrm{Tr}(X),\quad\text{s.t.}\quad X\ge 0 ,\, \|\mathbf{y}-\mathcal{A}(X)\|^2_2\le\epsilon,
\end{equation}
which resembles the Dantzig selector~\cite{Compressed2}, with $\epsilon>0$ being the error level. It is an estimator based on the
intuition derived from compressed sensing that under the restricted isometry property (RIP) \cite{Can08},
the positive semidefinite  trace norm minimizer compatible with the data is the actual state~\cite{Compressed}. This estimator can be cast as an SDP, which again means that $\mathcal{A}(X)$ has to be computed in matrix form, with all memory requirements that come along with it as mentioned above. However, for $L=7$ qubits this is again still feasible if the number of measurement settings is not too large (about less than a thousand on a standard workstation). 
The obvious shortcoming of this estimator---apart from the fact that it will not work for large systems---is the estimation of $\epsilon>0$, as often discussed
(see, e.g., ref.\ \cite{Compressed2}). 
This algorithm is designed to generally produce low-rank estimates.

\subsection*{Simulations on spectral thresholding}

\begin{figure*}[t]
\subfigure[\, 5 repetitions per measurement setting.]{%
  \includegraphics[width=.65\columnwidth]{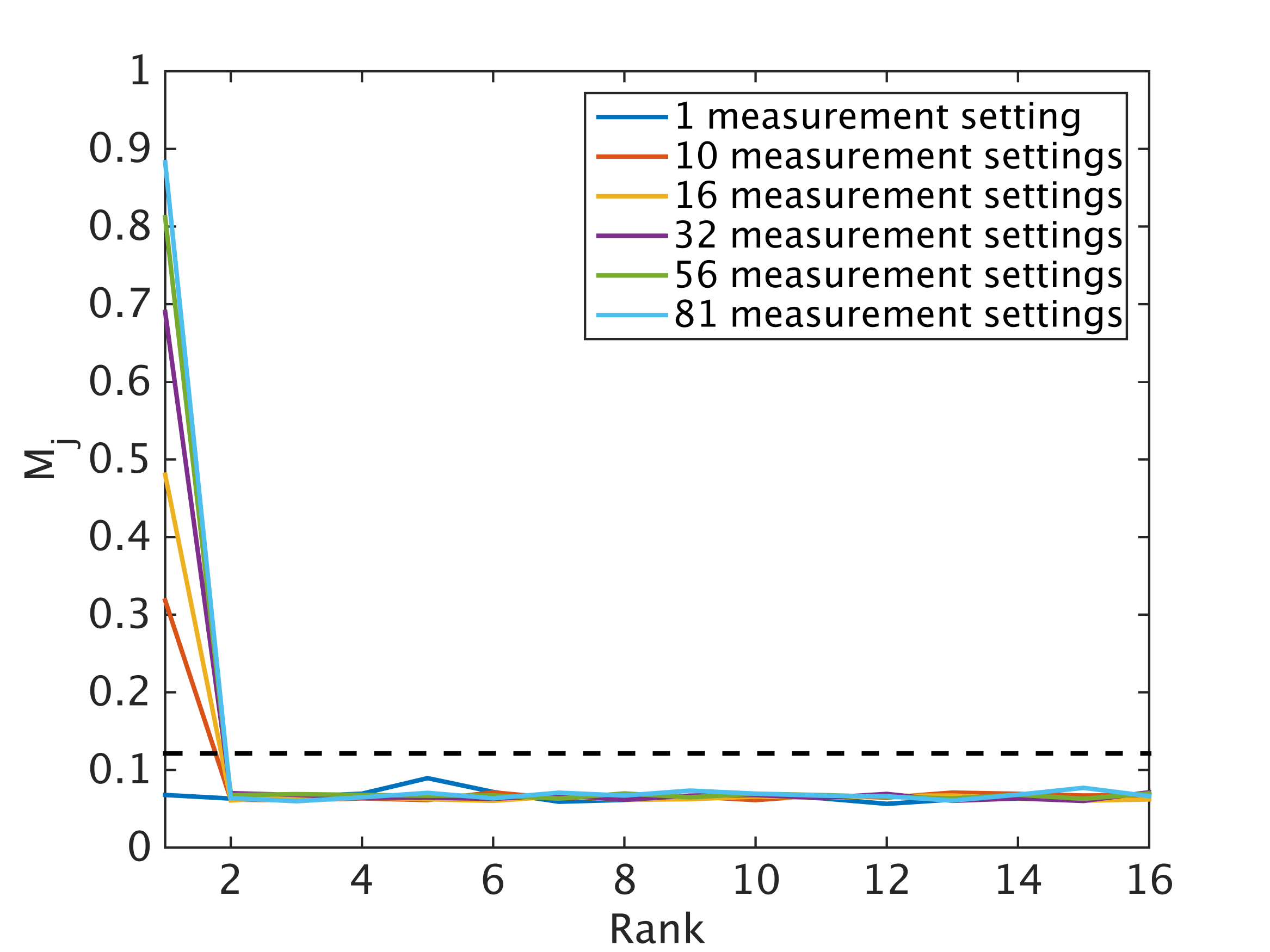}}
\quad
\subfigure[ \, 16 repetitions per measurement setting.]{%
  \includegraphics[width=.65\columnwidth]{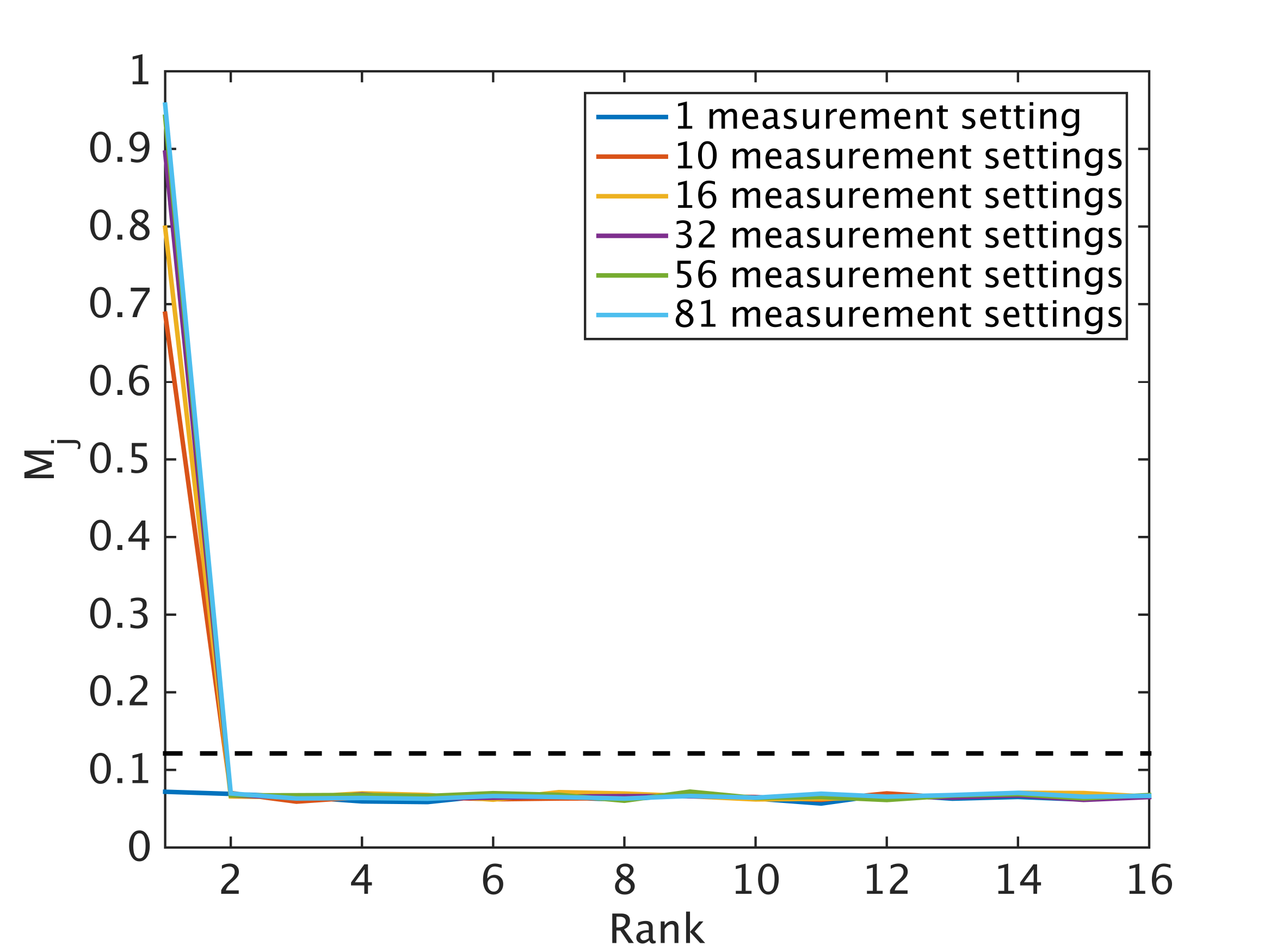}}
\quad
\subfigure[\, 100 repetitions per measurement setting.]{%
  \includegraphics[width=.65\columnwidth]{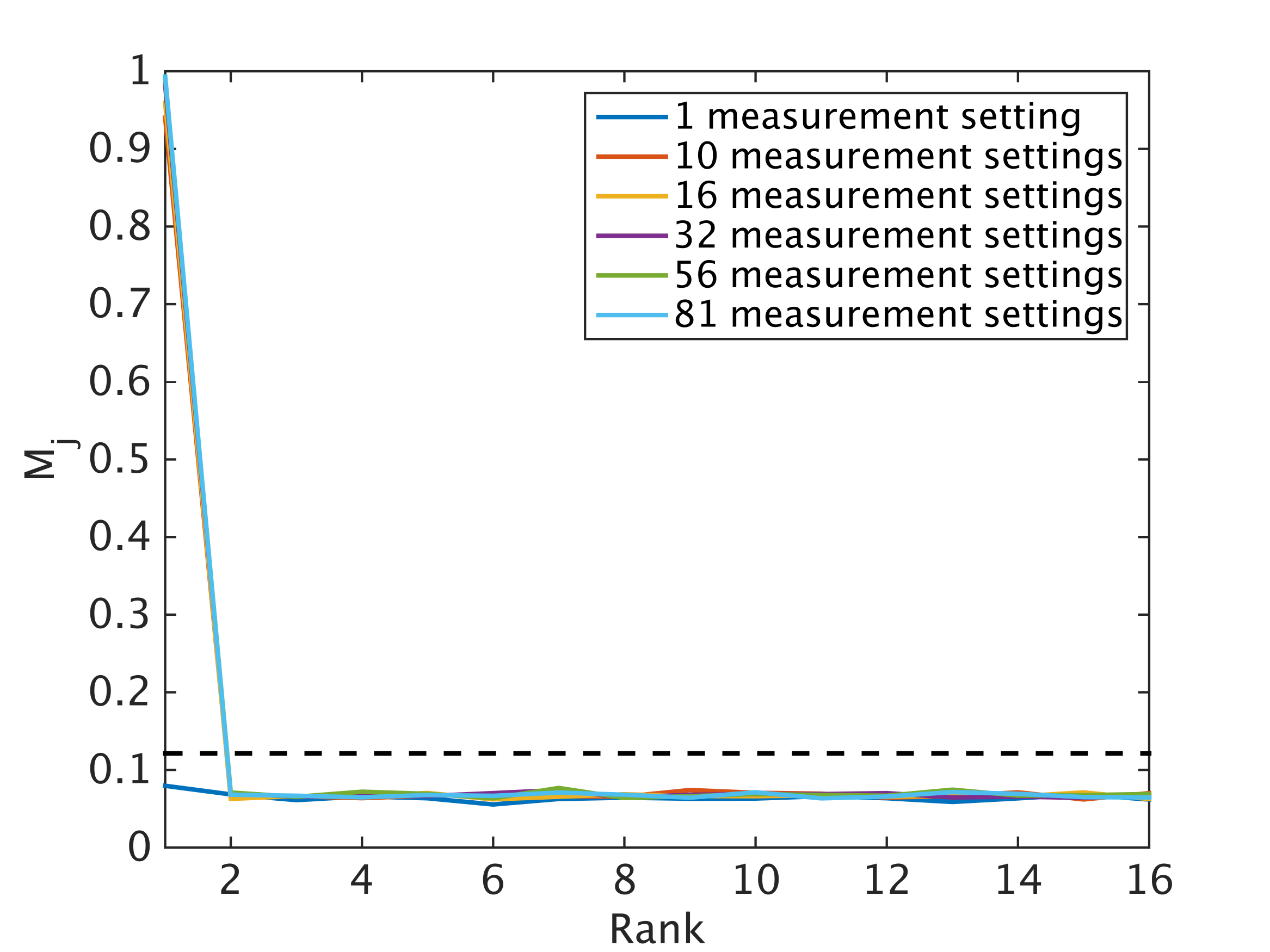}}
\caption{Mean eigenvector overlap as a function of the rank truncation of the recovered density matrix of a true state of rank $1$. The dotted line represents the threshold given in eq.\ \eqref{Eq:threshold}.}
\label{Rank1_Sim}
\end{figure*}

\begin{figure*}[t]
\subfigure[\, 5 repetitions per measurement setting.]{%
  \includegraphics[width=.65\columnwidth]{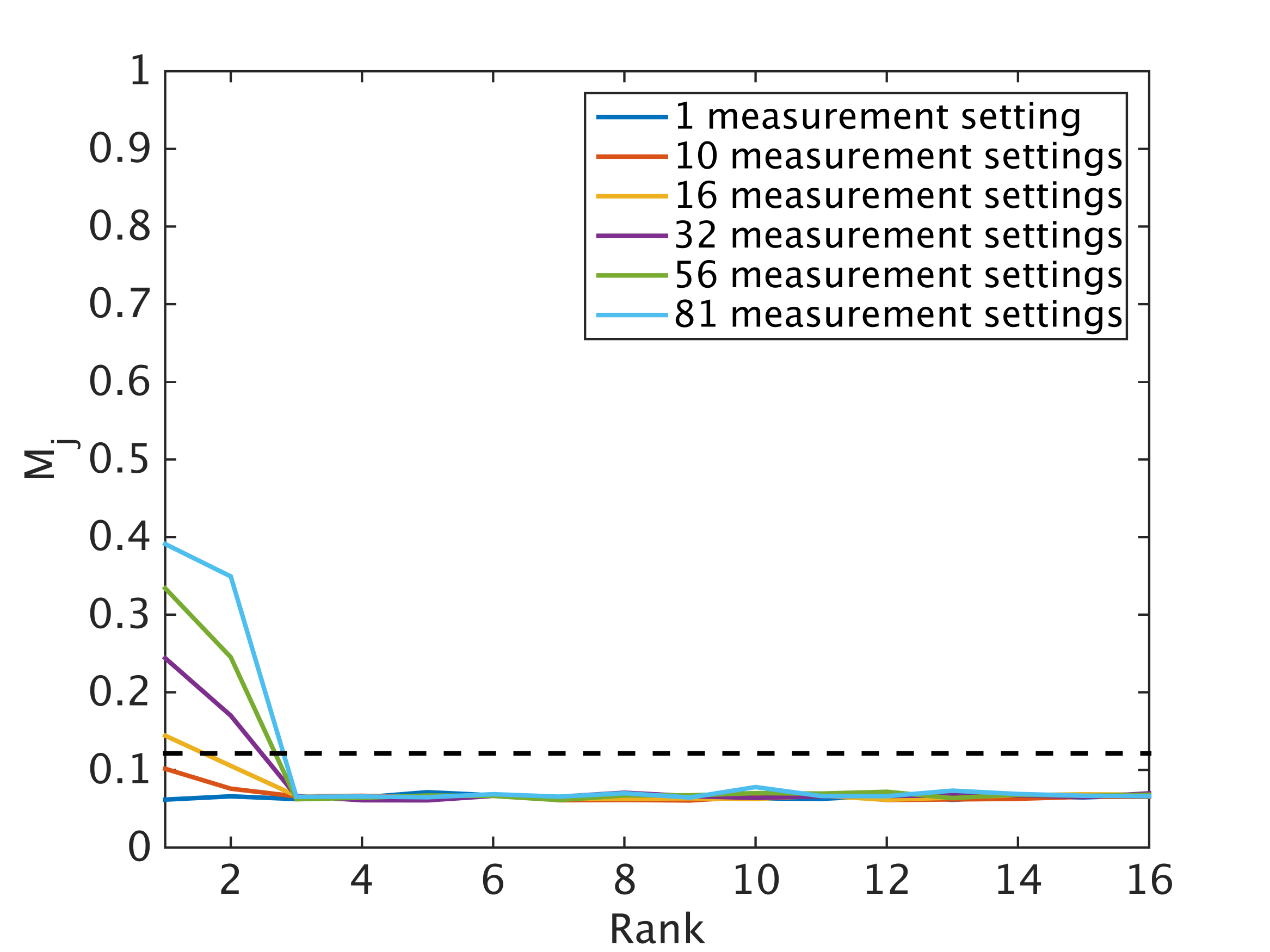}}
\quad
\subfigure[ \, 16 repetitions per measurement setting.]{%
  \includegraphics[width=.65\columnwidth]{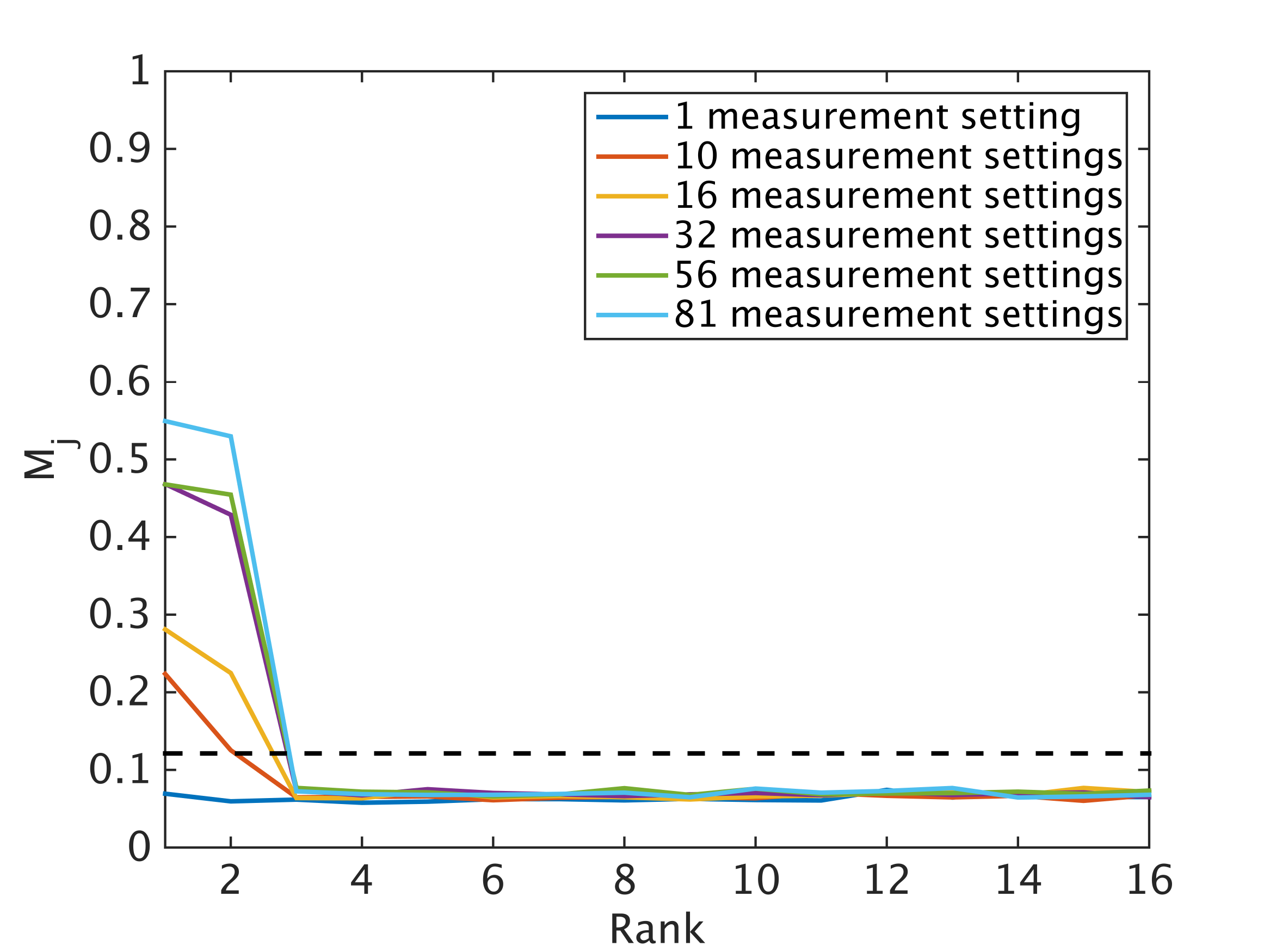}}
\quad
\subfigure[\, 100 repetitions per measurement setting.]{%
  \includegraphics[width=.65\columnwidth]{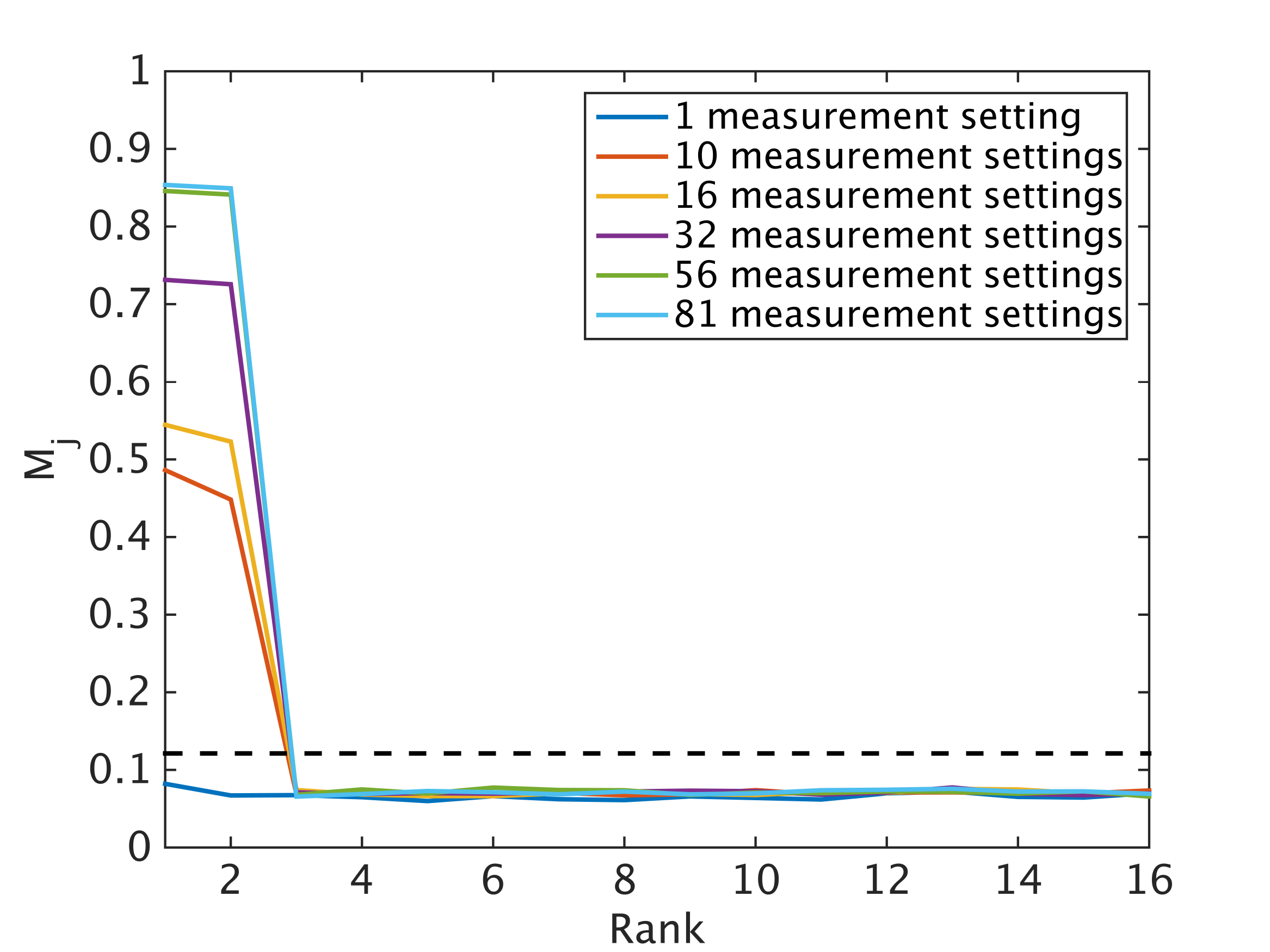}}
\caption{Mean eigenvector overlap as a function of the rank truncation of the recovered density matrix of a true state of rank $2$. The dotted line represents the threshold given in eq.\ \eqref{Eq:threshold}.}
\label{Rank2_Sim}
\end{figure*}

\begin{figure*}[t]
\subfigure[\, 5 repetitions per measurement setting.]{%
  \includegraphics[width=.65\columnwidth]{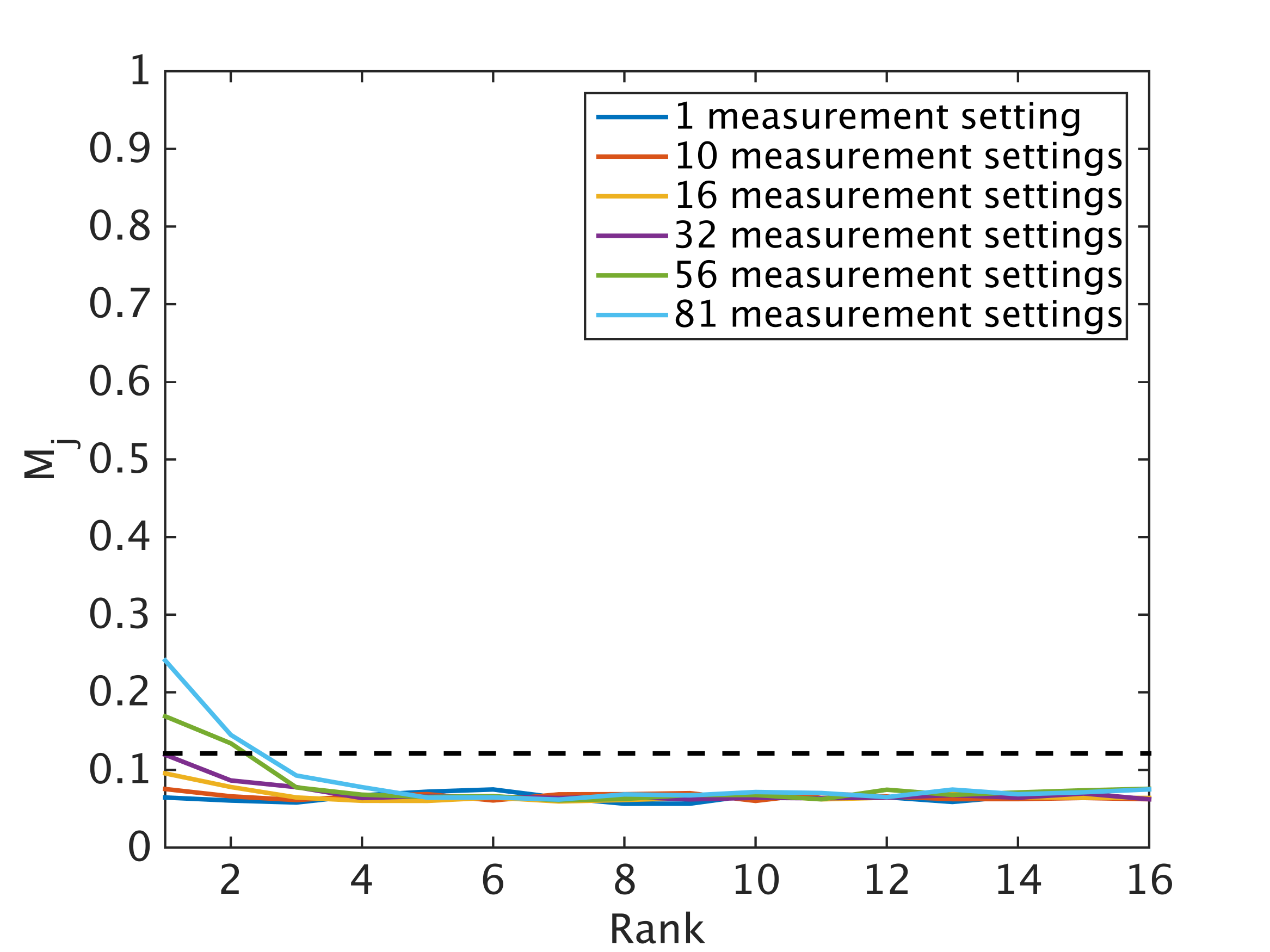}}
\quad
\subfigure[ \, 16 repetitions per measurement setting.]{%
  \includegraphics[width=.65\columnwidth]{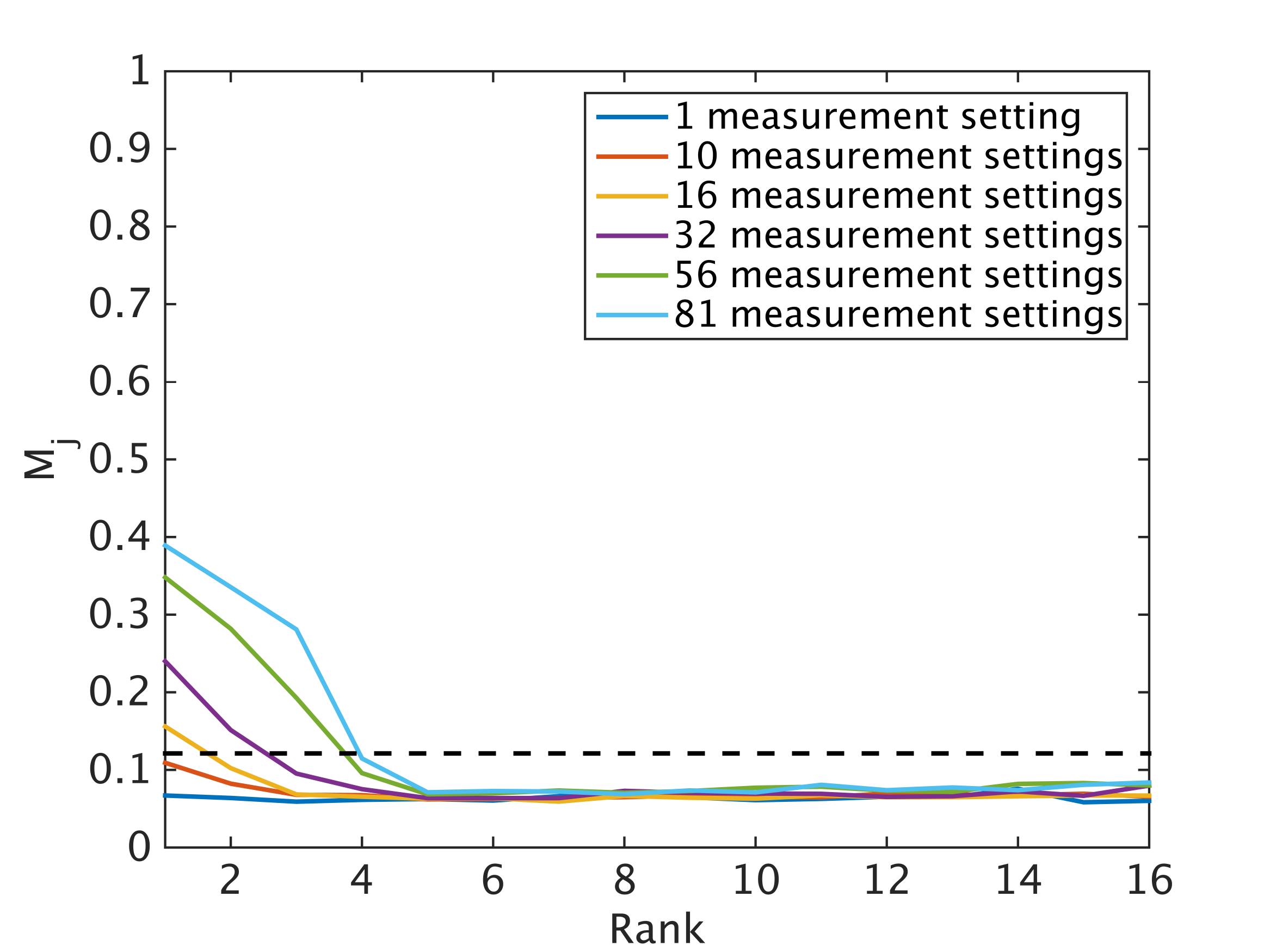}}
\quad
\subfigure[\, 100 repetitions per measurement setting.]{%
  \includegraphics[width=.65\columnwidth]{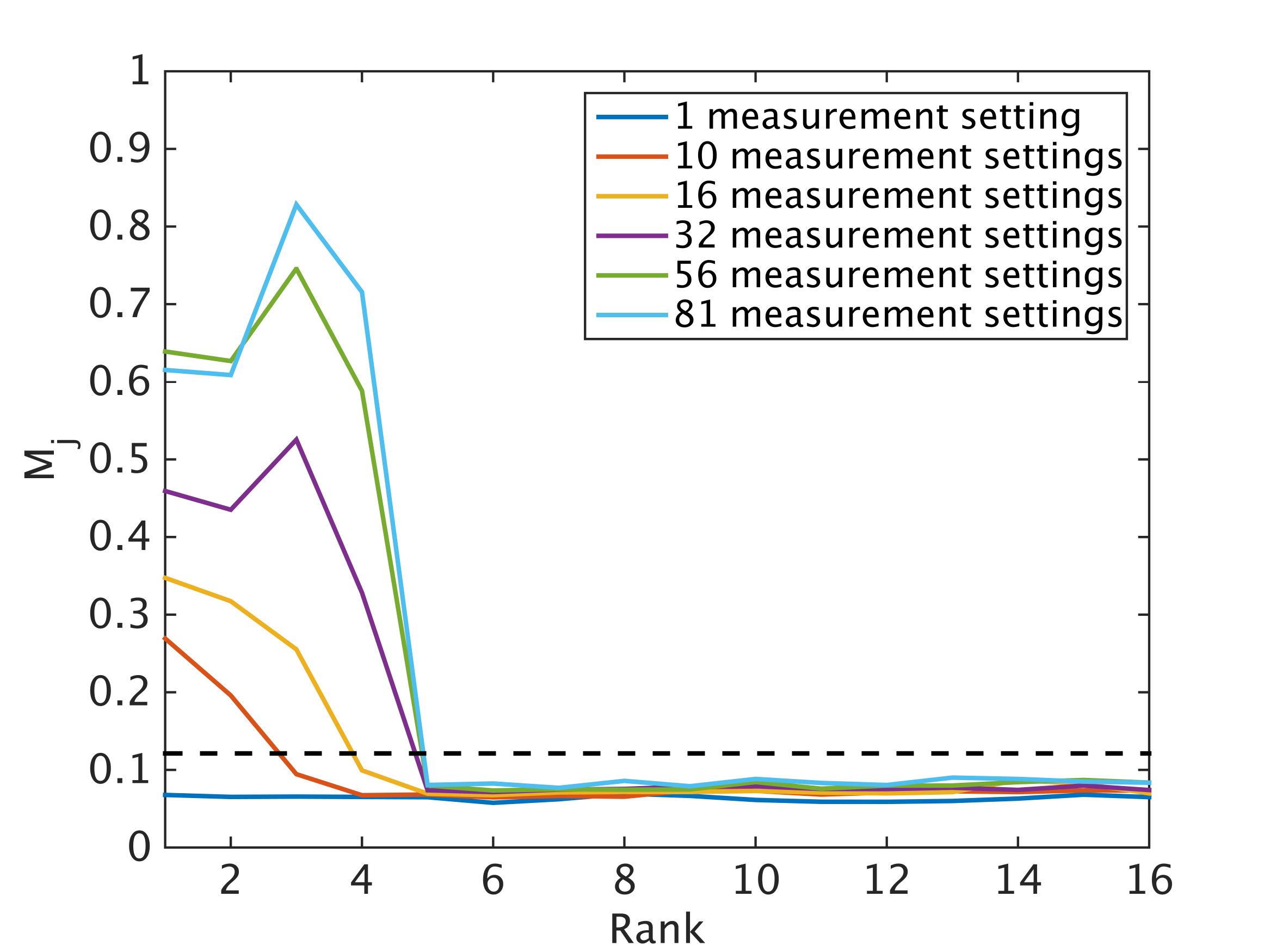}}
\caption{Mean eigenvector overlap as a function of the rank truncation of the recovered density matrix of a true state of rank $4$. The dotted line represents the threshold given in eq.\ \eqref{Eq:threshold}.}
\label{Rank4_Sim}
\end{figure*}

\begin{figure*}[t]
\subfigure[\, 5 repetitions per measurement setting.]{%
  \includegraphics[width=.65\columnwidth]{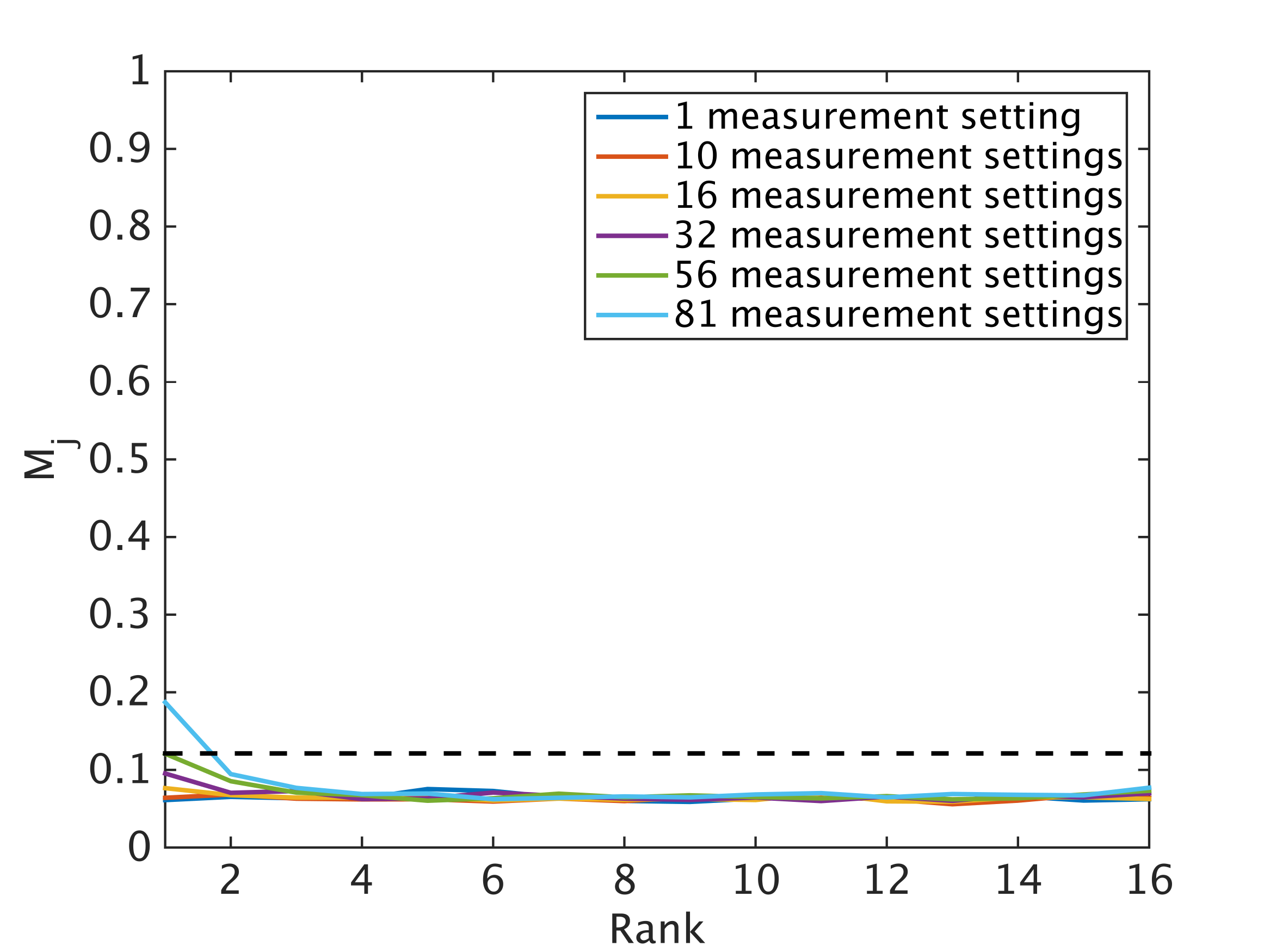}}
\quad
\subfigure[ \, 16 repetitions per measurement setting.]{%
  \includegraphics[width=.65\columnwidth]{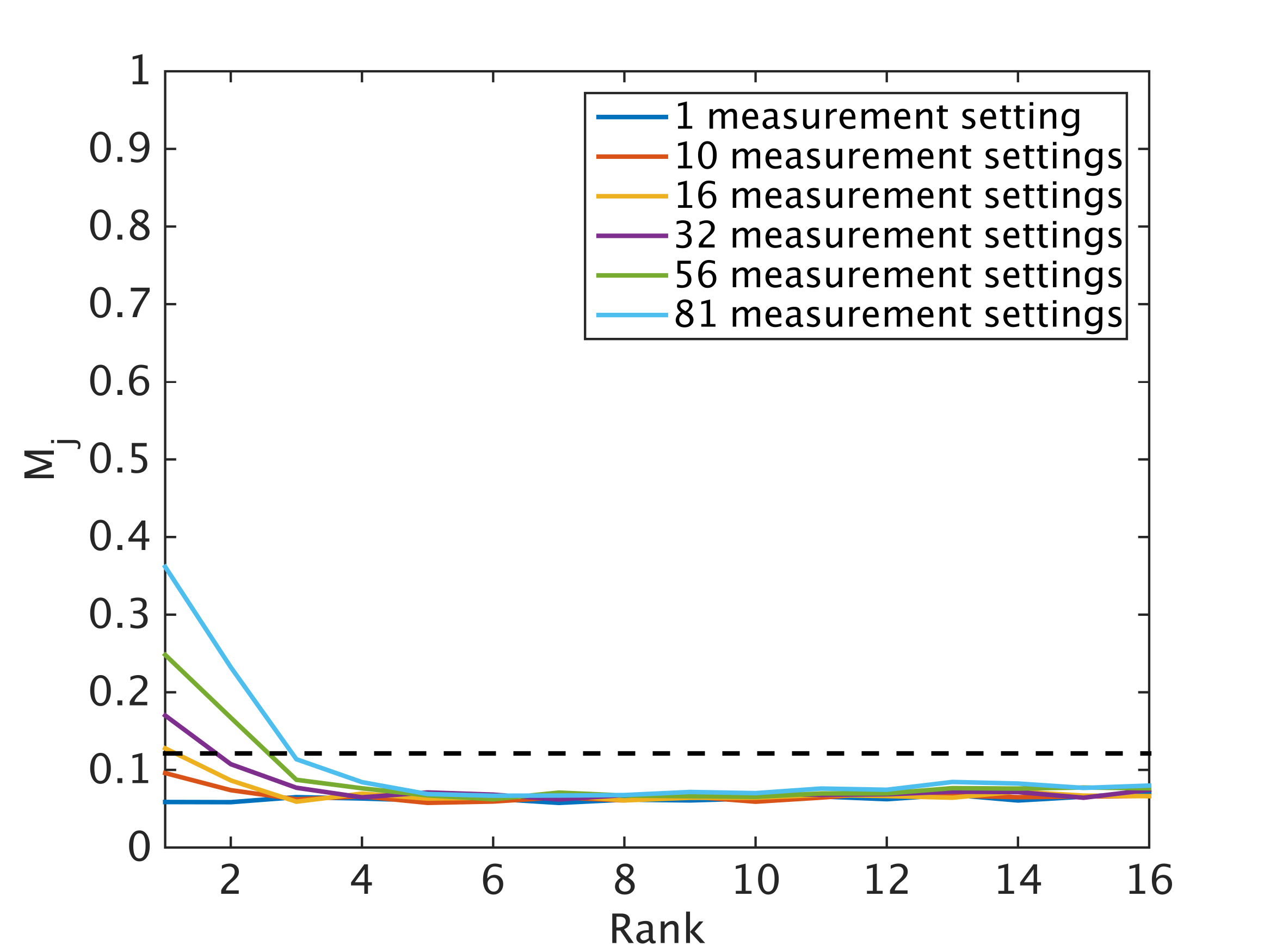}}
\quad
\subfigure[\, 100 repetitions per measurement setting.]{%
  \includegraphics[width=.65\columnwidth]{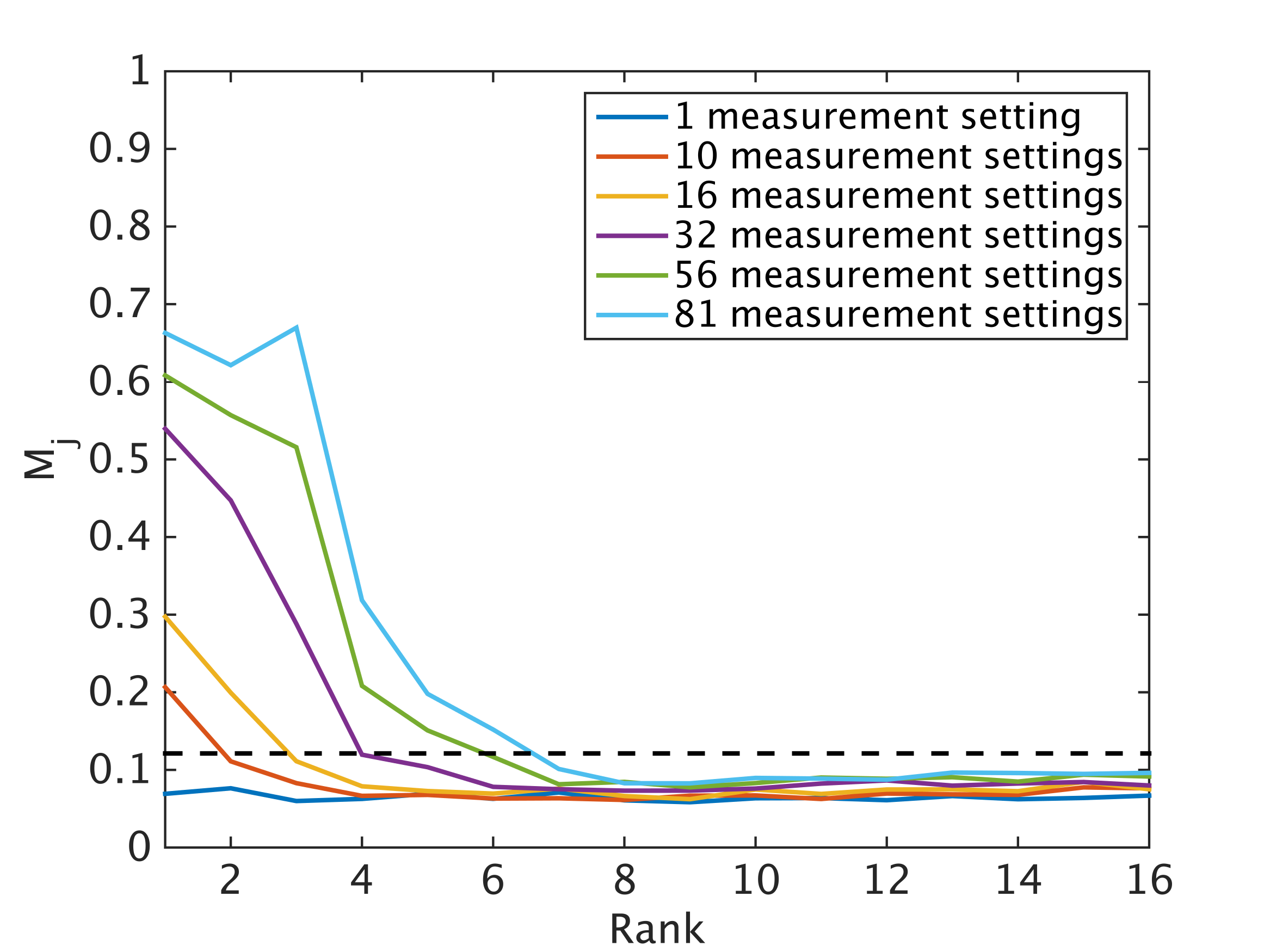}}
\caption{Mean eigenvector overlap as a function of the rank truncation of the recovered density matrix of a true state of rank $8$. The dotted line represents the threshold given in eq.\ \eqref{Eq:threshold}.}
\label{Rank8_Sim}
\end{figure*}

\begin{figure*}[t]
\subfigure[\, Rank 1 true state.]{%
  \includegraphics[width=.65\columnwidth]{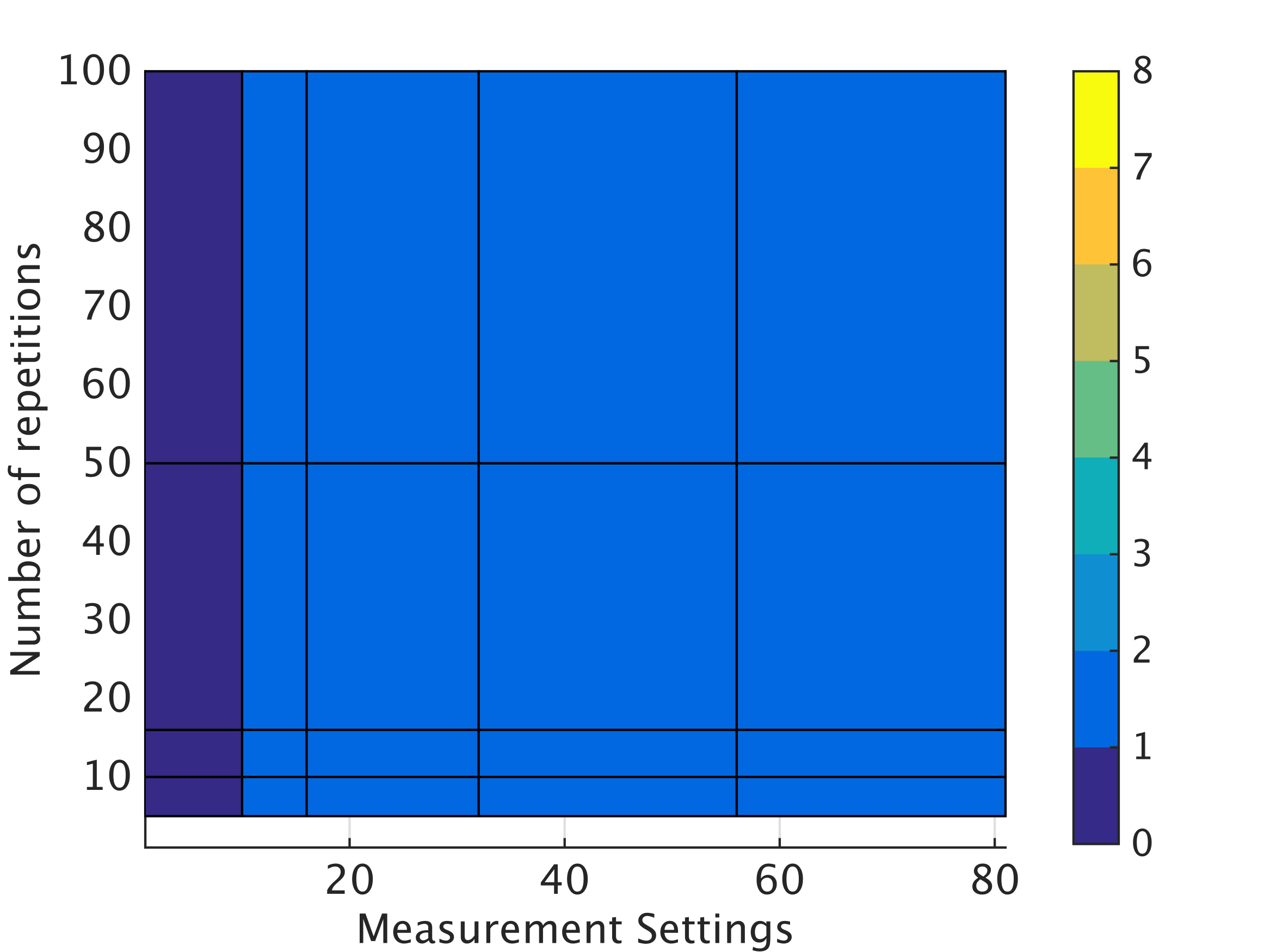}}
\quad
\subfigure[ \, Rank 4 true state.]{%
  \includegraphics[width=.65\columnwidth]{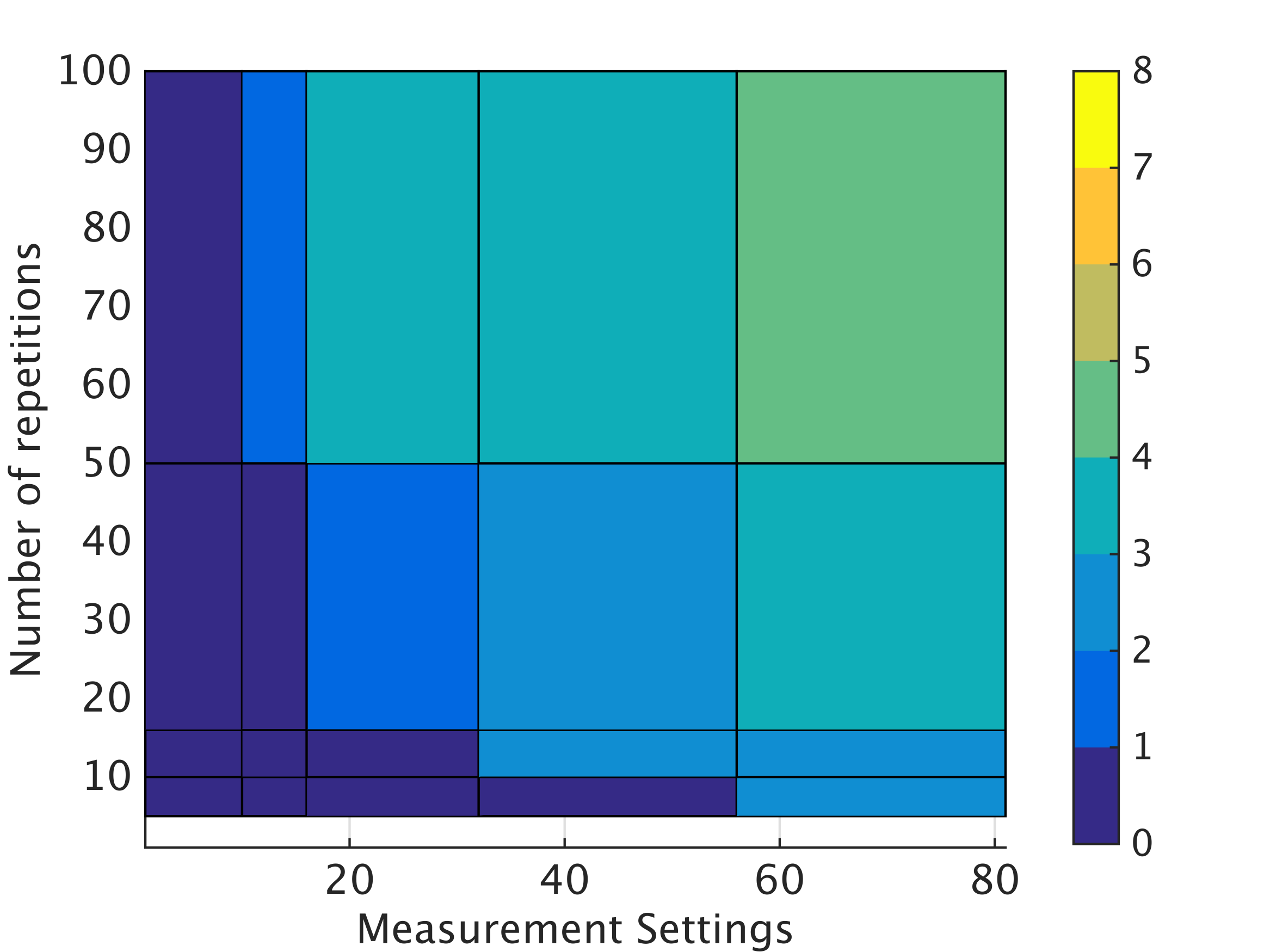}}
\quad
\subfigure[\, Rank 8 true state.]{%
  \includegraphics[width=.65\columnwidth]{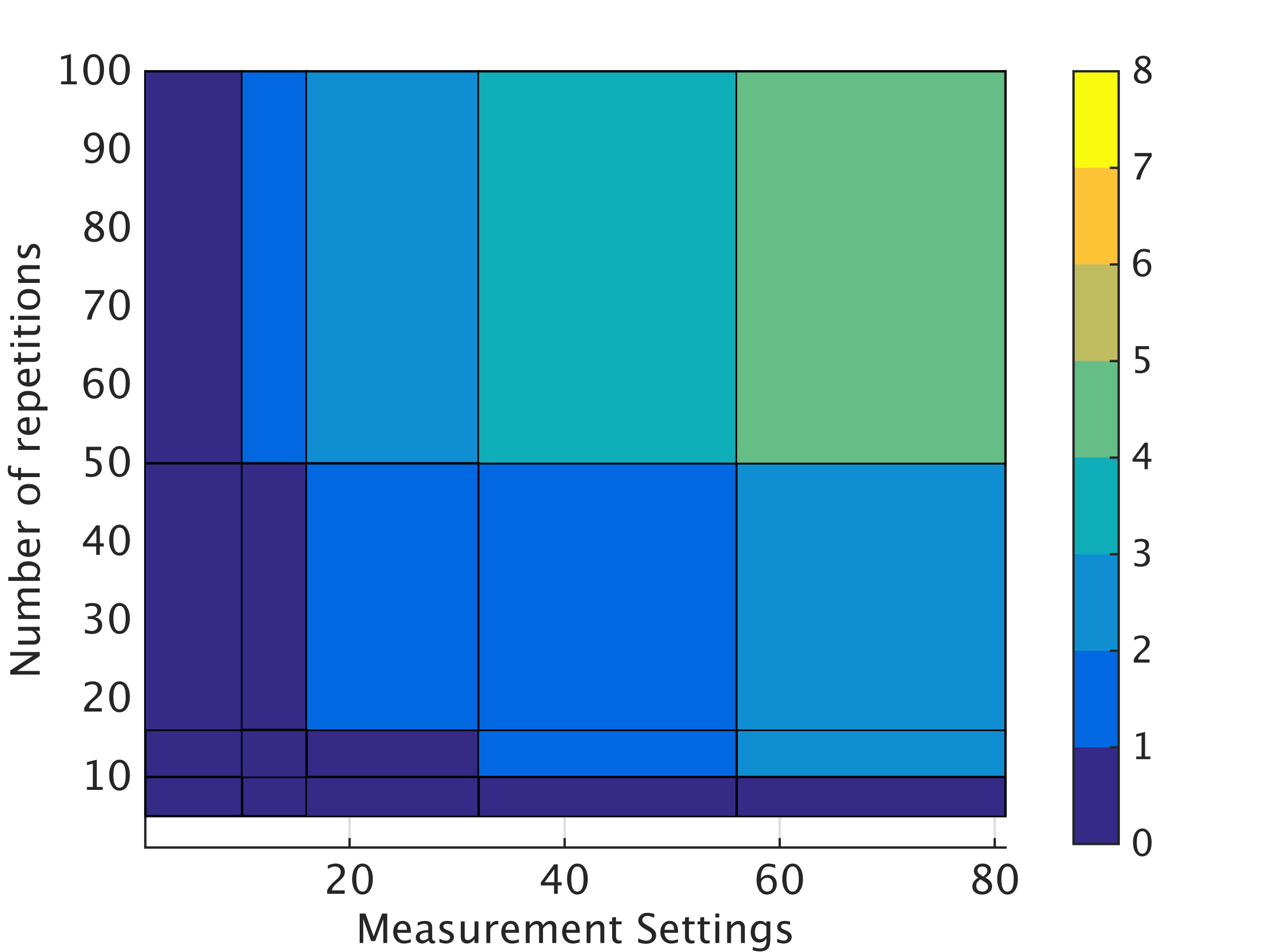}}
\caption{Estimated rank according to eq.\ \eqref{Eq:SpectralThresholding} as a function of the number of measurement settings and number of repetitions per measurement. When insufficient information is available for our criterion to provide a non-zero rank, we choose a rank one estimate by default.}
\label{Rank_Sim}
\end{figure*}

\begin{figure*}[t]
\subfigure[\, Rank 1 true state.]{%
  \includegraphics[width=.65\columnwidth]{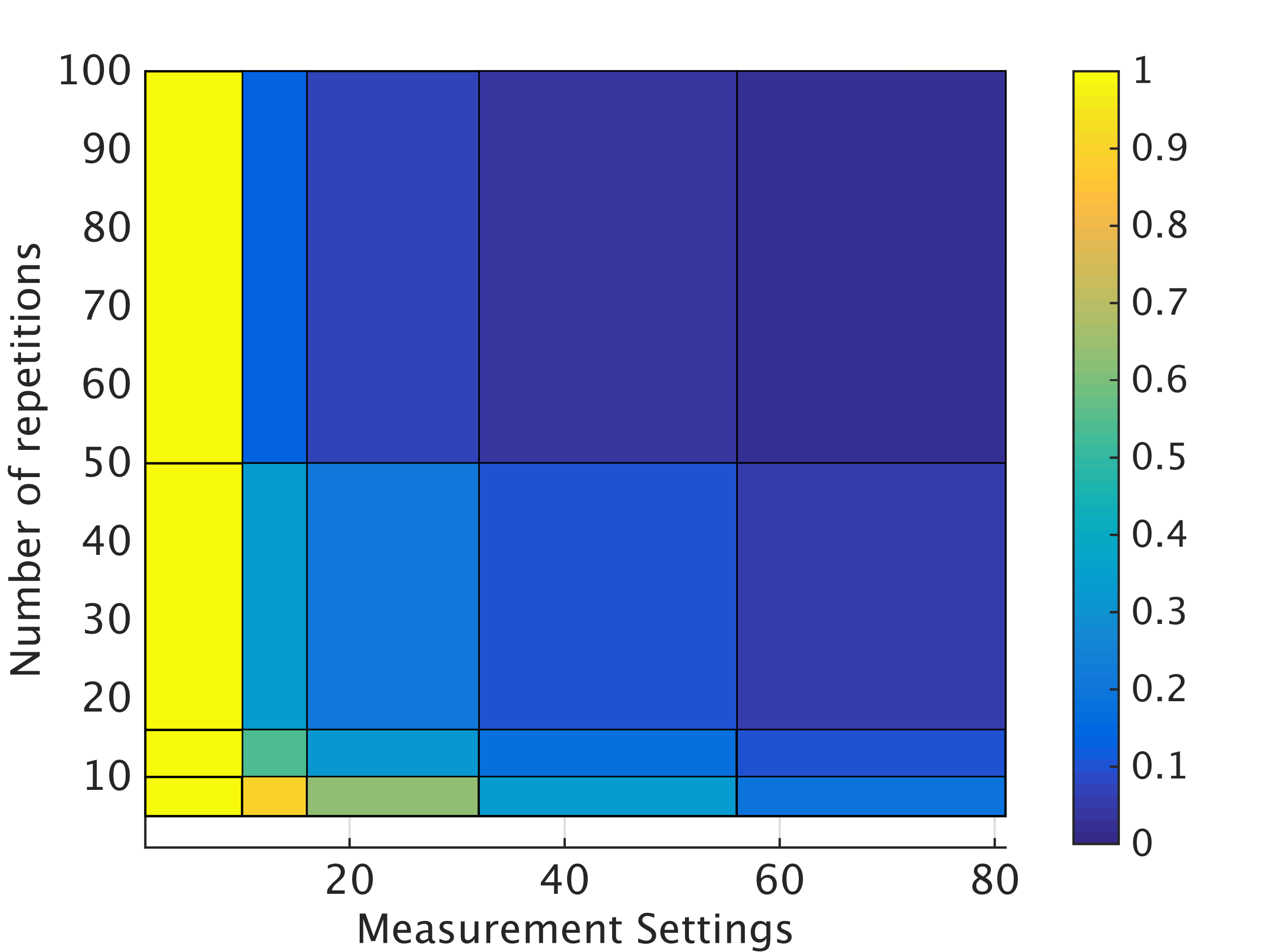}}
\quad
\subfigure[ \, Rank 4 true state.]{%
  \includegraphics[width=.65\columnwidth]{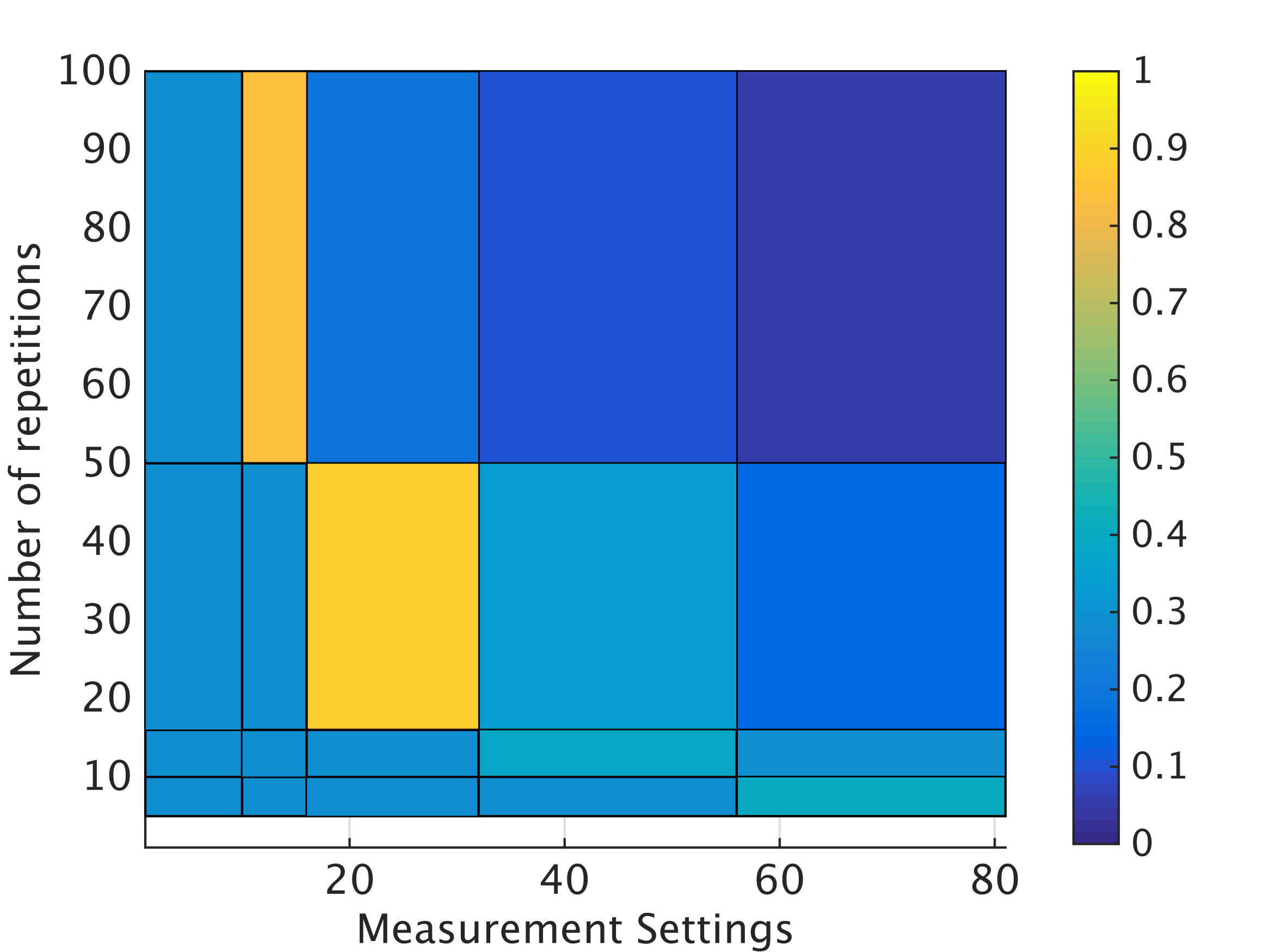}}
\quad
\subfigure[\, Rank 8 true state.]{%
  \includegraphics[width=.65\columnwidth]{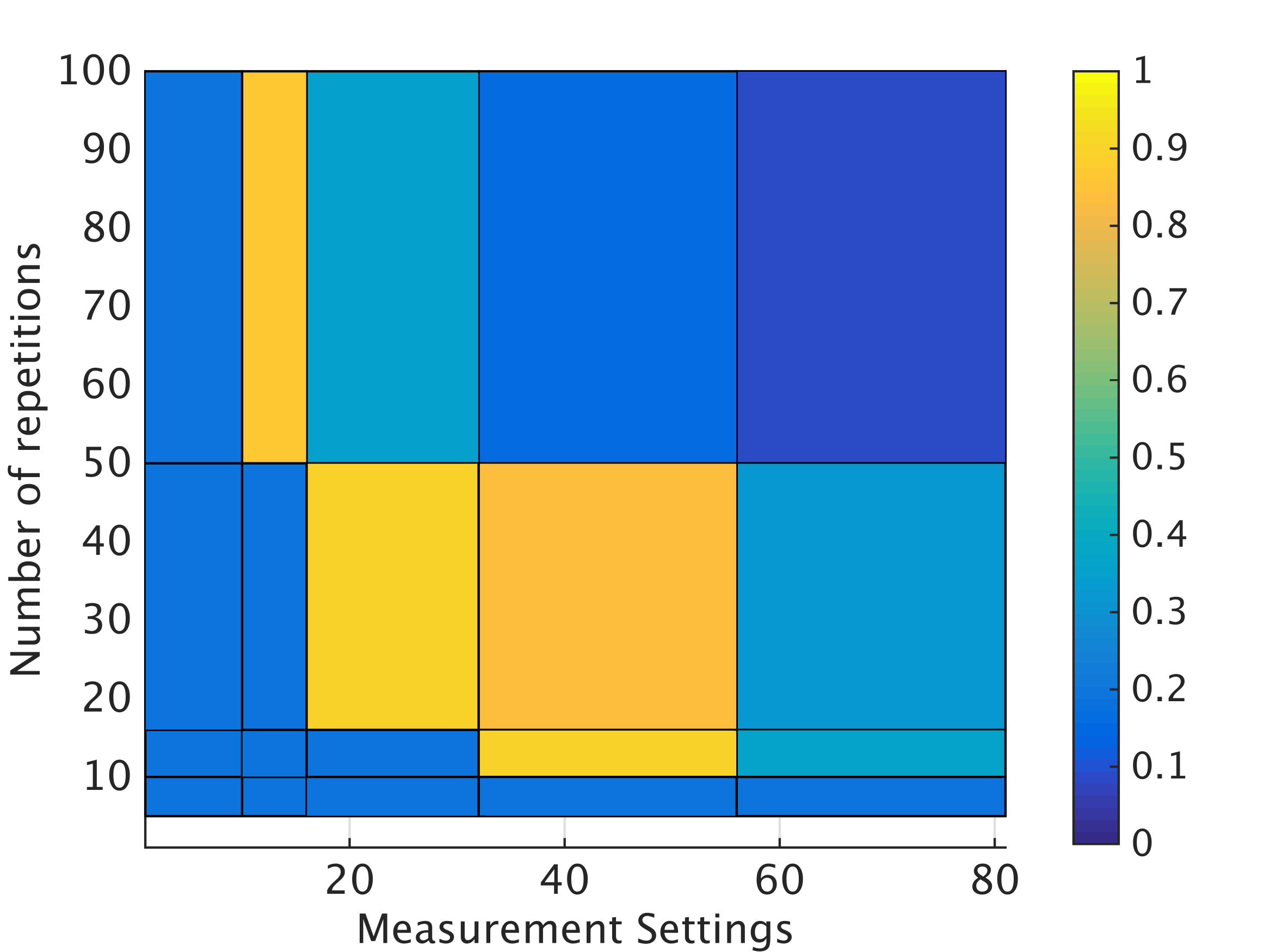}}
\caption{Mean risk, eq.\ \eqref{Eq:MeanRisk}, computed for the thresholding procedure
eq.\ \eqref{Eq:SpectralThresholding} of the main text.}
\label{MeanRisk}
\end{figure*}

In order to test our model selection protocol, we perform numerical simulations in a relatively small system of $L=4$ qubits. We proceed by generating random states of fixed rank: in this example, we choose ranks $1$, $2$, $4$, and $8$. These states, which we refer to as the true states $\rho$, are used to simulate the outcomes of a set of $1$, $10$, $16$, $32$, $56$, and $81$ local random measurements. Each simulated measurement, in turn, is done for different repetitions per observable. We choose, for comparison, $5$, $10$, $16$, and $100$ repetitions per measurement setting. More repetitions means less noise in the observed outcomes. Each numerical experiment is repeated $100$ times and the reconstruction of the density matrix is done via eq.~\eqref{eq:lasso} with $\mu=0$. Then, we compute our figure of merit $M_j$ as indicated in eq.~\eqref{Eq:Overlap}.
The results are presented in figs.~\ref{Rank1_Sim}-\ref{Rank8_Sim}. It is clear from the plots that the method
described in the main text---when given enough measurements---can, in principle, distinguish the correct rank of the true state (see fig.~\ref{Rank_Sim}). However, in the informationally incomplete
low-data regime that we are interested in, it tends to give a lower rank than the true rank. This is indeed a desirable
feature since the amount of collected data is not enough to justify a higher rank fit. Additionally, for benchmarking we compute the expected risk defined by
\begin{equation}\label{Eq:MeanRisk}
	\mathbbm{E} \|\rho-\rho_k \|_2^2,
\end{equation}
where $\rho$ is the true state and $\rho_k$ is the reconstructed and truncated state
from spectral thresholding. This is shown in fig.~\ref{MeanRisk}.

\begin{figure*}[t]
\subfigure[\, Rank 1 true state.]{%
  \includegraphics[width=.65\columnwidth]{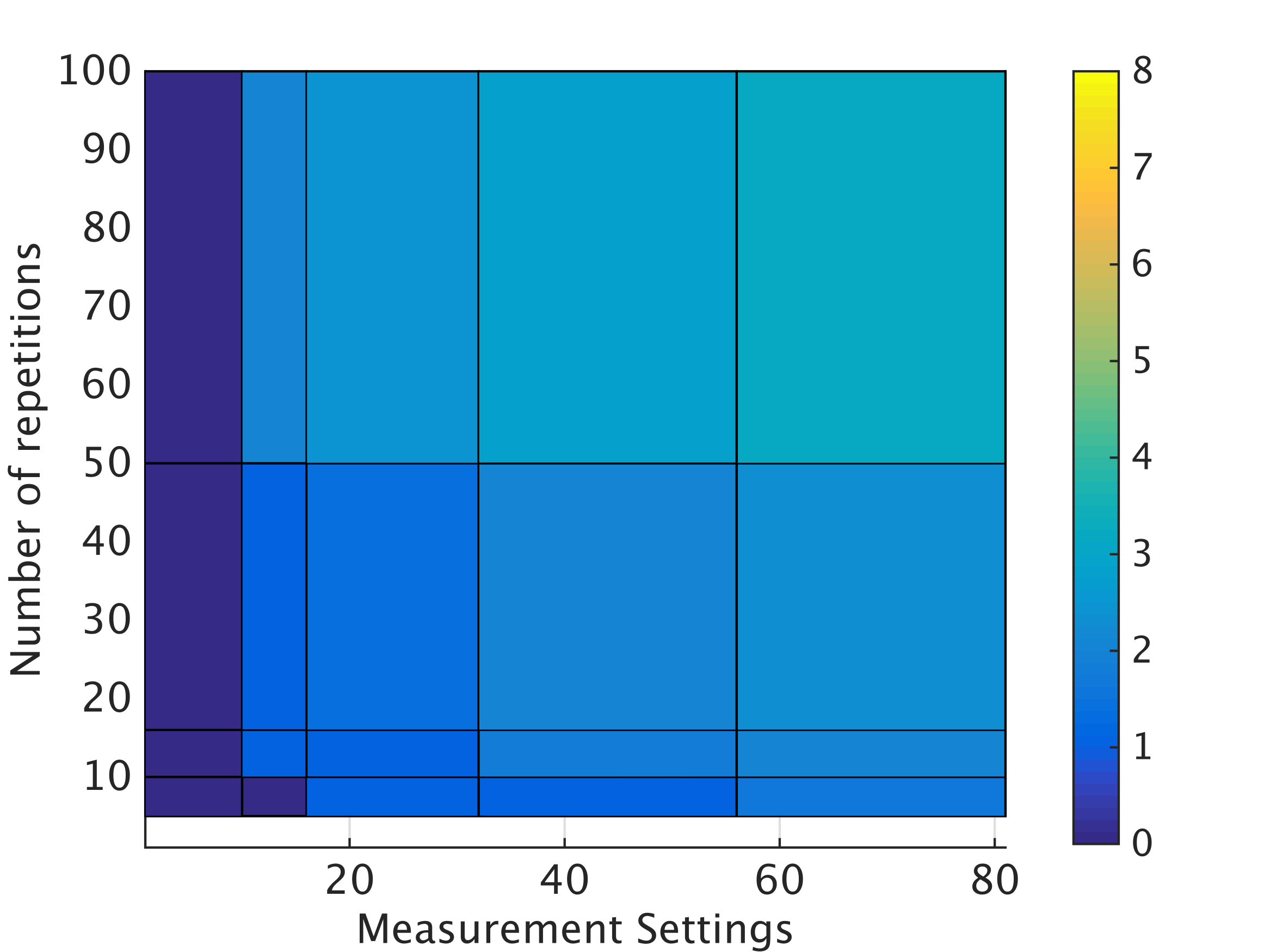}}
\quad
\subfigure[ \, Rank 4 true state.]{%
  \includegraphics[width=.65\columnwidth]{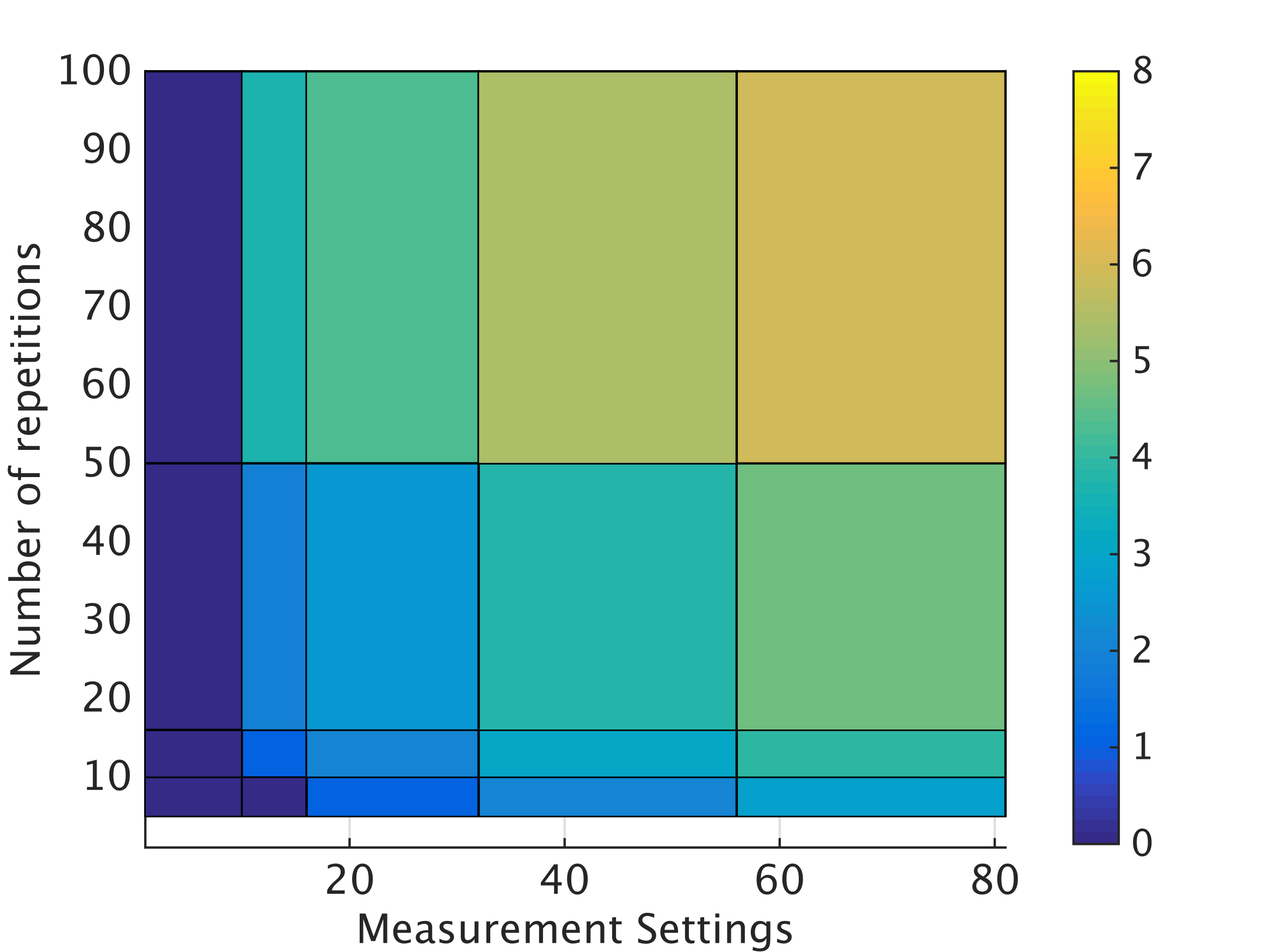}}
\quad
\subfigure[\, Rank 8 true state.]{%
  \includegraphics[width=.65\columnwidth]{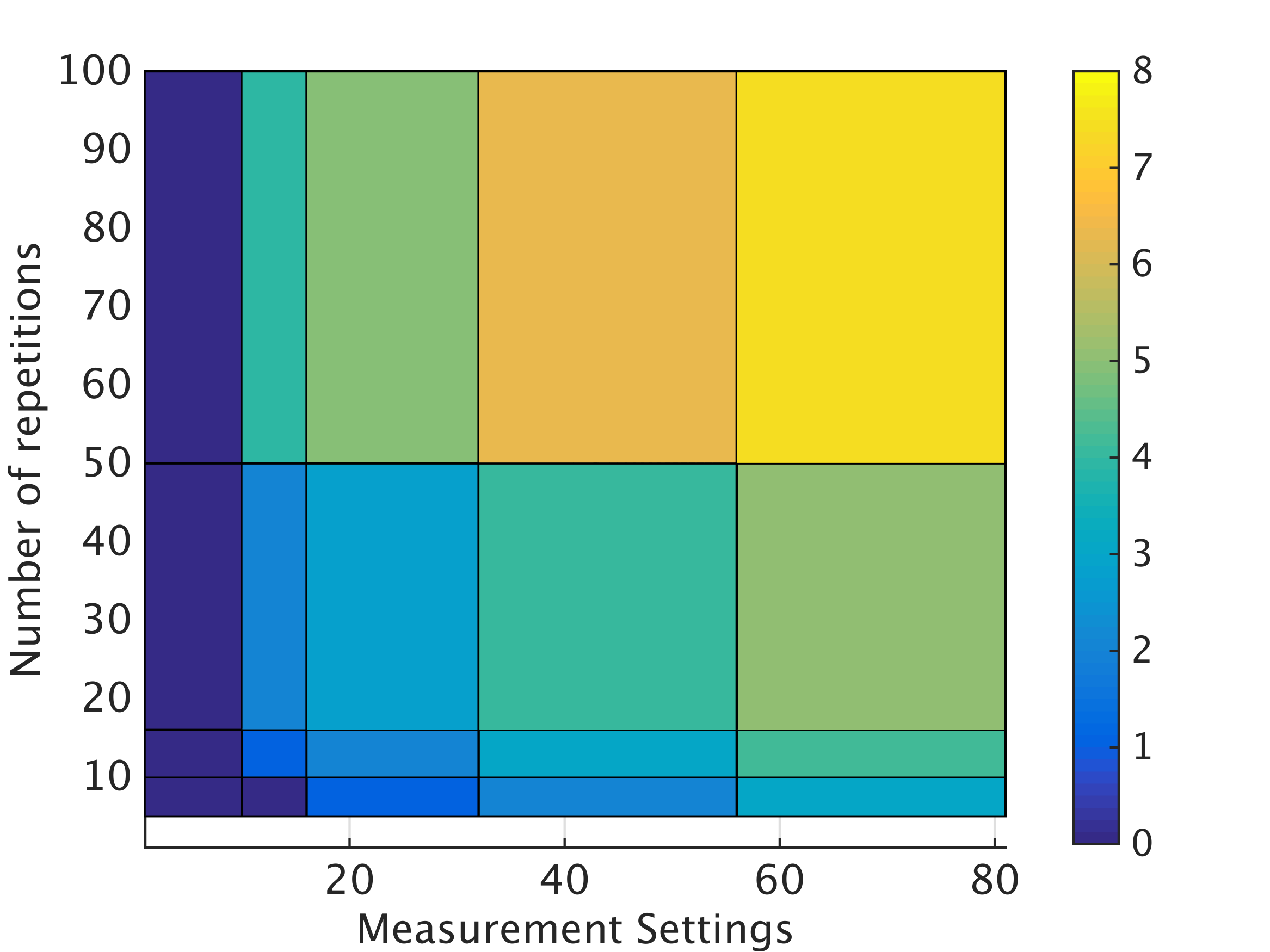}}
\caption{Estimated mean rank according to the spectral thresholding method developed in ref.~\citenum{Guta2}
with positivity constraint. The method seems to overfit a little when not enough measurements are used in the reconstruction. Also, note that we are plotting the mean reconstructed rank, which does not take discrete values.}
\label{RankGuta}
\end{figure*}

\begin{figure*}[t]
\subfigure[\, Rank 1 true state.]{%
  \includegraphics[width=.65\columnwidth]{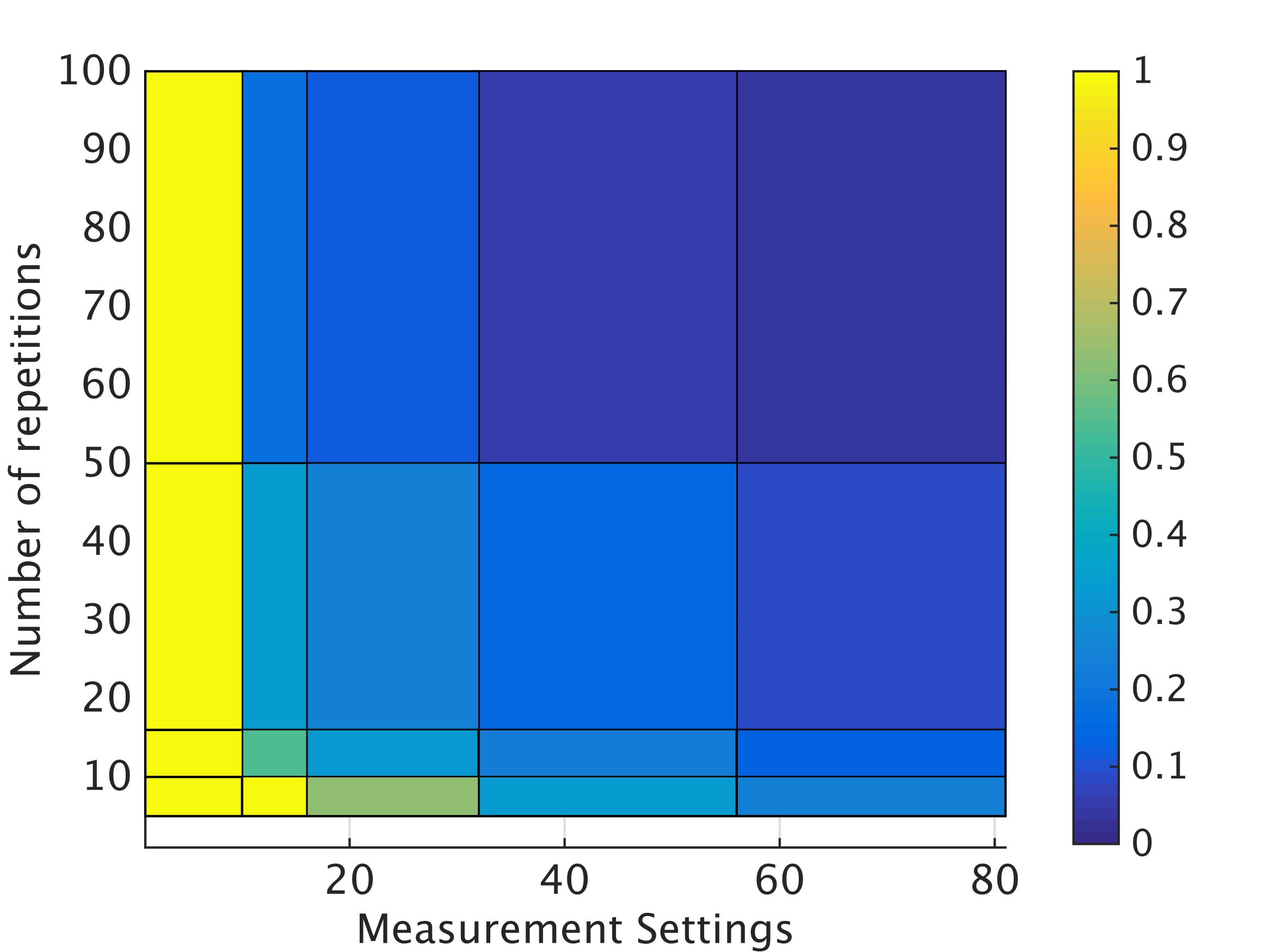}}
\quad
\subfigure[ \, Rank 4 true state.]{%
  \includegraphics[width=.65\columnwidth]{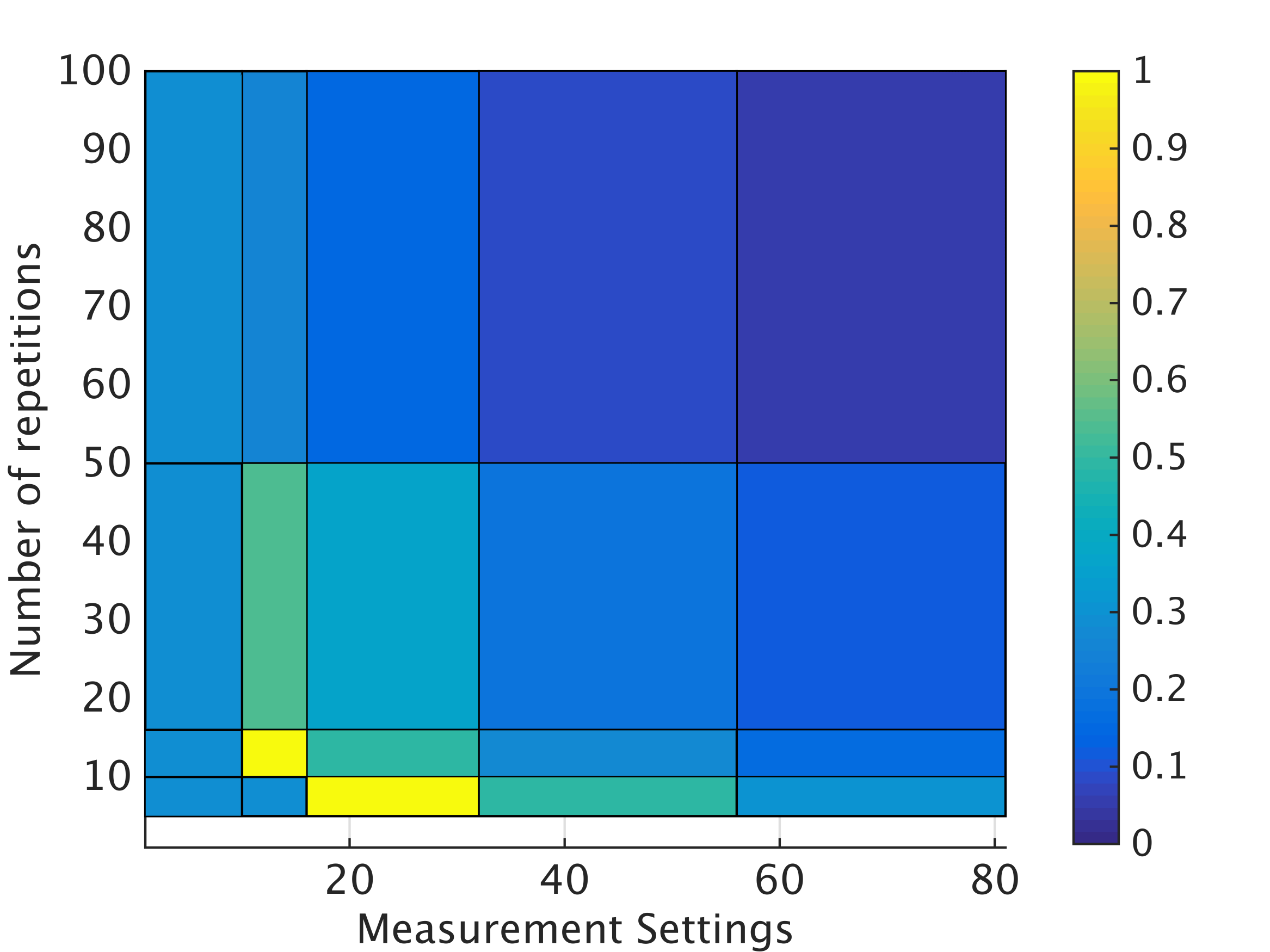}}
\quad
\subfigure[\, Rank 8 true state.]{%
  \includegraphics[width=.65\columnwidth]{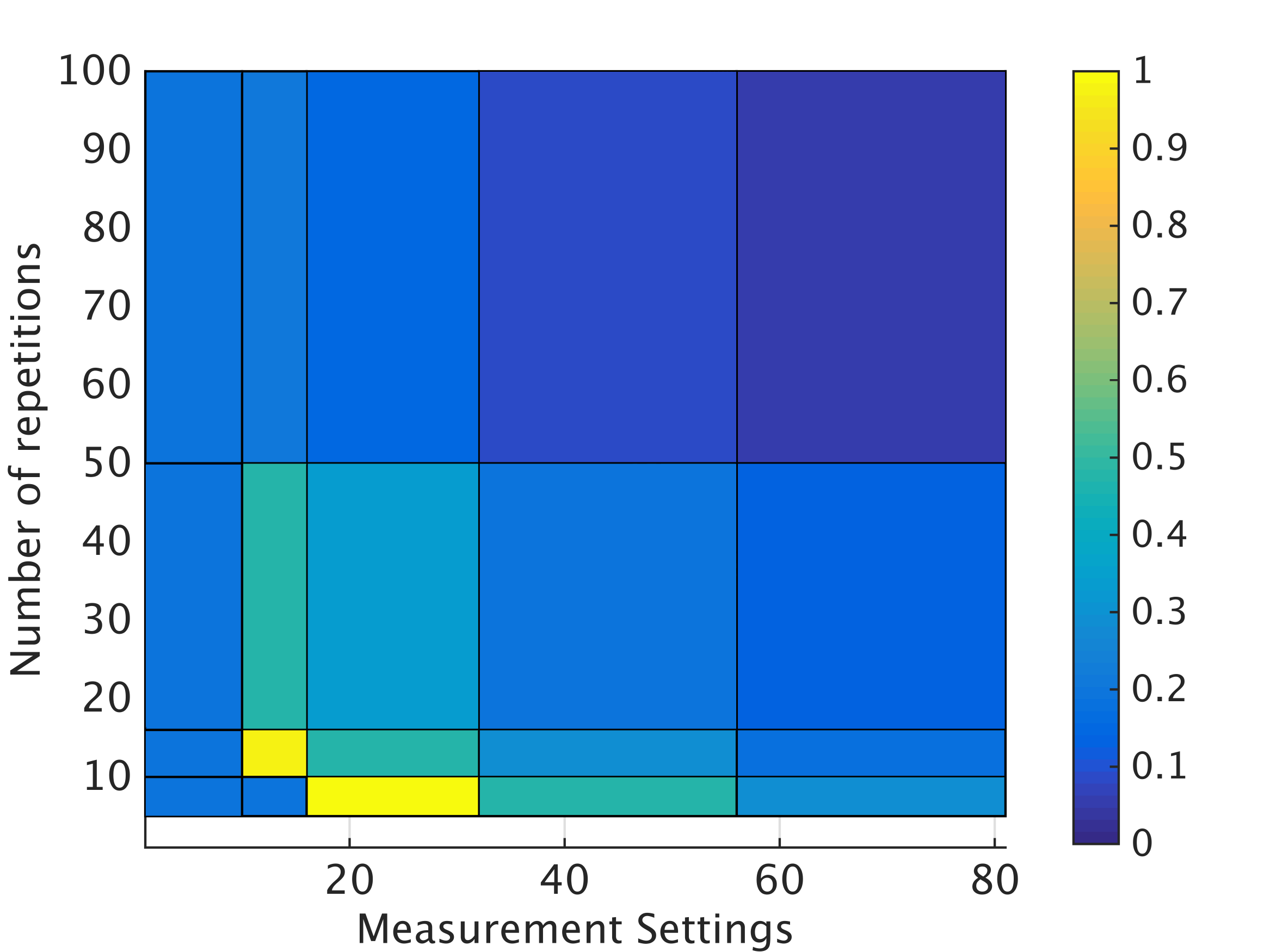}}
\caption{Mean risk, eq.\ \eqref{Eq:MeanRisk}, computed for the thresholding procedure in ref.~\citenum{Guta2}
with positivity constraint.}
\label{MeanRiskGuta}
\end{figure*}

Since the procedure outlined in this work is related to model selection, for completeness we compare our method to the one proposed in ref.~\citenum{Guta2}. There the authors have developed a particular spectral thresholding algorithm that is applied to a reconstructed density matrix via a plain least squares estimator that does not impose the positivity constraint. Even more, their approach is valid for informationally complete measurements and rigorous proofs are given for its performance. In our test, we have modified the approach in ref.~\citenum{Guta2} to include the positivity constraint (via the \texttt{LS-SDP} estimator) and have naively applied it to the regime of informationally incomplete measurements. As suggested in ref.~\citenum{Guta2}, we choose the threshold parameter as
\begin{equation}
	\gamma(\epsilon)^2 = \frac{2d}{N}\log{\frac{2d}{\epsilon}}.
\end{equation}
In our example, we select $\epsilon = 0.05$ and set $N$ to be the total number of prepared quantum states used in the simulation of a particular experiment.
For every reconstructed density matrix, computed via eq.\ \eqref{eq:lasso} with $\mu=0$, we calculate its spectral decomposition and set all eigenvalues smaller than $4\gamma(\epsilon)$ equal to zero as prescribed in ref.~\citenum{Guta2}. After this thresholding procedure, the spectrum is no longer normalized, and we correct this by shifting all eigenvalues by the same quantity (as opposed to dividing by its sum). The average rank, over 100 experiments, obtained by this method is shown in fig.~\ref{RankGuta}. It is clear that the method performs well, as expected, for the informationally complete case. However, when not enough measurements are used in the reconstruction, it seems to (on average) estimate a rank that is greater than what our thresholding method gives (see fig.~\ref{Rank_Sim}). In terms of risk, as defined in eq.~\eqref{Eq:MeanRisk}, and as shown in fig.~\ref{MeanRiskGuta}, we cannot see a very clear trend that distinguishes both methods, except that on average they seem to be similar in risk. Furthermore, we should notice that, for the informationally incomplete regime we are interested in analyzing,
to date there is not a rigorously proven method for spectral thresholding available yet.

\subsection*{Spectral thresholding with experimental data: \\Additional results}
\begin{figure*}[t]
\subfigure[\,Trace norm minimization estimate ($F = 0.91$), corresponding to the
result of the matrix Lasso in eq.\ \eqref{eq:lasso} in the limit $\mu\rightarrow \infty$.]{%
  \includegraphics[width=.65\columnwidth]{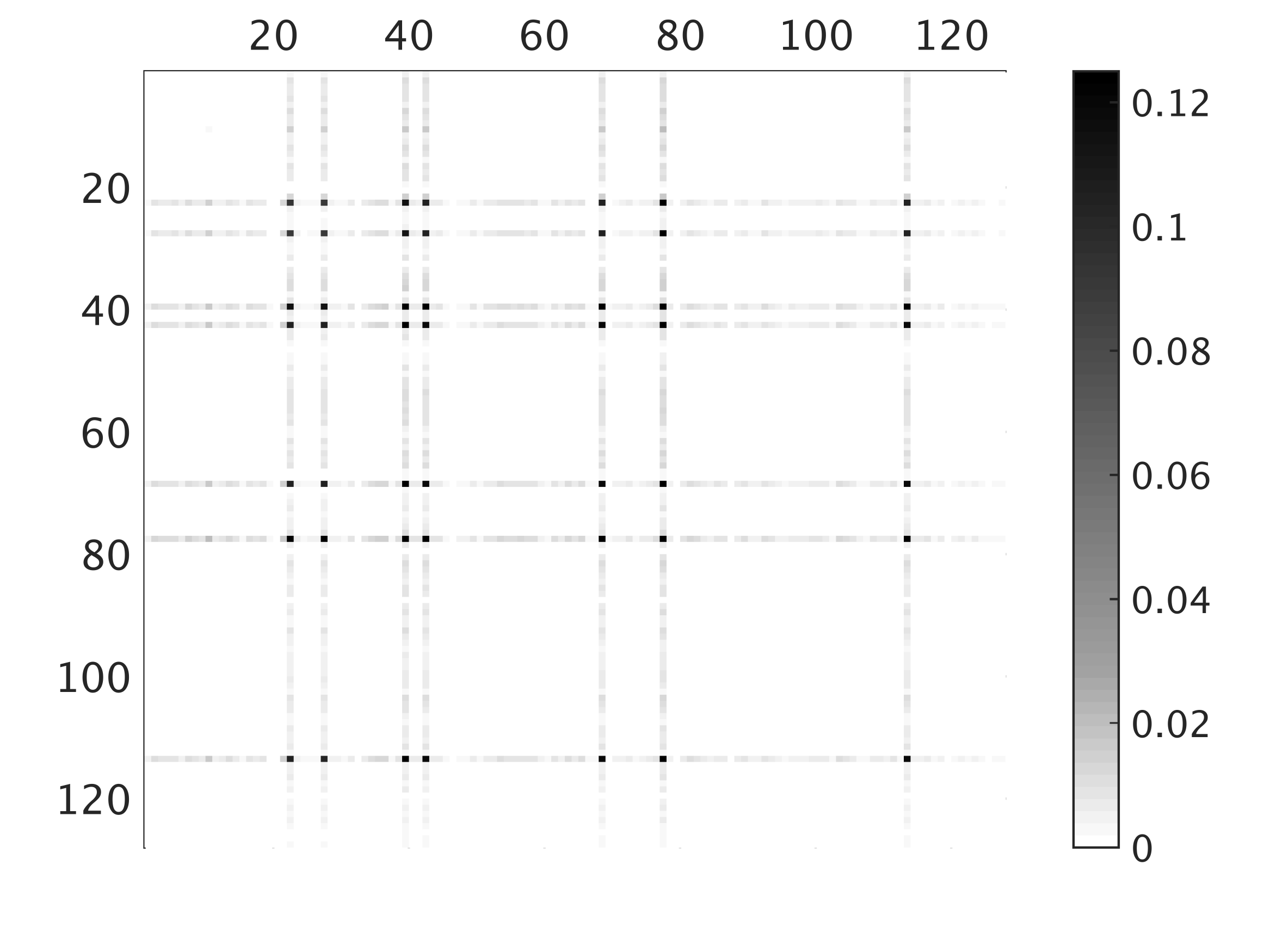}}
\subfigure[ \, Least squares estimate ($F = 0.23$), corresponding to $\mu=0$.]{%
  \includegraphics[width=.65\columnwidth]{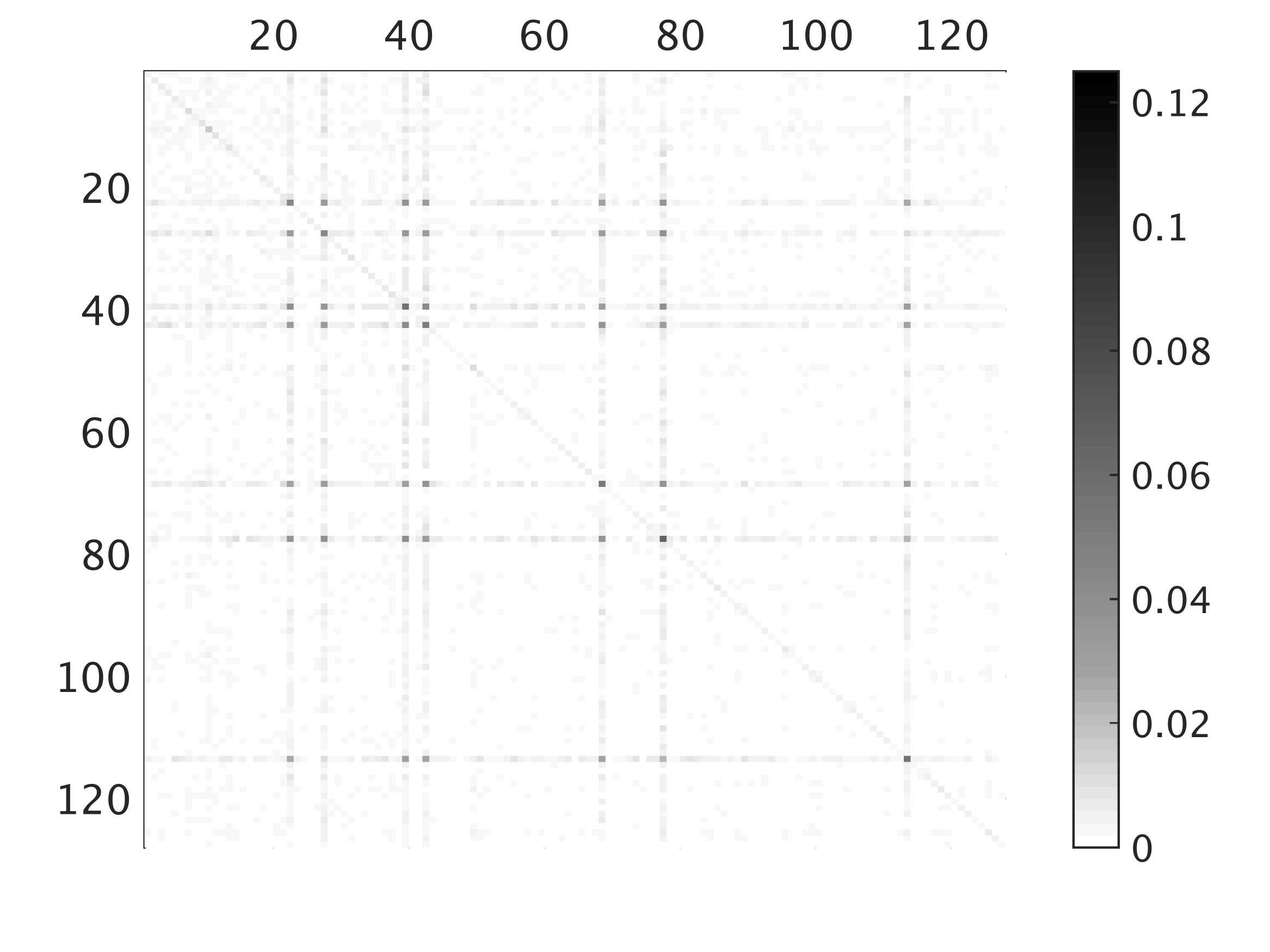}}
\quad
\subfigure[\, Rank $37$ leading subspace projection of the least squares estimate ($F = 0.25$) obtained by our spectral thresholding method.]{%
  \includegraphics[width=.65\columnwidth]{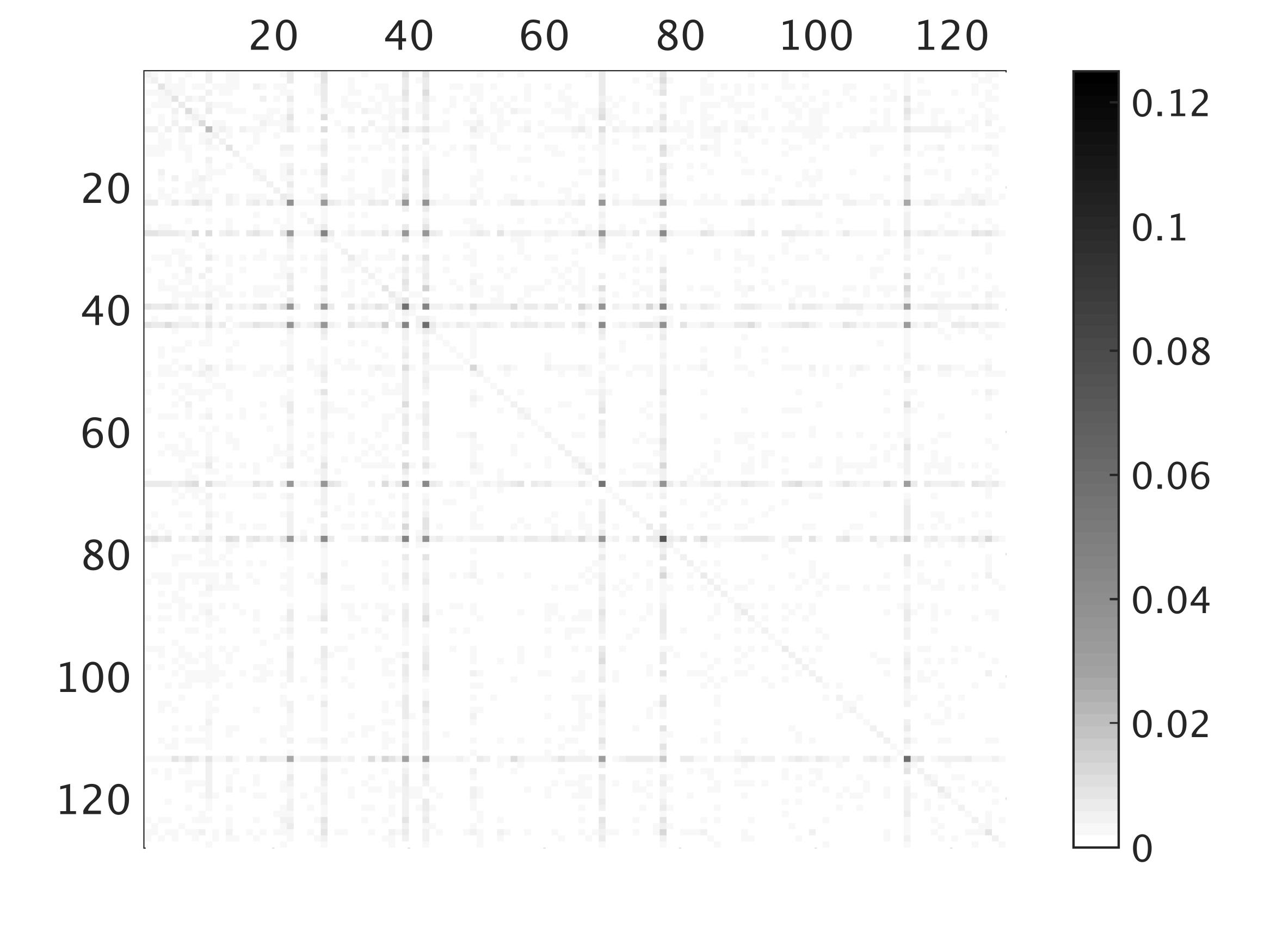}}
\caption{Example of quantum state reconstruction for the logical $|\bar{1}\rangle $
state vector. A 2-D plot of the absolute values of the entries of the density matrix is presented. The same notation as
in fig.\ \ref{Data1} in the main text is used.}
\label{Data2}
\end{figure*}

\begin{figure*}[t]
\subfigure[\,Trace norm minimization estimate ($F = 0.94$),
corresponding to the
result of the matrix Lasso in eq.\ \eqref{eq:lasso} in the limit $\mu\rightarrow \infty$.]{%
  \includegraphics[width=.65\columnwidth]{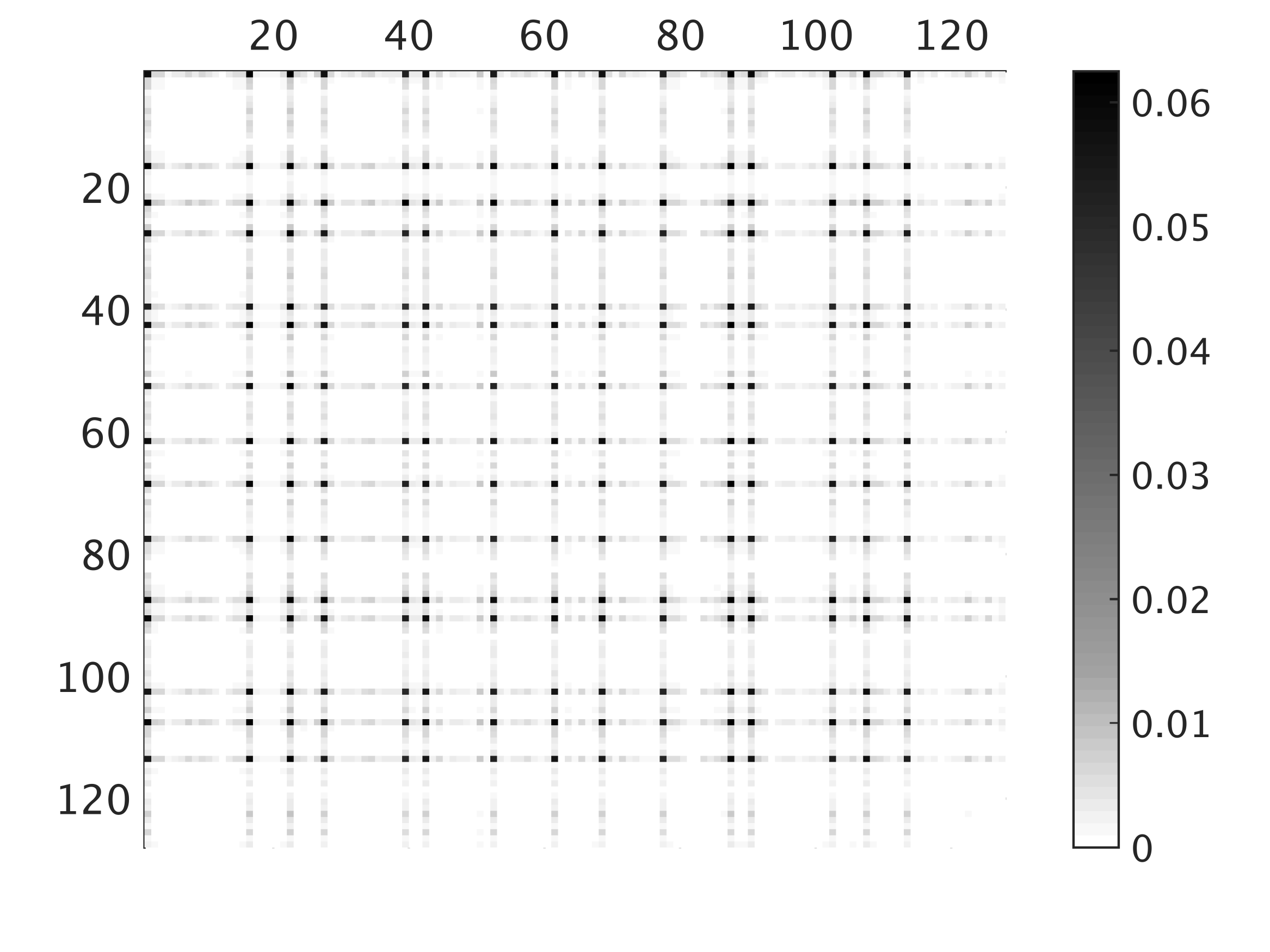}}
\subfigure[ \, Least squares estimate ($F = 0.29$), corresponding to $\mu=0$.]{%
  \includegraphics[width=.65\columnwidth]{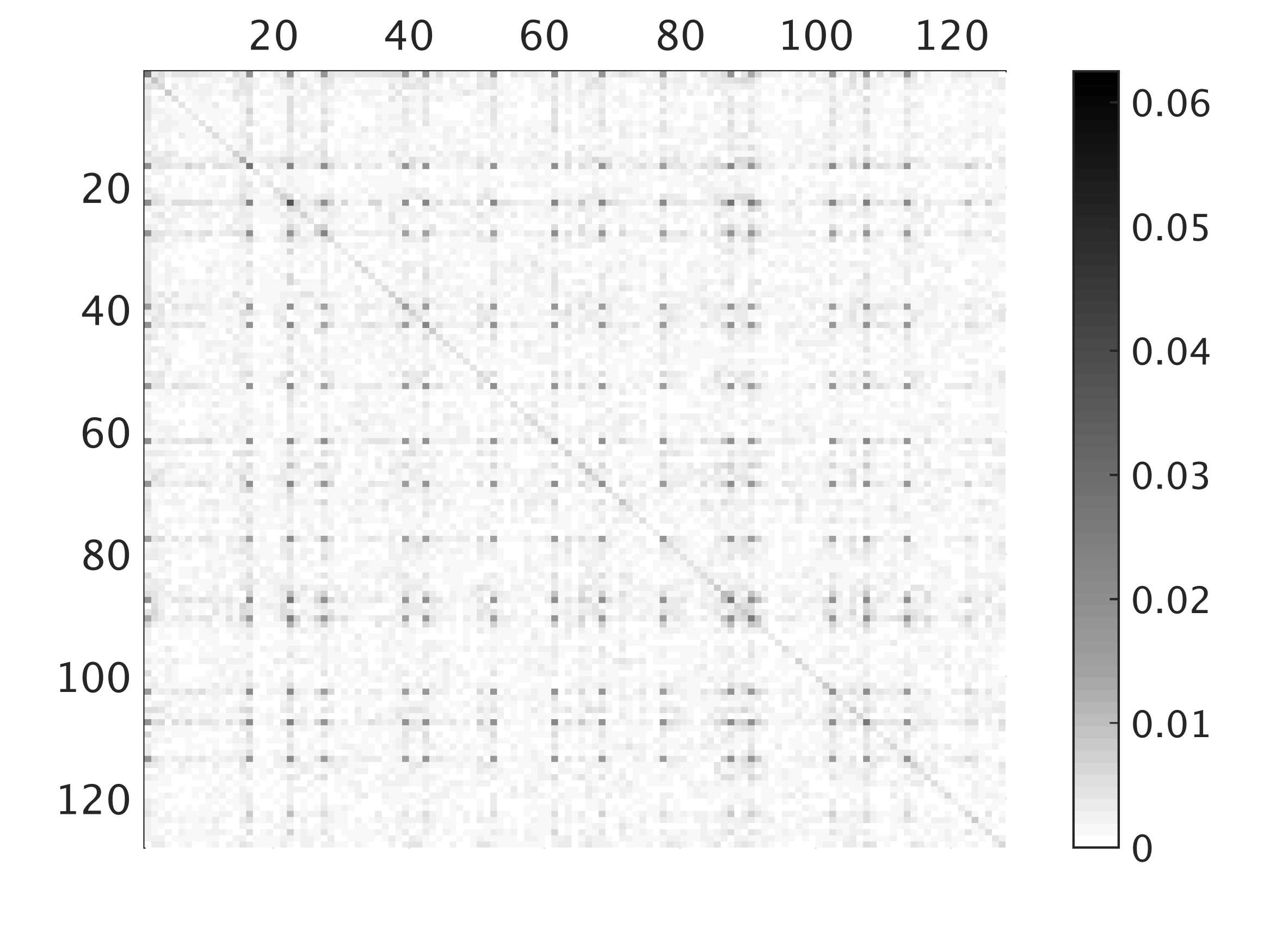}}
\quad
\subfigure[\, Rank $41$ leading subspace projection of the least squares estimate ($F = 0.32$) obtained by our spectral thresholding method.]{%
  \includegraphics[width=.65\columnwidth]{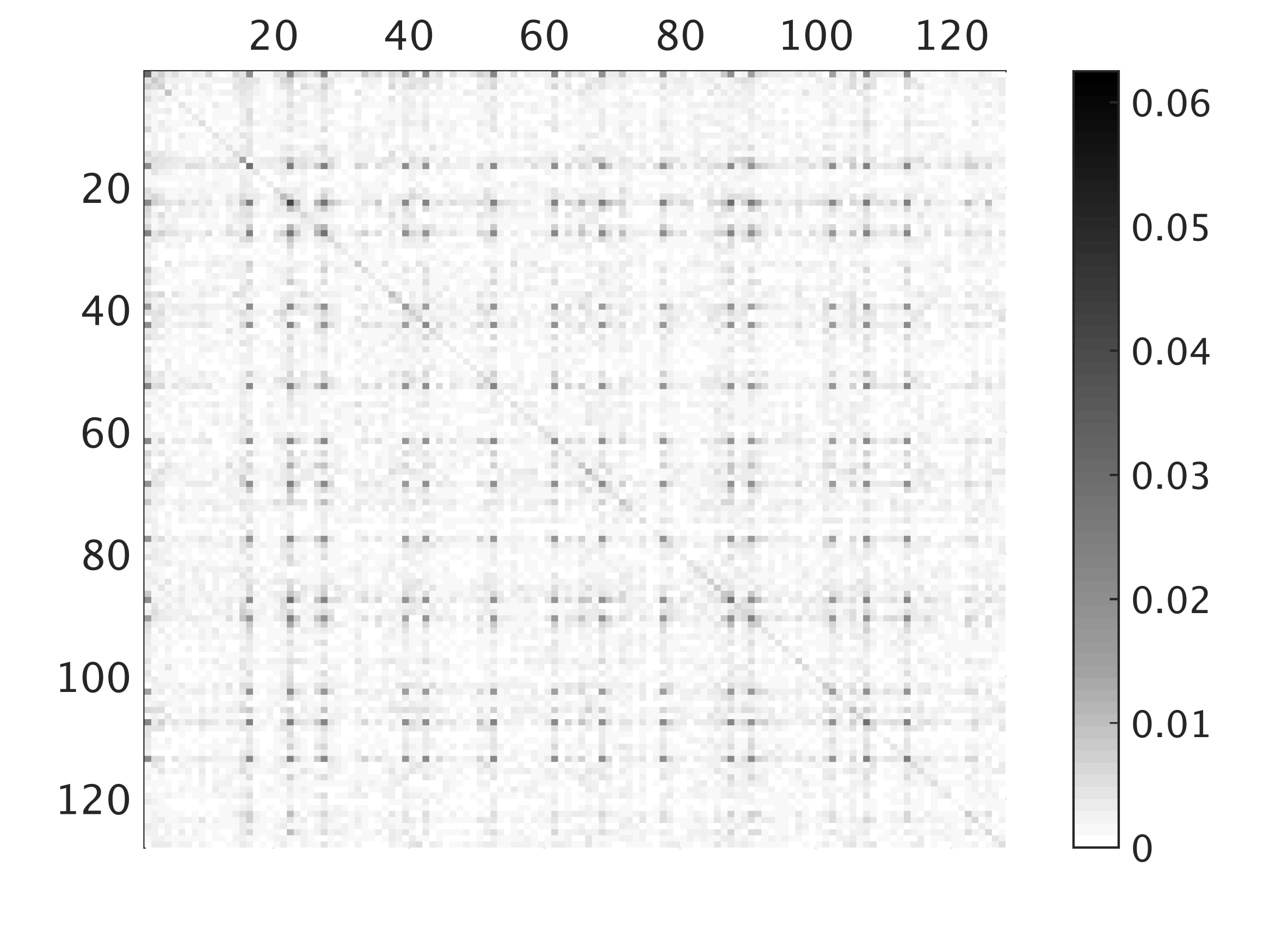}}
\caption{Example of quantum state reconstruction for the logical $(|\bar{0}\rangle+|\bar{1}\rangle)/\sqrt{2}$
state vector. A 2-D plot of the absolute values of the entries of the density matrix is presented.
The same notation as
in fig.\ \ref{Data1} in the main text is used.}
\label{Data3}
\end{figure*}

In this section, we add the results for 2 additional states that were prepared in the laboratory. In figs.~\ref{Data2} and \ref{Data3} we compare the reconstructed states for the anticipated $|\bar{1}\rangle$ and $(|\bar{0}\rangle+|\bar{1}\rangle)/\sqrt{2}$ encoded state vectors, respectively. As before, we can see how the trace norm minimization can deliver an almost pure state while the least squares estimator does not, which is revealed in the computed fidelities with respect to the anticipated states. Also, note that our spectral thresholding method is also applied successfully.

 \end{document}